\documentclass[a4paper,11pt]{article}
\pdfoutput=1
\usepackage{jheppub} 

\usepackage{amsthm}
\usepackage{amssymb}
\usepackage{amsmath} 
\usepackage{color}
\usepackage{array} 
\usepackage{graphicx}

\usepackage{hyperref}

\newcommand{\be}{\begin{equation}}
\newcommand{\ee}{\end{equation}}

\allowdisplaybreaks

\begin{document}
\begin{titlepage}

\begin{center}

\hfill LTH 1129 \\
\vskip 6mm

 {\LARGE\bfseries  A Riemann-Hilbert approach  \\    \vskip 2mm 
 to rotating attractors }
\\[10mm]

\textbf{M.C. C\^amara$^1$, G.L.~Cardoso$^1$, T. Mohaupt$^2$ and S.~Nampuri$^1$}

\vskip 6mm
{\em  $^1$Center for Mathematical Analysis, Geometry and Dynamical Systems,\\
  Department of Mathematics, 
  Instituto Superior T\'ecnico,\\ Universidade de Lisboa,
  Av. Rovisco Pais, 1049-001 Lisboa, Portugal}\\[4mm]
  
{\em $^2$Department of Mathematical Sciences, \\
University of Liverpool, Peach Street, Liverpool L69 7ZL, UK}  

{\tt ccamara@math.tecnico.ulisboa.pt} \;,\;\,
{\tt gcardoso@math.tecnico.ulisboa.pt}\;,\;\,
{\tt  Thomas.Mohaupt@liverpool.ac.uk }\;,\;\,{\tt
  nampuri@gmail.com}\\[10mm]

\end{center}

\vskip .2in
\begin{center} {\bf ABSTRACT } \end{center}
\begin{quotation}\noindent 

\noindent
We construct rotating extremal black hole and 
attractor solutions in gravity theories by solving a
Rie\-mann-Hilbert problem associated with
the Breitenlohner-Maison linear system. By employing a 
vectorial Riemann-Hilbert factorization method 
we explicitly factorize the corresponding monodromy
matrices, which have second order poles in the spectral parameter.
In the underrotating case we identify elements of the Geroch group
which implement Harrison-type transformations which map the 
attractor geometries to interpolating rotating
black hole solutions. The factorization method we use yields an
explicit solution to the linear system, from which we do not only
obtain the spacetime solution, but also an explicit expression for
the master potential encoding the potentials of
the infinitely many  conserved currents which make this sector 
of gravity integrable.

\end{quotation}
\vfill
\today
\end{titlepage}

\tableofcontents

\section{Introduction}

Exploring the space of solutions is an essential part of deepening our
understanding of 
gravity. While a large variety of methods is available, a full
classification of solutions is far out of reach. Studies therefore
concentrate on special classes of solutions which exhibit
symmetries. Black holes, i.e. spacetimes with event horizons, are a
particularly important class of solutions as they provide a laboratory for
testing ideas about quantum gravity. Birkhoff's theorem and its
generalizations provide a complete classification of
stationary axisymmetric black holes in Einstein and Einstein-Maxwell
theory. Among matter coupled gravitational theories, effective supergravity
theories arising from string theory are particularly important, as
they have a potential UV completion. Therefore
their stationary black hole solutions have been studied extensively. 
The subclass of static, spherically symmetric 
extremal black holes exhibits special properties: the near horizon
geometry contains an $AdS_2$-factor, and the attractor mechanism 
\cite{Ferrara:1996dd,Ferrara:1996um,Moore:1998pn}
forces scalar fields to take specific values at the horizon. This effectively
halfs the number of degrees of freedom, and allows to find extremal
black holes by solving first order gradient flow equations. While the
attractor mechanism generalises to rotating, stationary axisymmetric 
extremal black holes \cite{Astefanesei:2006dd}, finding explicit
solutions becomes much harder, as the solutions
depend non-trivially on two variables and one has to solve PDEs
rather than ODEs. Therefore one needs to explore alternative systematic methods for
constructing rotating solutions. One such method was pioneered in the seminal
work of Breitenlohner and Maison \cite{Breitenlohner:1986um}, 
and has since then been explored further
by various authors, including
\cite{Nicolai:1991tt,Katsimpouri:2012ky,Katsimpouri:2013wka,Katsimpouri:2014ara}.
 It is based on the observation that the stationary
axisymmetric sector of four-dimensional gravity is integrable, and
that the problem of solving the Einstein equations can be replaced by
solving a linear system depending on an additional variable, the
`spectral parameter.' Solving the linear system is in turn equivalent
to solving a Riemann-Hilbert (R-H) problem. Apart from
four-dimensional pure gravity, the method can be applied to
matter-coupled and higher-dimensional gravitational theories, subject
to two conditions: (i) solutions must admit as many commuting
isometries as needed for the consistent reduction to a two-dimensional
theory, (ii) the effective two-dimensional theory must be a scalar
sigma model with a symmetric target space, coupled to two-dimensional
gravity. Thus, so far, the method is restricted to actions with no more
than  two derivatives, and
without a cosmological constant or scalar potential.

\subsection{Concepts}

We now sketch the working philosophy and concepts of the R-H
formulation. Firstly, by imposing sufficiently many commuting
isometries the theory is reduced to a two-dimensional theory, which
can have Euclidean or Minkowski signature. Since we are
interested in stationary solutions, we focus on the Euclidean
case. The two-dimensional effective action contains an
Einstein-Hilbert term and a scalar sigma model, whose target space is
assumed to be a symmetric space $G/H$. The essential part of
solving the equations of motion is solving the scalar equation of
motion, which can be shown to be the integrability condition of an
auxiliary linear system, depending on an additional variable, the
spectral parameter $\tau\in \mathbb{C}$ 
\cite{Breitenlohner:1986um}. 
The linear system and the spectral parameter are
related to the extension of the manifest rigid $G$-symmetry of the two-dimensional
theory, to a hidden rigid symmetry under an infinite-dimensional 
group $\tilde{G}$, which is known as the Geroch group 
\cite{Breitenlohner:1986um, Nicolai:1991tt}. 
Quantities depending on the spectral parameter
can be interpreted as elements of $\tilde{G}$ or of its Lie algebra 
$\tilde{\mathfrak{g}}$. Moreover, the linear system can be interpreted
as relating a master current $J$, which
incorporates the infinitely many conserved currents of the theory, to
an associated master potential $X$, by $J=\star dX$. To each solution of
the linear system, one can associate a so-called monodromy matrix
${\cal M}(\omega)$, which depends on another spectral parameter
$\omega \in \mathbb{C}$. 
Due to the coupling to gravity, the spectral parameter $\tau$ is a
function of the two-dimensional spacetime, and is subject to a
differential equation. The other spectral parameter $\omega\in\mathbb{C}$ arises as an
integration constant of this equation and is independent of spacetime. The relation
between $\tau$ and $\omega$ defines an algebraic curve, called the
spectral curve, which has the two-dimensional 
spacetime coordinates $x$ as parameters.
Given the monodromy matrix the solution of the linear system can be
recovered by solving a R-H factorization problem 
${\cal M}(\omega) = M_-(\tau, x) M_+(\tau, x)$, where $M_\pm$ have
certain analyticity properties, to be reviewed in due course.  For our 
purposes it is essential that the factorization is of the specific,
so-called canonical form given above. This implies that the solution 
of the equations of motion can be extracted from $M_-$, while $M_+$
encodes the infinite set of conserved currents.

\subsection{Methodology}

We will restrict ourselves to three gravitational theories: pure
four-dimensonal gravity, four-dimensional Einstein-Maxwell theory and
the four-dimensional Einstein-Maxwell-Dilaton theory obtained by
reducing pure five-dimensional gravity on a circle. Having converted
the equations of motion to a R-H problem, one faces two problems:
proving the existence and possibly uniqueness of a solution, and
obtaining the solution explicitly. Most of the literature is focussed
on existence, and employs factorization algorithms that quickly become
very cumbersome when applied in practice. Moreover
\cite{Breitenlohner:1986um} and the subsequent literature, including 
\cite{Nicolai:1991tt, Katsimpouri:2012ky,Katsimpouri:2013wka,Katsimpouri:2014ara},
impose a particular ansatz which requires the 
spacetime to be asymptotically flat and the monodromy matrix to have
only first order poles in $\omega$. These are severe limitations as they exclude
extremal solutions, and the attractors solutions which are their
near-horizon limits. 

We will present a different method, where an explicit factorization of
the monodromy matrix is obtained by solving auxiliary vectorial R-H
problems by making systematic use of Liouville's theorem of complex
analysis. A main
advantage of our method is that it is completely general, and in particular
does not require that the underlying spacetime solution
is asymptotically flat, or that the monodromy matrix has only first
order poles in $\omega$.  The practicability of the method is
demonstrated by a variety of examples where we obtain the
factorizations of monodromy matrices explicitly, that is by specifying
the factors $M_{\pm }$ as matrices of rational functions, with all parameters
expressed in terms of physical quantities such as mass, angular
momentum, and charges. We have also included the proofs of various
theorems for the sake of clarity and completeness.

We also study the action of the Geroch group
on spacetime solutions,
solutions of the linear system, and the associated monodromy matrices.
Our formulation of the linear system makes the relation
of its variables to the infinitely many conserved currents and associated potentials
transparent. This allows to not only use the finite-dimensional group
$G$ but the full Geroch group $\tilde{G}$ to generate new solutions
from seed solutions.

\subsection{Results}

We present a method for factorizing monodromy matrices, and apply it
to a variety of examples with first and second order poles in
$\omega$, and with flat as well as non-flat asymptotics.
We observe that while non-extremal black
holes have monodromy 
matrices with first order poles associated to them,
extremal black holes and attractor geometries have monodromy 
matrices with second order poles.
This solves the
long-standing problem of constructing extremal black hole solutions
within the R-H formulation.
We remark that the factorization problem
for second order poles is different from the one for first order
poles, and cannot be obtained by taking a limit where first order
poles coalesce. 
This can be understood by drawing an analogy with what happens when performing
an additive decomposition of a rational function into a sum of terms with one pole. A decomposition
involving terms with double poles can never be viewed as a limiting case of another decomposition
involving only terms with simple poles.
We also observe that, when using our universal method in the context of rational monodromy matrices,
factorization problems are actually easier 
to solve for second order poles than
for first order poles. This is how it should be, since solving the
Einstein equations  is easier for extremal than for non-extremal 
solutions.

From the explicit R-H factorizations we perform we obtain the following 
backgrounds which have hitherto been inaccessible:
\begin{enumerate}
\item
Static attractor geometries for
four-dimensional Einstein-Maxwell and five-dimensional pure Einstein 
theory reduced on a circle. The double poles of
the monodromy matrices correspond to the double zeros of the temporal
metric warp factor at the horizon.
\item
Rotating attractor geometries \cite{Astefanesei:2006dd}, which are the
near-horizon limits of extremal rotating black holes. The monodromy matrices
again exhibit double poles.  In the underrotating case the monodromy
matrix can be chosen triangular, which simplifies
factorization. In fact, the monodromy matrix takes the same form as
for the static attractor, the difference being just the value of
certain coefficients. Therefore obtaining underrotating solutions is
as difficult as obtaining non-rotating solutions, and it is manifest
that underrotating extremal solutions, which are distinguished by
 the absence of an ergo-region from overrotating solutions, are 
generated by `spinning up' non-rotating charged extremal solutions. 
In the overrotating case the monodromy matrix has a different form
which indicates that such solutions, which have an ergoregion
 and generalize the Kerr 
solution, form a separate branch in the space of stationary solutions.
Despite that the monodromy matrix cannot be brought to triangular
form, we still achieve to factorize it, and present the explicit
solution for the uncharged subcase.
\end{enumerate}

We also study the operation of the Geroch group on solutions of the
field equations and of the linear system. This allows to obtain new
solutions from a seed solution.
We distinguish two cases:
\begin{enumerate}
\item 
`Constant' Geroch group elements, that is elements of $\tilde{G}$
which do not depend on the spectral parameter $\omega$, and thus
can be interpreted as elements of the finite-dimensional group $G$ 
underlying the symmetric space $G/H$. We identify group elements which
implement Harrison-type transformations that can change the asymptotic
behaviour at infinity.  In particular we show that the Schwarzschild
solution can be mapped to either a non-extremal black hole in $AdS_2$
or to a non-extremal asymptotically flat charged black hole. We also show
that underrotating dyonic attractors can be mapped to underrotating
extremal asymptotically flat black hole solutions. In these cases the 
solution of the R-H factorization problem is preserved by the group
action, so that it is not necessary to solve a new R-H problem.
\item
`Non-constant' Geroch group elements, which depend on the spectral 
parameter $\omega$. These correspond to consistent deformations of 
the monodromy matrix which necessitates to solve a new R-H problem. We
present one explicit example where the static attractor geometry is
deformed into a new solution with the same $AdS_2\times S^2$ asymptotics.
The deformed solution is complicated but can be shown to be completely
regular, except at the origin of the coordinate system, where the
factorization breaks down. 
\end{enumerate}

To summarize, we give a proof of concept that the R-H method can be
applied to the full range of stationary axisymmetric black hole
solutions occuring in gravitational theories with symmetric scalar
target spaces, including dyonic, under- and overrotating extremal
solutions and their attractor geometries. One issue which we study
throughout the paper is the relation between the monodromy matrix and
the associated spacetime solution. Ideally one would like to have a
criterion which tells one in advance which monodromy matrices will
correspond to interesting spacetime solutions. It was already
observed in \cite{Breitenlohner:1986um} that under certain regularity 
assumptions the solution on the axis of rotation determines the
solution completely. Thus one should be able to obtain the full 
monodromy matrix from knowing the solution on the axis, only.
Subsequently it has been argued \cite{Chakrabarty:2014ora}
that there is in fact a rather direct and simple relation, namely 
that the monodromy matrix ${\cal M}(\omega)$, evaluated at real
points, is equal to the matrix representing the spacetime
solution in Weyl coordinates, evaluated on the axis. 
We will refer to this statement as the `substitution rule.' 
While plausible, and supported by
explicit examples, this statement involves a highly delicate limit.
Our explicit factorization method allows us to clearly characterize
the assumptions underlying the substitution rule, namely that specific
factors in the decomposition of the monodromy matrix have a regular
limit on the axis of rotation. We investigate the validity of the
substitution rule in all examples we consider, and encounter
subtleties that have not been noted previously. 
One such observation is that
the parameter space decomposes into subregions, for which the
validity of the substitution rule needs to be verified case by
case. Moreover, while the substitution rule is found to be valid in
almost all cases, there is one region in the parameter space of the
Schwarzschild solution, where the required limit does not exist, so
that no monodromy matrix can be assigned using the substitution 
rule. However, proceeding in
the opposite direction one can factorize the `Schwarzschild' monodromy
matrix in this region of parameter space and actually obtain a spacetime
solution different from the Schwarzschild solution.

\subsection{Outline}

This paper is organised as follows. In section 2 we review sigma
models with symmetric target spaces, relegating specific details about
the parametrisation of the symmetric spaces used in later sections to
Appendix A. In Section 3 we review  R-H problems, prove the necessary
and sufficent conditions for the existence of a canonical
factorization and formulate a lemma based on Liouville's theorem which
we use to carry out factorization. In Section 4 we show that the
field equations, after reduction to two dimensions, can be obtained as
an integrability condition for the Breitenlohner-Maison linear system,
and introduce the spectral curve.  In Section 5 we explain how a monodromy
matrix can be associated to solutions of the linear system, and
carefully analyze the conditions under which the substitution rule is
valid. The subtleties involved in applying the substitution rule are
illustrated using two examples. In Section 6 we show how to obtain
a solution to the field equations by performing the canonical
factorization of a given monodromy matrix. In Section 7 we
review the relation between the integrability of the field equations
and the existence of infinitely many hidden symmetries, and observe
that the solution $X$ of the linear system, which we obtain by solving the
R-H problem, is a master potential for the infinitely many conserved 
currents related to these
symmetries.  Section 8 provides the explicit factorizations of 
static and rotating attractor geometries, and as well of the
Schwarzschild solutions, in all regions of their respective parameter
spaces.  In Sections 9 and 10 we study the action of the Geroch
group and obtain new solutions to the field equations and to the
linear system from seed monodromies. In Section 9 we
employ Geroch group elements which are independent of the spectral
parameter and act as Harrison-like transformations, and give examples
where attractor geometries are mapped to full interpolating black hole
solutions. In Section 10 we use Geroch group elements depending on the
spectral parameter to deform attractor solutions.

\subsection{Future directions}

There are several open problems, which we leave to future research.
This  includes a better
understanding of analyticity properties, in particular the status of the 
substitution rule. Another question is to which extent the R-H
approach can be generalized to theories with a cosmological constant or
a scalar potential, as they occur in gauged supergravity. As discussed
for example in \cite{Klemm:2015uba}, integrability is not necessarily
lost completely. Another interesting extension would be theories with
higher derivative terms. While very involved in general, one might at
first concentrate on attractor geometries in theories where the higher
derivatives are restricted by supersymmetry. Finally, the
infinite-dimensional symmetries that become visible upon reduction to
less than three dimensions are not only symmetries of a subsector of
the solution space, but also hidden symmetries of the full theory. In
the case of eleven-dimensional supergravity, this might hold clues for
formulating the theory non-perturbatively, for example using 
hyperbolic affine Lie algebras \cite{Nicolai:1991tt}, the
infinite-dimensional Lie algebra $E_{11}$ \cite{West:2001as}, 
or tensor hierarchy algebras \cite{Bossard:2017wxl}, which all
can be viewed as extensions of the Lie algebra of the Geroch group. 
It will be interesting to investigate which role  R-H problems might
play in this context.

\section{Sigma models with symmetric target spaces \label{sigmodrev}}

In this section we review the relevant background material
on sigma models with symmetric target spaces. Standard 
references for symmetric spaces are \cite{Gilmore,Helgason},
and for sigma models resulting from dimensional reduction of
gravitational theories \cite{Breitenlohner:1987dg}. 
In appendix \ref{backgroundLie} we give a summary 
of the
explicit realizations of various Lie algebras
found in \cite{Houart:2009ed,Chakrabarty:2014ora}. 

The dimensional reduction 
of gravity coupled to uncharged vector and scalar fields
from four (or higher) to three (or lower) dimensions can 
be brought to the form of a non-linear scalar sigma model
coupled to gravity,
\[
S[\Phi] \propto
\int_N d^p x \, e \, \left( \frac{1}{2} R_g - g^{mn} h_{ab}(\phi) 
\partial_m \phi^a \partial_n \phi^b \right)\;.
\]
Since nothing we will say depends on $p=3$, we will keep
the dimension arbitrary.
In the above,  $e$ denotes the vielbein determinant associated with the
spacetime metric $g$, and $R_g$ is the Ricci scalar. 
The scalar fields $\phi^a(x)$ can be interpreted
as the components of a map
\[
\Phi: (N,g) \rightarrow (M,h) \;,\;\;\;x\mapsto \Phi(x) = (\phi^a(x^m)) \;,
\]
between two semi-Riemannian manifolds, spacetime
$N$ with metric $g=g_{mn} dx^m dx^n$
and the scalar target space $M$ with metric $h=h_{ab} d\phi^a d\phi^b$.
The action is the Dirichlet energy functional for maps between
semi-Riemannian spaces, whose stationary points are harmonic maps.

We will only be interested in the special case where
$M$ is a symmetric space $G/H$, which covers the dimensional
reductions of pure gravity, Einstein-Maxwell theory and 
maximal supergravity theories, as well as many models with symmetric
target spaces arising in string theory compactifications. Let us
then review the relevant ingredients from group theory. Consider
first the special case where $M$ is a simple, real, connected Lie group $G$
with Lie algebra $\mathfrak{g}$. We take $G$ to be given concretely
as a matrix group, so that $\Phi(x)$ is a matrix-valued function on 
spacetime $N$, which depends analytically on fields
$\phi^a(x)$, which provide coordinates on $G$. The natural way to define a
semi-Riemannian metric on $G$ is to pick a possibly indefinite
scalar product on $\mathfrak{g}$, and extend it by the group action
to a left- or right-invariant metric on $G$. If the scalar product
is invariant under the adjoint action of $G$, such a metric
is in fact bi-invariant. One natural choice, which turns out to be 
relevant in our case, is to take a scalar product which is proportional
to the Killing form:
\[
B_{\mathfrak{g}}(Y,Z) \propto \mbox{Tr}(\mbox{ad}(Y) \mbox{ad}(Z)) \;,\;\;\;
Y,Z\in \mathfrak{g} \;.
\]
If $R$ is any irreducible representation of $G$, 
then $\mbox{Tr}(R(Y) R(Z))$ is precisely of this form.
The resulting metric on $G$ can be expressed using the Maurer-Cartan
form. Since $G$ is realized as a matrix group, the left-invariant
Maurer-Cartan form is $\omega_L=g^{-1} dg$, while the right-invariant
Maurer-Cartan form is $\omega_R = dg g^{-1}$. The bi-invariant line
element obtained by extending $B_{\mathfrak{g}}(Y,Z)$ by left or
right group action is
\[
ds^2_G = B_{\mathfrak{g}}(g^{-1}dg, g^{-1} dg)  
= B_{\mathfrak{g}} ( dg g^{-1}, dg g^{-1} ) = 
h_{ab} d\phi^a d\phi^b \;,
\]
where $h_{ab} = B_{\mathfrak{g}} (g^{-1} \partial_a g, g^{-1} \partial_b g)$
are the metric components with respect to the coordinates $\phi^a$
on $G$. 

To obtain the sigma model metric, we replace $g\in G$ by 
the group-valued function $\Phi(x)$, and pull back the metric
from $G$ to spacetime $N$:
\[
B_{\mathfrak{g}}(\Phi^{-1} d\Phi, \Phi^{-1} d\Phi ) = 
h_{ab}(\Phi(x)) \partial_m \phi^a \partial_n \phi^b dx^m dx^n \;,
\]
where we used that $d\Phi=\partial_m \phi^a \partial_a \Phi dx^m$. 
Variation of the action
\[
S \propto \int d^3 x e \left( \frac{1}{2} R_g - g^{mn} 
\mbox{Tr}\left( \Phi^{-1} \partial_m \Phi \Phi^{-1} \partial_n \Phi 
\right) \right)
\]
with respect to $\Phi$ yields the scalar equation of motion
\[
\nabla^m \left( \Phi^{-1} \partial_m \Phi \right) = 0\;,
\]
where $\nabla$ is the Levi-Civita connection for the spacetime metric
$g$. Since $J_m = \Phi^{-1}\partial_m \Phi$ is the conserved current
associated to the invariance of the action under left action of $G$,
the equation of motion can be interpreted as a conservation equation.

We now turn to the case where $M=G/H$ is a symmetric space.
Denoting the Lie algebras of $G$ and $H$ by $\mathfrak{g}$ and
$\mathfrak{h}$, symmetric spaces are locally in one to one
correspondence with symmetric decompositions
\[
\mathfrak{g} = \mathfrak{h} \oplus \mathfrak{p}\;
\]
where $\mathfrak{p}$ is a linear subspace of $\mathfrak{g}$
complementary to $\mathfrak{h}$ such that
\begin{equation}
\label{Sym_Decomp}
[\mathfrak{h}, \mathfrak{h}] \subset \mathfrak{h} \;,\;\;\;
[\mathfrak{h}, \mathfrak{p}] \subset \mathfrak{p} \;,\;\;\;
[\mathfrak{p}, \mathfrak{p}] \subset \mathfrak{h} \;.
\end{equation}
The symmetric space $G/H$ can be identified with $\exp \mathfrak{p}$.
Any symmetric decomposition arises as the eigenspace decomposition 
of an involutive
Lie algebra automorphism $\theta \in \mbox{Aut}(\mathfrak{g})$, 
$\theta^2 = \mbox{Id}_{\mathfrak{g}}$:
\[
\theta(Z) = Z \;,\forall Z \in \mathfrak{h} \;,\;\;\;
\theta(Z) = -Z \;,\forall Z \in \mathfrak{p} \;.
\]
We denote the associated involutive Lie group automorphism by $\Theta$:
\[
\Theta(\exp(Z)) = \exp(\theta(Z)) \;,
\]
which implies
\[
\Theta(h) = h \;,\forall h \in H \;,\;\;\;
\Theta(g) = g^{-1} \;,\forall g \in \exp \mathfrak{p} \simeq G/H \;.
\]
Instead of the automorphisms $\theta$, $\Theta$, one often works with
the corresponding `generalized transposition', which is defined by
\be
Z^\natural = - \theta(Z) \Rightarrow
g^\natural = \Theta(g^{-1})  \;.
\label{gentrans}
\ee
Note that this acts anti-homomorphically, $(g_1 g_2)^\natural= g_2^\natural
g_1^\natural$, and that
\[
h^\natural = h^{-1} \;,\forall h \in H \;,\;\;\;
g^\natural = g \;,\forall g \in \exp \mathfrak{p} \simeq G/H \;.
\]
The generalized transposition has the convenient feature that
$\mathfrak{p}$ and $G/H \simeq \exp \mathfrak{p}$ are invariant
(rather than $\mathfrak{h}$ and $H$).
From the first relation it is clear that for $H\subset O(n)$ the generalized
transposition acts by matrix transposition while for $H \subset U(n)$
it acts by Hermitian conjugation. As we will see in explicit examples,
for pseudo-orthogonal and pseudo-unitary groups this relation is 
modified by the Gram matrix of the bilinear or sesquilinear
form left invariant by the group. 

To obtain a sigma model valued in $G/H$ rather than $G$,
we can restrict the $G$-valued field $\Phi(x)$ to a $\natural$-invariant
field $M(x)$. The corresponding right-invariant Maurer-Cartan
form is $A=dM M^{-1}$, and by pulling back the resulting metric
on $G/H \simeq \exp \mathfrak{p}$ to $N$ we obtain\footnote{Here 
$B_{\mathfrak{g}}$ is understood to be restricted to $\mathfrak{p}$.
By bi-invariance of the underlying metric on $G$, one obtains the
same result when using the left-invariant Maurer-Cartan form. This
is also verified explicitly using that $B_{\mathfrak{g}}$ is proportional
to the trace evaluated in an irreducible representation.}
\begin{equation}
\label{AA}
B_{\mathfrak{g}} (A_m, A_n ) dx^m dx^n  =
B_{\mathfrak{g}} (\partial_m M M^{-1}, \partial_n M M^{-1} ) dx^m dx^n \;. 
\end{equation}

Another way to describe $G/H$ is as an orbit space by identifying
$g\simeq h g$ for all $h\in H$, $g\in G$. Note that we take $H$ to act
from the left, but write the resulting coset space as $G/H$ 
rather than the more accurate $H\backslash G$. A sigma model
with target space $G/H$ is obtained from a sigma model
with target space $G$ by gauging $H$, that is by identifying
$\Phi(x) \simeq h(x) \Phi(x)$ and introducing $H$-covariant
derivatives
\begin{equation}
\label{DPhi}
D_m \Phi = \partial_m \Phi - Q_m \Phi \;,
\end{equation}
where $Q_m$ is a $\mathfrak{h}$-valued gauge field. 
One way to proceed is to fix a gauge and to describe elements
of $G/H$ by choosing one element of $G$ from each orbit as
representative. We will denote the fields corresponding
to such representatives by $V(x)$. One common choice is to 
represent elements of $G/H$ by triangular matrices. This is
called the triangular gauge, or Borel gauge. If $G$ is non-compact
and $H$ a maximal compact subgroup, then such a gauge choice
can be made globally, by using the Iwasawa decomposition 
$G=HL$, where $L$ is a triangular subgroup. In the special case
where $G$ is the normal real form (or split real form, that is
the real form with a maximal number of non-compact generators)
the subgroup $L$ is a maximal triangular subgroup and called
the Iwasawa subgroup. For other real forms, $L$ can be taken to
be the standard Borel subgroup (associated with the positive
roots) together with the non-compact Cartan generators \cite{Houart:2009ed}. 
If $H$ is a non-compact maximal subgroup of $G$, then the 
Iwasawa decomposition does not exist globally. However, one might
still be able to find a triangular subgroup acting with 
open orbit on $G/H$, so that $L$ can be identified with an open
part of $G/H$.\footnote{An explicit example is provided by the symmetric space 
$G_{2(2)}/(SL(2,\mathbb{R})\times SL(2,\mathbb{R}))$, which 
occurs in the dimensional reduction of minimal five-dimensional
supergravity to three dimensions, see \cite{Cortes:2014jha}.} 

Since the group $G$ acts on $G/H$, fields $V(x)$ are subject
to both local $H$-transformations, which we take to act from 
the left, and rigid $G$-transformations, which then act from
the right:
\[
H\times G \;: \;\;\;V(x) \mapsto h(x) V(x) g^{-1} \;.
\]
Rigid $G$-transformations will in general not preserve the gauge
imposed to gauge-fix  the $H$-transformations. Therefore $G$-transformations
need to be accompanied by an $H$-transformation which restores the
gauge:
\[
G\;:\;\;\;V(x) \mapsto h(x,g) V(x) g^{-1} \;.
\]
To obtain an expression for the metric on $G/H$, we first note that the
right-invariant Maurer Cartan form $dV V^{-1}$ transforms under
$H\times G$ as
\[
dV V^{-1} \mapsto d(hVg^{-1}) (hVg^{-1})^{-1} = h dV V^{-1} h^{-1} + dh h^{-1} \;.
\]
Thus, while by construction $G$-invariant, it transforms under $H$
gauge transformations. Since $dV V^{-1}$ is $\mathfrak{g}$-valued
we can use the projector $\Pi: \mathfrak{g} \rightarrow \mathfrak{p}$,
which is orthogonal with respect to the Killing form, to decompose
it:
\be
dV V^{-1} = P + Q \;,\;\;\;P = \Pi(dV V^{-1}) \in \mathfrak{p} \;\;\;,\;\;\;
Q = (\mbox{Id} - \Pi) (dV V^{-1}) \in \mathfrak{h} \;.
\label{VPQdecomp}
\ee
The forms $P$, $Q$ are $G$-invariant, and under local $H$ transformations
$P$ transforms in the adjoint representation, while $Q$ transforms like
a $H$-connection:
\[
P \mapsto h P h^{-1} \;,\;\;\;Q \mapsto h Q h^{-1} + dh h^{-1}\;.
\]
Under the generalized transposition $P^\natural=P \in \mathfrak{p}$
and $Q^\natural = - Q \in \mathfrak{h}$, so that we can write
the decomposition explicitly as
\[
P = \frac{1}{2} \left( dV V^{-1} + (dV V^{-1})^\natural \right) \;, \;\;\;
Q = \frac{1}{2} \left( dV V^{-1} - (dV V^{-1})^\natural \right) \;.
\]
For later use also note that $dV V^{-1} = P+Q$ implies
$(dV V^{-1})^\natural = P-Q$. 
Since $P$ is the projection of the right-invariant Maurer-Cartan
form to $\exp \mathfrak{p} \simeq G/H$, the resulting sigma model
metric is
\begin{equation}
\label{PP}
ds^2 = B_{\mathfrak{g}}(P_m, P_n) dx^m dx^n  \;,
\end{equation}
where $P_m$ is the pull-back to $N$ of the projection $P$ of the 
right-invariant Maurer-Cartan form to $\mathfrak{p}$. 

Alternatively, 
since $Q$ can be interpreted as a $H$-connection one-form, we can use it
to gauge a $G$ sigma model to obtain a sigma
model with target space $G/H$. The corresponding gauge connection
$Q_m$ is obtained by pulling back the connection form $Q$ to $N$.
The $H$-covariant derivative of a $G$-valued field is:
\[
D_m V = \partial_m V - Q_m V = \frac{1}{2} \left( \partial_m V 
+ (\partial_m V V^{-1})^\natural V \right) \;,
\]
where compared to (\ref{DPhi}) we have written $V(x)$ instead of
$\Phi(x)$, to indicate that this field is a coset representative. 
Replacing partial by covariant derivatives in a $G$-valued sigma
model gives:
\begin{equation}
\label{DVV-1}
ds^2 = B_{\mathfrak{g}}(D_m V V^{-1}, D_n V V^{-1} ) dx^m dx^n \;.
\end{equation}
Since $D_m V V^{-1}= P_m$ the expressions (\ref{PP}) and 
(\ref{DVV-1}) define the same $G/H$ sigma model metric. 

Finally, let us compare the description of $G/H$ in terms of
gauge-fixed representatives $V(x)$ with the description 
in terms of $\natural$-symmetric representatives $M(x)$. Given 
any choice of gauge, we can map $V(x)$ one-to-one to an element
$M(x)\in \exp \mathfrak{p}$ by
\be
V(x) \mapsto M(x) = V^\natural(x) V(x) \;. 
\label{defMV}
\ee
Note that since $h^\natural(x) = h^{-1}(x)$ for $h\in H$, the field
$M(x)$ is manifestly invariant under $H$-gauge transformations, while
under rigid $G$-transformations
\[
M(x) \mapsto g^{\natural, -1} M(x) g^{-1} \;.
\]
To compare the expressions for the metric, we write $A=dM M^{-1}$ 
in terms of $V$, and observe that it is related to $P$:
\[
A = V^\natural (dV V^{-1}) V^{\natural,-1} + 
dV^\natural V^{\natural, -1} \Rightarrow 
P = \frac{1}{2} V^{\natural, -1} A V^\natural \;.
\]
Therefore
\[
B_{\mathfrak{g}}(\partial_m M M^{-1}, \partial_n M M^{-1} ) dx^m dx^n
= 4 B_{\mathfrak{g}}(P_m, P_n) dx^m dx^n 
\]
and the sigma model metric (\ref{AA}) agrees with 
(\ref{PP}) and (\ref{DVV-1}) up to a positive factor. 

Using left-invariant instead of right-invariant Maurer-Cartan 
forms results in the same metric on $G/H$, since the underlying
metric on $G$ is bi-invariant.\footnote{Some formulae change of course.
In particular $A' = M^{-1} dM$ is related to the projection
$P$ of $dV V^{-1} $ to $\mathfrak{p}$ by $2P = V A' V^{-1}$. 
\label{PpAa}}  

Using the $\natural$-symmetric representation, the action is 
\begin{equation}
\label{ActionSigma3d}
S \propto \int d^px \, e \, \left( \frac{1}{2} R_g - g^{mn} \mbox{Tr}
\left(  \partial_m M M^{-1} \partial_n M M^{-1} \right) \right) \;.
\end{equation}
Variation of $M$ gives the equation of motion\footnote{Since
we use the $H$-invariant representative $M(x)$, the covariant
derivative is the Levi-Civita connection with respect to 
spacetime. When using $G$-valued representatives $V(x)$, derivatives
also involve the $H$-connection $Q_m$.}
\begin{equation}
\label{ConserveM}
\nabla^m ( \partial_m M M^{-1}) = \nabla^m A_m = 0 \;,
\end{equation}
which can be viewed as current conservation equations corresponding
to the invariance of the action under the rigid $G$ action on 
$G/H$. Note that there are only $\dim G - \dim H$ independent
currents, since $M$ only depends on as many parameters. This must 
of course be the case 
since for every point in $G/H$ the isotropy group is isomorphic to $H$. 

\section{Factorization and Riemann-Hilbert problems \label{intromath} }

In this section we briefly describe the class of Riemann-Hilbert factorization problems that we will study
in this paper. We explain the notion of canonical factorization, summarize necessary and sufficient conditions 
for its existence, and we describe the vectorial Riemann-Hilbert problem that we will be solving throughout 
in order to obtain {\sl explicit} solutions to 
matrix factorization problems.

Riemann-Hilbert (R-H) problems have a variety of applications in mathematics and physics.  
One important application, with which this paper will be concerned, is the application to integrable systems that arise when
dimensionally reducing the gravitational field equations in $D$ spacetime dimensions to field equations in two dimensions 
by means of $D-2$ commuting isometries (see \cite{Alekseev:2010mx} for a review of integrable reductions of Einstein's field equations).
In this context, we will be interested in the following class of R-H factorization problems.

Consider a closed simple contour $\Gamma$ in the complex plane, which here we assume to be the unit circle centred around the origin, dividing the complex plane $\mathbb{C}$ into two regions
denoted by $D_{\pm}$. 
We take $D_+ (D_-)$ to denote the interior (exterior) region of the unit circle.  

Now consider an $n\times n$ matrix function 
$\cal M$, defined on the contour $\Gamma$,
such that
both $\cal M$ and ${\cal M}^{-1}$ are continuous on $\Gamma$.
We seek matrices $M_{\pm}$ satisfying the following conditions:
\begin{itemize}
\item $M_+$ is analytic and bounded in $D_+$, and its inverse $M^{-1}_+$ is also analytic and bounded in $D_+$;

\item $M_-$ is analytic and bounded in $D_-$, and its inverse  $M^{-1}_-$ is also analytic and bounded in $D_-$;

\item as one approaches $\Gamma$ in a non-tangential manner, $M_{\pm}$ and $M^{-1}_{\pm}$ tend to limiting matrices which we also denote by $M_{\pm}$ and  $M^{-1}_{\pm}$,
respectively;

\item on the contour $\Gamma$, $\cal M$ has the Birkhoff decomposition (also called Wiener-Hopf factorization)
\begin{equation}
{\cal M} = M_- \, D \, M_+ \;, \label{RHjump}
\end{equation}
where $D$ is a diagonal matrix of the form $diag (\tau^{k_j})_{j=1,2,...n}$, where $k_j\in \mathbb{Z}$ and $\tau 
\in \Gamma$.

\end{itemize}

The above problem, namely seeking matrices $M_{\pm}$ with the above properties and satisfying the jump condition ${\cal M} \, M_+^{-1}= M_- \, D $ across $\Gamma$, is an example of a matricial Riemann-Hilbert factorization problem.

The integers $k_j$ appearing as exponents of the elements of the diagonal matrix $D$ are called the partial indices of the Birkhoff factorization, and $k=k_1+k_2+...+k_n$ is its total index. When $k_1=k_2=...=k_n=0$, i.e., 
when the matrix $D$ that appears in the Birkhoff decomposition is the unit matrix, $D = \mathbb{I}$, the factorization is called  canonical.

A Birkhoff decomposition is guaranteed to exist under rather general circumstances, namely for all invertible matrix functions whose elements are H\"older continuous functions on $\Gamma$ \cite{MP}.
A continuous  function $f : \Gamma \rightarrow \mathbb{C}$  is H\"older continuous with exponent $\alpha\in ]0,1[$ if there exists a  non-negative constant $\Lambda >0$ such that
\begin{equation}
| f(x) - f(y)|  \leq \Lambda \, |x-y|^{\alpha}  \;\;\; \forall \; x, y  \in \Gamma \;.
\end{equation}
The algebra of all such functions is denoted by $C_\alpha (\Gamma)$, or simply $C_\alpha $. Its subalgebras consisting of functions which admit a bounded analytic extension to $D_\pm$ are denoted by $C_\alpha^\pm$, respectively. For $M_\pm \in C_\alpha^\pm$ the corresponding analytic extensions are uniquely
determined by their values on the unit circle in terms of singular integrals with Cauchy kernel \cite{MP} :
\begin{equation}
\pm 2 \pi i \, M_\pm(z) =  \oint_{\Gamma} \frac{M_\pm(\tau)}{\tau - z} d\tau \;\;\;,\;\;\; \forall z \in D_\pm \;.
\end{equation}
Since ${\cal M}^{\pm 1}\in (C_\alpha)^{n\times n}$, 
also the factors $M_\pm$, as well as their inverses, are in $(C_\alpha^\pm)^{n\times n}$ \cite{MP}.
However, the factorization is not canonical, in general. In the case of triangular matrices the
factorization is canonical if the diagonal elements admit a canonical (scalar) factorization
\cite{LS}.
For the general case, 
necessary and sufficient conditions for the existence of a canonical factorization of ${\cal M} \in (C_\alpha)^{n \times n}$ are the following.

\vskip 3mm

{\sl Theorem:}  ${\cal M} \in (C_\alpha)^{n \times n}$  admits a canonical factorization ${\cal M} = M_- \, M_+$ if and only if $\det {\cal M}$ admits a (scalar) canonical factorization $\det {\cal M}=\gamma_-\gamma_+$ with $\gamma_-^{\pm 1}\in C_\alpha^-, \gamma_+^{\pm 1}\in C_\alpha^+$ and
the vectorial Riemann-Hilbert problem 
\begin{equation}\label{A}
{\cal M} \phi_+ = \phi_-
\end{equation}
on $\Gamma$, with $\phi_\pm\in (C_\alpha^\pm)^{n\times 1} $ satisfying the boundedness condition $\phi_- ( \infty) = 0$,  has only the trivial
solution.

\begin{proof}
If ${\cal M} = M_- \, M_+$ is a canonical factorization, then $\det {\cal M}=\det M_-\det M_+$ is also a canonical factorization. On the other hand, for $\phi_{\pm} \in 
(C_\alpha^{\pm})^{n\times 1}$,
\begin{equation}
{\cal M} \phi_+ = \phi_- \Leftrightarrow M_-M_+\phi_+=\phi_- \Leftrightarrow M_+\phi_+=M_-^{-1}\phi_-\,.
\end{equation}
Since the left-hand side of the last equality belongs to $(C_\alpha^+)^{n\times 1}$ and the right-hand side belongs to $(C_\alpha^-)^{n\times 1}$, by Liouville's theorem we conclude that both sides are equal to a constant which, taking into account the condition at $\infty$, is zero.

Conversely, assume that $\det {\cal M}=\gamma_-\gamma_+$, with $\gamma_-^{\pm 1}\in C_\alpha^-, \gamma_+^{\pm 1}\in C_\alpha^+$, and \eqref{A} admits only the trivial solution. Since ${\cal M}\in (C_\alpha)^{n\times n}$, there exists a Birkhoff factorization ${\cal M} = M_- \, D \, M_+ $ as in \eqref{RHjump}, with $M_+^{\pm 1}\in (C_\alpha^+)^{n\times n}\,,\,M_-^{\pm 1}\in (C_\alpha^-)^{n\times n}$ (see \cite{MP}, for instance). Thus we have, taking into account that $D=diag (\tau^{k_j})_{j=1,2,...n}$,
\[
\gamma_-\gamma_+=\det M_-\,\tau^k\,\det M_+\;\;\;\;\;\;(k=k_1+k_2+...k_n) \;,
\]
which is equivalent to
\begin{equation}\label{B}
\gamma_- \det M_-^{-1}=\tau^k\gamma_+^{-1}\det M_+\,.
\end{equation}
Let us now show that $k=0$. Indeed, if $k>0$ we see that both sides of \eqref{B} represent a constant, which must be zero since the right-hand side vanishes for $\tau=0$. As the left-hand side is invertible, 
we conclude that this is impossible and $k$ cannot be positive. On the other hand, if $k<0$ then by the lemma below we conclude that both sides of \eqref{B} are equal to $\frac{P}{\tau^{|k|}} $, where $P$ is a polynomial. So we have
\[
\gamma_+^{-1}\det M_+=P\,\,\,,\,\,\,\\\gamma_- \det M_-^{-1}=\frac{P}{\tau^{|k|}}\,.
\] 
From the first equality we conclude that $P$ cannot have zeroes in $D_+\cup\Gamma$, while from the second equality we see that $P$ cannot have zeroes in $D_-$ either; therefore $P$ is a constant and it follows that $\frac{P}{\tau^{|k|}}$ has a zero at $\infty$, which is impossible because $\gamma_-^{-1}\det M_-$ is invertible with bounded inverse.

Now, since $k=k_1+k_2+...k_n=0$, we have either $k_1=k_2=...=k_n=0$ or there is some negative $k_j$. Suppose that $k_1<0$. Taking $\phi_+$ as the first column of $M_+^{-1}$ and $\phi_-$ as the first column of $M_-$ multiplied by $\tau^{k_1}$, we have
\[
M_+\phi_+=[1\,0\,...\,0]^T\;\;,\;\;M_-^{-1}\phi_-=[\tau^{k_1}\,0\,...\,0]^T
\]
and therefore
\[
DM_+\phi_+=[\tau^{k_1}\,0\,...\,0]^T=M_-^{-1}\phi_-\,.
\]
So we would have a non-trivial solution to \eqref{A}, and we conclude that we must have  $k_1=k_2=...=k_n=0$.
\end{proof}

We note that the condition $\phi_- ( \infty) = 0$ entering in theorem \eqref{A}
can be substituted by the condition $\phi_+ ( 0) = 0$.

In the proof we used the following lemma, which will be used repeatedly. It is a simple generalisation of Liouville's theorem, to which it can be reduced by cross multiplication with suitable rational functions.

\vskip 3mm

{\sl Lemma:}
Let $\phi_1=r_1 \phi_+\,,\,\phi_2=r_2 \phi_-$ where $r_1\,,\,r_2$ are rational functions, bounded on $\Gamma$, and $\phi_+\,,\,\phi_-\in C_\alpha^\pm$ do not vanish at any of the poles of $r_1$  in $D_+$, $r_2$ in $D_-$, respectively. If $\phi_1=\phi_2$ on $\Gamma$, then both $\phi_1$ and $\phi_2$ are equal to a rational function whose poles are the poles of $r_1$  in $D_+$ and $r_2$ in $D_-$ (including $\infty$), counting their multiplicity.

\vskip 3mm

It is easy to see that, if a canonical factorization of ${\cal M}$ does exist, then it is unique up to a constant matrix factor. This freedom of the constant matrix can be removed by imposing a normalization condition, as follows.
 Given two canonical factorizations, 
\begin{equation}
{\cal M} = M_- M_+ = {\tilde M}_- {\tilde M}_+ \;,
\end{equation}
we have $ M_+ {\tilde M}_+^{-1} = M_-^{-1} {\tilde M}_-  \;.$
Since the matrices $M_+ {\tilde M}_+^{-1}$ and $M_-^{-1} {\tilde M}_- $ are bounded and analytic in $D_{\pm}$, respectively, and equal to one another
on $\Gamma$, they define an entire matrix function when taken together. Hence, by Liouville's theorem, they must be constant, 
\begin{equation}
M_+ {\tilde M}_+^{-1} = M_-^{-1} {\tilde M}_-  = K  \;.
\label{MpmK}
\end{equation}
Since the factors on the left hand side are invertible, so is $K$. It then follows from \eqref{MpmK}, 
\begin{equation}
M_- =  {\tilde M}_- K^{-1} \;\;\;,\;\;\; {M}_+ = K {\tilde  M}_+ \;.
\end{equation}
Choosing $K = {\tilde M}^{-1}_+ (0)$, we obtain a matrix ${M}_+$ that equals the identity at the origin $0 \in \mathbb{C}$,
$ {M}_+ (0) = \mathbb{I}$. This results in the canonical factorization 
\begin{equation}\label{F}
{\cal M} = M_- M_+  \;\;\;,\;\;\; M_+ (0)  = \mathbb{I} \:.
\end{equation}
In the following, we will use the notation $M_+ = X$ in order to stress that the factor $M_+$ is normalized to
$M_+ (0)  = \mathbb{I}$. We accordingly write the normalized canonical factorization as
\begin{equation}
{\cal M} = M_- X  \;.
\end{equation}

Note that while $M_{\pm}$ are defined in a certain region of the complex plane, 
the jump condition \eqref{F} is only defined on the contour $\Gamma$. It may not make any sense for $z \notin \Gamma$.

The matrix valued functions we will encounter later in this paper take values in a matrix group, which is naturally equipped with a `generalized transposition.' This operation acts anti-homomorphically, and maps $(C_\alpha^\pm)^{n\times n}$ to $(C_\alpha^\mp)^{n\times n}$.
 We will be concerned with the canonical factorization of matrices ${\cal M}$ which are invariant under this operation composed with the map $\tau \to  - 1/\tau$, i.e.,
\begin{equation}
{\cal M}^{\natural}(-1/\tau) = {\cal M}(\tau) \;.
\label{calMnatM}
\end{equation}
This results in
\begin{equation}
X^{\natural} (-1/\tau) \, M_-^{\natural} (-1/\tau) = M_-(\tau) \, X(\tau) \;\;\;,\;\;\; \tau \in \Gamma \;,
\end{equation}
and hence
\begin{equation}
M_-^{\natural} (-1/\tau)\, X^{-1}(\tau)=( X^{\natural})^{-1} (-1/\tau) \, M_-(\tau) \,  \;\;\;,\;\;\; \tau \in \Gamma \;.
\end{equation}
Since the left hand side of this equality represents a matrix in $(C_\alpha^+)^{n\times n}$ while the right hand side represents a matrix in $(C_\alpha^-)^{n\times n}$, it follows that both are equal to a constant matrix $M$. Therefore 
\begin{equation}\label{MXM}
M_-(\tau)=X^{\natural} (-1/\tau)\,M \;,
\end{equation}
where the equality holds for all $\tau\in D_-\cup\Gamma$ and the matrix $M$ is {\sl independent} of $\tau$ and satisfies
\begin{equation}
M^{\natural} = M \;.
\label{MnatM}
\end{equation}
Moreover, using $X(0) = \mathbb{I}$, we infer from \eqref{MXM} that
\begin{equation}\label {E}
M_- (\infty) = M \:.
\end{equation}
Thus we have on $\Gamma$
\begin{equation}\label {Fa}
{\cal M}(\tau) = X^{\natural} (-1/\tau) \, M  \, X(\tau) \;\;\;,\;\;\;\;\;\; M^{\natural} = M \;.
\end{equation}

In this paper, we will be interested in {\sl explicit} factorizations of the form \eqref{Fa} for matrices 
\begin{equation}\label{curve}
{\cal M}(\omega)\;\;\; {\rm with} \,\;\;\; \omega=\frac{1}{\tau}\,[v \, \tau  + \tfrac12 \, \rho \,  (1 - \tau^2) ]\;.
\end{equation}
Here $(v, \rho) \in \mathbb{R}^2$ are real parameters from the point of view of the
factorization. Below we will identify these parameters with so-called Weyl coordinates in spacetime.
  Locally, the algebraic curve relation relating $\omega$ and $\tau$ in \eqref{curve} expresses $\tau$ in terms
 of $\omega$ and the Weyl coordinates as (assuming $\rho \neq 0$)
\begin{equation}
\tau(\omega, v, \rho) = \frac{1}{\rho} \left( v -\omega  \pm \sqrt{\rho^2 + (v - \omega)^2} \right) \;.
\label{tauw}
\end{equation}
In view of the dependence on the Weyl coordinates, $\tau$ is often called the position dependent spectral parameter.
Thus,
the matrices $M_{\pm}$ in \eqref{F} (and hence $X$) will depend
on both $\tau$ and the Weyl coordinates, so that in effect,
\begin{equation}
{\cal M} (\omega) = X^{\natural} (-1/\tau, v, \rho) \, M(v, \rho)  \, X(\tau, v, \rho) \;\;\;,\;\;\; \tau \in \Gamma \;
\label{MXMX}
\end{equation}
for $\omega=\frac{1}{\tau}\,[v \, \tau  + \tfrac12 \, \rho \,  (1 - \tau^2) ]$. Note that although $M$ does not depend on $\tau$, it may depend on the Weyl coordinates.

The factorization \eqref{F},
 if it exists, can be obtained by solving $n$ vectorial Riemann-Hilbert problems of the form
\begin{equation}\label {vect}
{\cal M} (\Phi_k)_+= (\Phi_k)_- \;\;\;,\;\;\;(\Phi_k)_+(0)=(\delta_{1,k},\delta_{2,k},...,\delta_{n,k})
\;\;\;,\;\;\; k = 1, \dots, n \;,
\end{equation}
where $(\Phi_k)_\pm$ is the $k$th column of $M_+^{-1}$ and of $M_-$, respectively.
This will be illustrated later in the paper.


\section{Reducing to two dimensions: the Lax pair \label{sec-Lax} }

In this section, we briefly review the reformulation of the dimensionally reduced gravity/matter field equations in terms of a Lax pair, called
Breitenlohner-Maison linear system, that makes use of the algebraic curve
\eqref{curve}. We follow \cite{Breitenlohner:1986um,Schwarz:1995af,Lu:2007jc}.

We consider the reduction of four-dimensional gravitational theories at the two-derivative level (and in the absence of a cosmological constant) down to two dimensions. We do this in a two-step procedure.
First we reduce to three dimensions along an isometry of the 
four-dimensional metric $G_{MN}$, 
\begin{eqnarray}
ds^2_4 &=& G_{MN} dx^M dx^N = s \, \Delta (dy + B_m dx^m)^2 + \Delta^{-1} ds^2_3 \;, \nonumber\\
ds^2_3 &=& g_{mn} dx^m dx^n \;,
\label{metric43}
\end{eqnarray}
where $y$ denotes the direction over which we are reducing.
The reduction may be space-like ($s=1$) or time-like ($s=-1$). 
In this paper we consider the case $s = -1$.
We dualize one-forms in three dimensions to scalars. Next, we reduce down to two dimensions along a second
isometry, 
\begin{equation}
ds_3^2 = e^{\psi} \, ds_2^2 + \rho^2 \, d\phi^2 \;,
\label{line32}
\end{equation}
where $\phi$ denotes the direction along which we are reducing.  Here, $\rho$ and $\psi$ are functions of the coordinates in two dimensions. In the following, for convenience, we use complex coordinates $(z, \bar z)$ in two dimensions which
we collectively denote by $x \equiv (z, \bar z)$. 
Later we will specialize to Weyl coordinates, which will be denoted by  $x \equiv (\rho, v)$.

The resulting equations of motion in two dimensions take the form 
\begin{eqnarray}
d \left( \rho \star A \right) &=&0 \;,
\label{eqrA} \\
d \star d \rho &=&0 \;,
\label{eqrho} 
\end{eqnarray}
where the one-form $A$
equals\footnote{We denoted this one-form by $A'$ in footnote \ref{PpAa}.
In the following, we will simply denote it by $A$.}
\begin{eqnarray}
A = M^{-1} \, d M \;,
\end{eqnarray}
with $M$ defined in \eqref{defMV}.
Thus, it satisfies
\begin{equation}
d A + A \wedge A = 0 \;.
\label{dAAA}
\end{equation}
The operation $\star$ denotes the Hodge dual of the one-form, which in two dimensions yields a one-form.
We have
\begin{equation}
(\star)^2 = - {\rm id} \;.
\end{equation}
In particular,
\begin{equation}
 \star d z = -i  d z \;\;\;,\;\;\;  \star d \bar z = i  d \bar z \;.
 \label{funcstdz}
 \end{equation}

The equations of motion \eqref{eqrA} and \eqref{eqrho} have to be suplemented by the equations of motion for the
conformal factor $\psi$.
Given a solution to \eqref{eqrA} and \eqref{eqrho}, $\psi$ is then determined by integration \cite{Schwarz:1995af,Lu:2007jc}.  

Locally, the general solution to the equation of motion \eqref{eqrho} is
\be
\rho(z, \bar z) = f(z) + \overline{ f(z)} \;,
\label{rhoffb}
\ee
where $f$ is an analytic function.
Thus, the non-trivial step in solving the equations of motion consists in
determining an $A$ of the form $A= M^{-1} d M$ that satisfies \eqref{eqrA}.
This is a non-linear differential equation for $M$ that will be solved
by means of a Lax pair and an associated Riemann-Hilbert matrix factorization problem.

To solve the equation of motion \eqref{eqrA} for $A = M^{-1} dM$, we 
introduce an auxiliary
system of equations that are linear, and whose solvability implies the equation of motion \eqref{eqrA}  of the reduced system.
Such a linear system is called a Lax pair. The linear system depends on an additional complex
parameter $\tau$, called the spectral parameter. Thus, resorting to a Lax pair introduces a dependence on
a spectral parameter, which is then subsequently used to reformulate the problem of finding solutions to the 
two-dimensional equations of motion 
as a matrix factorization problem 
in the complex $\tau$-plane (more generally, on a Riemann surface of genus $g$).

For the gravitational system at hand, 
the Lax pair is given by \cite{Schwarz:1995af,Lu:2007jc}
\begin{equation}
\tau \, (d + A) X = \star d X \;,
\label{lax1}
\end{equation}
which is linear in $X$. 
This linear system is called the Breitenlohner-Maison (BM) linear system \cite{Breitenlohner:1986um}. 
At first sight, it appears that given a solution $X$, we can obtain other solutions
by simply rescaling a given $X$ by a function of $\tau$. This is not the case, however, since
in \eqref{lax1} $\tau$ will turn out to be a function, to be determined, of the spacetime coordinates $x$,
and
$X$ is a matrix that will depend on $\tau$ as well
as on the spacetime coordinates $x$: 
$X=[X(\tau,x)]_{\tau = \tau(x)}$, where we demand $X(\tau,x)$ to be analytic in $\tau$, for $\tau$ in the interior of a closed contour that we assume to be the unit circle.

Applying $\star$ to \eqref{lax1} we obtain
\begin{equation}
\tau \, (\star d + \star A) X = - d X \;.
\label{lax2}
\end{equation}
In what follows, we will assume that \eqref{lax2} has a solution $X$ which is analytic and bounded
as a function of $\tau$ for $\tau$ in the unit disc, and with an inverse $X^{-1}$ satisfying
the same properties.
Then, multiplying \eqref{lax1} and \eqref{lax2} with $X^{-1}$ results in
\begin{eqnarray}
\tau \, d X \, X^{-1}  + \tau A &=& \star d X \, X^{-1}\;, \nonumber\\
\tau \, \star d X \, X^{-1}  + \tau \star A &=& - d X \, X^{-1}\;.
\label{lax3}
\end{eqnarray}

Now multiplying the first of these equations by $\tau$ and using the second equation results in
\begin{equation}
d X \, X^{-1} = - \frac{\tau}{1 + \tau^2} \, \star A - 
\frac{\tau^2}{1 + \tau^2} \, A  \;,
\label{Xeq}
\end{equation}
where we assume $\tau^2 \neq - 1$ when dividing through $(1 + \tau^2)$.
The purpose of these manipulations is to obtain an equation where $X$ enters only on the left hand side,
while the right hand side only depends on the one-form $A$ and on the spectral parameter $\tau$.

Now we study the solvability of \eqref{Xeq}.
Following \cite{Breitenlohner:1986um,Schwarz:1995af,Lu:2007jc}, 
we show that \eqref{Xeq} implies the equation of motion \eqref{eqrA},
if we take the spectral parameter  $\tau$  to depend
on the spacetime coordinates $x$
in a specific manner.

Acting with the differential on the left hand side yields
\begin{equation}
d \left( d X \, X^{-1} \right) = - dX \wedge dX^{-1} = dX X^{-1} \wedge dX X^{-1} = \left( dX X^{-1} \right)
\wedge \left( dX X^{-1} \right) \;.
\label{intLax}
\end{equation}
Then, using \eqref{Xeq}, we obtain 
\begin{eqnarray}
&&- 2 \tau \, d \tau \wedge A -(1 - \tau^2) \, d \tau \wedge \star A - \tau(1 + \tau^2) \, (\tau dA + d (\star A)) = 
\nonumber\\
&& \qquad \tau^2
\left[ (\star A) \wedge (\star A) + \tau^2 A \wedge A
+ \tau ((\star A) \wedge A 
+ A \wedge \star A )\right] \;, 
\end{eqnarray}
which we write as
\begin{eqnarray}
&&- 2 \tau \, d \tau \wedge A -(1 - \tau^2) \, d \tau \wedge \star A 
- \tau(1 + \tau^2) \left[ \tau (dA + A \wedge A) + 
 d (\star A) \right]= 
\nonumber\\
&& \qquad \tau^2
\left[ (\star A) \wedge (\star A) 
+ \tau ((\star A) \wedge A 
+ A \wedge \star A ) - A \wedge A\right] \;.
\end{eqnarray}
Now observe that the right hand side of this expression vanishes by virtue of
\begin{eqnarray}
(\star A) \wedge (\star A) - A \wedge A &=& 0 \;, \nonumber\\
(\star A) \wedge A  + A \wedge \star A &=& 0 \;.
\end{eqnarray}
Introducing
\begin{equation}
C \equiv \frac{1- \tau^2}{1 + \tau^2} \;\;\;,\;\;\; S \equiv \frac{2 \tau}{1 + \tau^2} \;\;\;,\;\;\; C^2 + S^2 = 1
\;\;\;,\;\;\; \tau^2 \neq - 1 \;,
\end{equation}
we have
\begin{eqnarray}
 S \, d \tau \wedge A +C  \, d \tau \wedge \star A + \tau\, \left[ \tau (dA + A \wedge A) + 
 d (\star A) \right]
  = 0 \;.
 \label{intLax2}
\end{eqnarray}
Assuming $\rho \neq 0$, we rewrite 
the term  $d (\star A)$ as
\begin{equation}
d (\star A) = \left(\frac{1}{\rho} d \left( \rho \star A \right) - \frac{1}{\rho} d\rho \wedge \star A
\right) \;.
\end{equation}
Inserting this into \eqref{intLax2} gives
\begin{eqnarray}\label{B.10}
S \, d \tau \wedge A + \left(C \, d \tau - \frac{\tau}{\rho} \, d \rho \right)  \wedge \star A 
+ \tau\, \left[ \tau (dA + A \wedge A) + \frac{1}{\rho} d \left( \rho \star A \right)  
  \right]
= 0 \;.
\end{eqnarray}
Now we recall that $A$ satisfies \eqref{dAAA}. Therefore the equation of motion \eqref{eqrA} must be satisfied if we impose
\begin{equation}
C \, d \tau - \frac{\tau}{\rho} \, d \rho = - S \, \star d \tau \;,
\label{spectral1}
\end{equation}
since this implies that the first two terms in \eqref{B.10} vanish, i.e., that $\tau$ satisfies
\begin{eqnarray}
S \, d \tau \wedge A + \left(C \, d \tau - \frac{\tau}{\rho} \, d \rho \right)  \wedge \star A =0 \;,
\label{intLax3}
\end{eqnarray}
taking into account that
\begin{equation}
 d \tau \wedge A = (\star d \tau) \wedge (\star A) \;.
\end{equation}

We proceed to analyze \eqref{spectral1}.
Taking $\star$ of \eqref{spectral1} we get
\begin{equation}
C \, \star d \tau - \frac{\tau}{\rho} \, \star d \rho = S \, d \tau \;.
\label{spectral1star}
\end{equation}
Multiplying \eqref{spectral1} with $C$, multiplying \eqref{spectral1star} with $S$ and subtracting the two equations, we have the equivalent equation
\begin{equation}
d \tau = \frac{\tau}{\rho} \left( C \, d \rho - S \star d \rho \right) \;.
\label{dtau}
\end{equation}

Assuming $\tau \neq 0$, we proceed to rewrite this equation,
\begin{equation}
2 \star d \rho = \frac{1 + \tau^2}{\tau} \left[  C \, d \rho - \frac{\rho}{\tau} \, d \tau \right] \;,
\end{equation}
which is equivalent to
\begin{equation}
2 \star d \rho
= d \left( \rho \left( \frac{1}{\tau} - \tau \right) \right) \;.
\label{reldrhotau}
\end{equation}
Now we use \eqref{rhoffb} and \eqref{funcstdz} to infer
\be
\star d \rho = \star d ( f + \bar f) = -i d (f - \bar f) \;,
\ee
so that 
\be 
d \left[ - 2 i (f - \bar f)  -  \left( f + \bar f \right) \left( \frac{1}{\tau} - \tau \right) \right] =0 \;.
\ee
Then, since this is the differential of a complex function, integrating we obtain
\begin{equation}
- 2 \left(i (f - \bar f) - \omega \right) =   \left( f + \bar f \right) \left( \frac{1}{\tau} - \tau \right) \;,
\end{equation}
where $\omega$ denotes a complex integration constant, called the constant spectral parameter 
 ($\omega \in \mathbb{C}$). 
This results in the relation
\begin{equation}
\omega = i (f - \bar f) + \frac{(f + \bar f)}{2 \tau} (1 - \tau^2) \;.
\label{wtau2}
\end{equation}
Recall that $\rho^2$ is the warp factor appearing in the reduction of the line element \eqref{line32}. Hence,
we can take $\rho = f + \bar f  >0$ without loss of generality.

We conclude that for $\tau(\omega, x)$ satisfying
\begin{equation}
\tau(\omega, x) = \frac{1}{f + \bar f} \left( i(f-\bar f) -\omega  \pm \sqrt{(f + \bar f)^2 + 
(i(f - \bar f) - \omega)^2} \right) \;\;\;,\;\; \omega \in
\mathbb{C} \;,
\label{tauw2inter}
\end{equation}
the solvability of the Lax pair \eqref{Xeq} indeed implies the equation of motion \eqref{eqrA}, for $A=M^{-1}dM$.

Now let us denote $f(z) = \frac12 (\rho - iv)$, hence
\begin{equation}
 \star d \rho = - d v \;.
 \label{funcv}
 \end{equation}
Then, \eqref{wtau2} becomes
\begin{equation}
\omega = v + \frac{\rho}{2 \tau} (1 - \tau^2) \;,
\label{wtau3}
\end{equation}
and \eqref{tauw2inter} becomes
\begin{equation}
\tau(\omega, x) = \frac{1}{\rho} \left( v -\omega  \pm \sqrt{\rho^2 + 
(v - \omega)^2} \right) \;\;\;,\;\; \omega \in
\mathbb{C} \;.
\label{tauw2}
\end{equation}
If we now assume that $f$ is bijective and has 
a non-vanishing derivative $f'(z)$, so that locally we have an analytic inverse function $f$, i.e. $z = f^{-1}( \rho - iv)$, then we can set
\begin{equation}
ds_2^2 = d\rho^2 + d v^2 
\label{weylrv}
\end{equation}
by a conformal
transformation.
The warp factor $\psi$ is then a function of $x \equiv (\rho, v)$, and the latter are called Weyl coordinates. In these coordinates, $\psi$ in \eqref{line32}
 is obtained by integrating \cite{Schwarz:1995af,Lu:2007jc}
\begin{eqnarray}
\partial_{\rho} \psi &=& \tfrac14 \rho \,  {\rm Tr} \left( A^2_{\rho} - A^2_v \right) \;, \nonumber\\
\partial_{v} \psi &=& \tfrac12 \rho \,  {\rm Tr} \left( A_{\rho}  A_v \right) \;.
\label{psi2}
\end{eqnarray}

In the following, we will assume the existence of Weyl cooordinates $x\equiv(\rho, v)$, thereby restricting to
gravitational solutions that can be described in terms of Weyl coordinates.
We will take $\rho >0$ throughout this paper.

\section{The monodromy matrix ${\cal M}$ \label{sec:monM}}

Given a solution $M(x)$ of the two-dimensional equations of motion \eqref{eqrA}
and a solution $X$ to the linear system \eqref{lax1}
satisfying the regularity conditions specified below \eqref{lax2}, 
one can assign to it
a so-called
monodromy matrix ${\cal M}$ \cite{Breitenlohner:1986um}
that possesses a canonical factorization.
We begin by reviewing its construction. In the next section we will show that the converse also holds:
given a monodromy matrix ${\cal M}$, its canonical factorization yields a solution to the
equations of motion \eqref{eqrA}.

We begin by showing that, given $M(x)$ and a solution $X$ to \eqref{Xeq} with $A=M^{-1}dM$,  we can define a corresponding (spacetime independent) monodromy matrix ${\cal M}(\omega)$ such that, substituting $\omega$ by the right hand side of \eqref{wtau3}, we have a factorization ${\cal M}\big(\omega(\tau,x)\big)=X^{\natural}(-1/\tau,x)M(x)X(\tau,x)$ such that $X\big(\tau(x),x\big)= X(x)$.

Using the decomposition \eqref{defMV}, we define the spectral deformation of $V(x)$,
\begin{equation}
{\cal P}(\tau, x) = V(x) \, X(\tau, x) \;
\label{defcalP}
\end{equation}
and we take ${\cal P}(x)={\cal P}\big(\tau(x),x\big)$ (where the dependence on $\omega$ is implicit).
We compute
\begin{eqnarray}
d {\cal P} {\cal P}^{-1} = dV V^{-1} + V \, dX X^{-1} \, V^{-1}
\end{eqnarray}
and, using \eqref{VPQdecomp}, \eqref{Xeq} and footnote \ref{PpAa}, we obtain
\begin{eqnarray}
d {\cal P} {\cal P}^{-1} &=& P + Q -  \frac{2 \tau}{1 + \tau^2} \, \star P - \frac{2 \tau^2}{1 + \tau^2} \, P
\nonumber\\
&=& Q + \frac{1 - \tau^2}{1 + \tau^2} \, P -  \frac{2 \tau}{1 + \tau^2} \, \star P\;.
\label{calPdp}
\end{eqnarray}

Now we define the spectral deformation of $M(x)$ by \cite{Breitenlohner:1986um}
\begin{equation}
{\cal M}(\tau, x) = {\cal P}^{\natural} (-1/\tau, x) \,  {\cal P}(\tau, x) = M_- (\tau, x) M_+(\tau, x) \;,
\end{equation}
where we used that $\natural$ acts as an anti-homomorphism on matrices, and where
\begin{eqnarray}
M_+(\tau, x) &=& X(\tau, x) \;, \nonumber\\
M_-(\tau,x) &=& X^{\natural}(-1/\tau, x) \, M(x) \;.
\end{eqnarray}
Computing  $d {\cal M}$, for ${\cal M}(x)={\cal M}\big(\tau(x),x\big)$, we have
\be
d {\cal M} = d {\cal P}^{(\natural)} \, {\cal P} + {\cal P}^{(\natural)} \, d {\cal P} = {\cal P}^{(\natural)} \left( \left(d {\cal P} {\cal P}^{-1} \right)^{(\natural)} + 
d {\cal P} {\cal P}^{-1} \right) {\cal P} \;,
\ee
where ${\cal P}^{(\natural)}(x)={\cal P}^{\natural} (-1/\tau, x)_{|_{\tau=\tau(x)}}\,,\,
{\cal P}(x)={\cal P}(\tau,x)_{|_{\tau=\tau(x)}}.$
Using  \eqref{calPdp} and $P^{(\natural)}= P$ and $Q^{(\natural)} = - Q$, it follows that
\begin{equation}
d {\cal M} =0 \;.
\end{equation}

This implies that the monodromy matrix ${\cal M} (\tau(\omega, x), x)$ is spacetime independent, i.e. ${\cal M} = {\cal M}(\omega)$
in view of \eqref{tauw2}.

The above procedure of assigning a monodromy matrix ${\cal M}(\omega)$ to a given $M(x)$
requires one 
to have an explicit solution $X$ to \eqref{lax1}.
Note, in particular, that when solving \eqref{lax1}, one will have
to pick the appropriate branch of $\tau (\omega, x)$ in \eqref{tauw2}  that corresponds to 
a given region $(\rho, v)$. Thus, obtaining an explicit solution  $X$ to \eqref{lax1}
may be difficult.
Therefore, to bypass this difficulty,
 it is most useful to have a rule providing a candidate monodromy matrix, which we call the substitution rule $M(x) \rightarrow {\cal M} (\omega)$. Whether this rule holds will then have to be verified
 a posteriori, on a case by case basis. 
Other approaches on how to obtain a candidate monodromy matrix can be found in
\cite{Breitenlohner:1986um,Chakrabarty:2014ora}.
 We proceed to explain the substitution rule.

Given $M(x)$, we seek a  matrix ${\cal M} (\omega)$ whose canonical factorization yields back $M(x)$,
\begin{equation}
{\cal M} (\omega(\tau, x)) = X^{\natural} (-1/\tau, x) \, M(x)  \, X(\tau, x) \;\;\;,\;\;\; \tau \in \Gamma \;,
\label{factorcan2}
\end{equation}
where  $x=(\rho, v)$. Here $\omega(\tau, x)$ is expressed 
in terms of $\tau$ by means of the algebraic curve \eqref{wtau3}.
Since $ \omega (\tau, x)$ is a continuous function of $\rho$ (at least as long as $\tau \in \mathbb{C}\backslash \{0\}$), so is
${\cal M} (\omega(\tau, x))$. It is then natural to {\sl assume} that the individual factors on the right hand side of \eqref{factorcan2} will
also be continuous in $\rho$, and that they can be continuously {\sl extended } to  $\rho =0$.  This is what we will be assuming in the following, in which case
\begin{equation}
\lim_{\rho \rightarrow 0^+} {\cal M} (\omega(\tau, x)) = \left( \lim_{\rho \rightarrow 0^+}  X^{\natural} (-1/\tau, x) \right) \,
\left(  \lim_{\rho \rightarrow 0^+}  M(x) \right) \, \left( \lim_{\rho \rightarrow 0^+}  X(\tau, x) \right) \;\;\;,\;\;\; \tau \in \Gamma \;.
\label{factorcan2lim}
\end{equation}
We have $\omega(\tau, x) \rightarrow v$ as $\rho \rightarrow 0^+$ and hence, the left hand
side of \eqref{factorcan2lim} tends to ${\cal M} (v)$ for all $\tau$, 
while $M(x)$ tends to $M(\rho=0, v)$.
Since the canonical factorization of a $\tau$-independent matrix ${\cal M} (v)$ is unique and trivial, i.e. $X = \mathbb{I}$ upon imposing a normalization condition, as explained before,
it follows by continuity that
\be
 \lim_{\rho \rightarrow 0^+}  X(\tau, x) = X(\tau, \rho = 0, v) = \mathbb{I} \;\;\;,\;\;\; \tau \in \Gamma \;,
 \label{Xcont}
 \ee
and similarly for $ \lim_{\rho \rightarrow 0^+}  X^{\natural} (-1/\tau, x) =  \mathbb{I} $.  Hence,
\be
{\cal M} (\omega = v) = M(\rho =0, v) \;,
\label{substruleMM}
\ee
which associates a matrix ${\cal M} (\omega)$ to a given matrix $M( x)$.

We now discuss two examples for which the substitution rule works.
In the first example,
we limit ourselves to verify that
assumption \eqref{Xcont} holds (we refer to subsection \ref{subsecsl3} for a full
verification of the substitution rule).
The example is
based on the monodromy matrix \eqref{underrotmonvg0} associated with the near-horizon
limit of an underrotating black hole.
Its canonical factorization yields (see \eqref{undefMM})
\begin{eqnarray}
{X}^{-1} (\tau, x) = 
\begin{pmatrix}
1 & \quad 0 & \quad 0 \\
\frac{2 B}{D \, \rho (\tau_0^+ - \tau_0^-)} \, \frac{\tau}{(\tau - \tau_0^-)}
 & \quad 1 &  \quad  0 
  \\
m_{31}  & \quad - \frac{2 B}{C \, \rho (\tau_0^+ - \tau_0^-)} \, \frac{\tau}{(\tau - \tau_0^-)}
   & \quad 1
\end{pmatrix} \;,
\label{undefMMX}
\end{eqnarray}
where the entry $m_{31}$ is given in \eqref{m11m31}.
Here, 
 $\tau_0^{\pm}$ are the two values of $\tau(\omega,x)$ evaluated at $\omega =0$. We denote the one that 
lies inside the unit circle in the $\tau$-plane by $\tau_0^+$, and the one
that lies outside of
the unit circle by $\tau_0^-$. They satisfy $\tau_0^+ \tau_0^- = -1$.
As already mentioned, we take $\rho > 0$. Then, when $v> 0$, we have
\be
\tau_0^+ = \frac{v - \sqrt{v^2 + \rho^2}}{\rho} \;\;\;,\;\;\; \tau_0^- = \frac{v + \sqrt{v^2 + \rho^2}}{\rho} \;,
\label{tzpos}
\ee
while when $v< 0$ we have
\be
\tau_0^+ = \frac{v + \sqrt{v^2 + \rho^2}}{\rho} \;\;\;,\;\;\; \tau_0^- = \frac{v - \sqrt{v^2 + \rho^2}}{\rho} \;.
\label{tzneg}
\ee
In the limit $\rho \rightarrow 0^+$, we may then approximate 
\be
\tau_0^- = \frac{2 v}{\rho} \;\;\,\;\;\; \tau_0^+ = - \frac{\rho}{2 v} \;,
\ee
which is valid for both $v>0$ and $v<0$. It follows that in the limit $\rho \rightarrow 0^+$,
\begin{eqnarray}
&& \frac{1}{
 \rho (\tau_0^+ - \tau_0^-)} \rightarrow - \frac{1}{2v} \;\;,\;\; \frac{\tau}{\tau - \tau_0^-} \rightarrow 0 \;\;,\;\;  \frac{\tau_0^+}{\tau_0^+ - \tau_0^-} \rightarrow 0 \;\;,\;\;
 \frac{\tau_0^-}{\tau - \tau_0^-}  \rightarrow - 1 \;, \nonumber\\
&& 
\frac{1}{(\tau_0^+ - \tau_0^-) (\tau - \tau_0^-)} \rightarrow 0  \;\;\;,\;\;\; \frac{(\tau_0^-)^2}{(\tau_0^+ - \tau_0^-) (\tau - \tau_0^-)} \rightarrow 1 \;.
 \end{eqnarray}
Using these relations in \eqref{undefMMX}, we indeed obtain
\be
 \lim_{\rho \rightarrow 0^+}  X^{-1} (\tau, x) = \mathbb{I} \;.
 \ee


The second example is 
the non-extremal Kerr black hole of General Relativity.
Here we find that the discussion of the substitution rule is
more subtle than the one that has appeared in the literature.
The associated $M(x)$, which takes the form \eqref{Msl2} of the appendix, is given in 
\cite{Katsimpouri:2012ky} in
prolate spheroidal coordinates $(u,y)$, and reads
\begin{equation}
M(u,y)= \frac{1}{u^2 - c^2 + a^2(y^2-1)} \begin{pmatrix}
(u - m)^2 + a^2 y^2& \quad 2 a m y   \\
2 a m y  & \quad (u + m)^2 + a^2 y^2 
\end{pmatrix} \;\;\;,\;\;\; \det {M} =1 \;.
\label{Mxkerr}
\end{equation}
These coordinates
are related to the Weyl coordinates $(\rho, v)$ by
\be
v = u \, y \;\;\;,\;\;\; \rho = \sqrt{(u^2 - c^2) (1-y^2)} \;\;\;,\;\;\; c < u < \infty \;\;\;,\;\;\; |y| < 1 \;.
\ee
Here,  $c = \sqrt{m^2 - a^2} > 0$ denotes the non-extremality parameter.

To obtain a candidate monodromy matrix, we have to formulate the substitution rule
\eqref{substruleMM} in the coordinates $(u, y)$. In the $(u,y)$-plane, 
we consider three  different ways to take the limit 
$\rho \rightarrow 0^+$. The first way consists in sending $y \rightarrow 1$ while keeping 
 $u \geq c_1 > c$. The second way consists in sending $y \rightarrow - 1$ while again keeping 
 $u \geq c_1 > c$. The third way consists in sending $u \rightarrow c$, keeping $|y| <1$.

Let us then consider the first way. Setting $y=1$ results in $v= u> c$.  Substituting $y=1$ and $u = \omega$ in 
\eqref{Mxkerr}, we obtain 
the following candidate monodromy matrix,
\begin{equation}
{\cal M}(\omega)= \frac{1}{\omega^2 - c^2} \begin{pmatrix}
(\omega - m)^2 + a^2 & \quad 2 a m   \\
2 a m  & \quad (\omega + m)^2 + a^2 
\end{pmatrix} \;\;\;,\;\;\; \det {\cal M} =1 \;.
\label{Monokerr}
\end{equation}
Consider now the factorization of the corresponding matrix ${\cal M}(\tau, x)$ that results by
substituting $\omega = v - \rho (\tau^2 -1)/( 2 \tau)$. Using \eqref{tauw2}, let $\tau_{\pm c}^+$ and $\tau_{\pm c}^-$
(with $\tau_{\pm c}^+$ inside the unit disc and $\tau_{\pm c}^-$ outside the unit disc)
denote the two values of $\tau$ corresponding to $\omega = \pm c$ (we assume $(\rho,v)$ such 
that
$|\tau_{\pm c}| \neq 1$). The scalar factor $W(\tau, x)$, corresponding
to $1/(\omega^2 - c^2)$ in \eqref{Monokerr} with $\omega = v - \rho (\tau^2 -1)/( 2 \tau)$,
admits the canonical factorization
\be
W = W_- W_+ = \left( \frac{\tau^2}{ (\tau - \tau_c^+)(\tau - \tau_{-c}^+)} \right)
\left(
\frac{4}{\rho^2 (\tau - \tau_c^-)(\tau - \tau_{-c}^-)} \right) \;.
\label{decompWpm}
\ee
Next we study the factorization of the matrix 
${\tilde {\cal M} }= W^{-1} {\cal M}$
on the right hand side of \eqref{Monokerr},
which we write as
\begin{equation}
{\tilde {\cal M}}(\\\tau,x)= \begin{pmatrix}
\frac{P(\tau)}{\tau^2} & \quad 2 a m   \\
2 a m  & \quad \frac{{\tilde P}(\tau)}{\tau^2}
\end{pmatrix} \;,
\label{Monokerr2}
\end{equation}
where $P$ and ${\tilde P}$ are polynomials of degree 4 in $\tau$, and that depend on the
parameters $x=(\rho, v)$. Since 
$\det {\cal M}(\tau, x) = 1$, it follows that 
\be
\frac{P(\tau) \, {\tilde P}(\tau) - 4 a^2 m^2 \tau^4}{\tau^4} = (\omega^2 - c^2)^2\vert_{\omega = v - \rho (\tau^2 -1)/( 2 \tau)} \;.
\ee
Therefore, $\frac{P(\tau) \, {\tilde P}(\tau) - 4 a^2 m^2 \tau^4}{\tau^4}$ has four double zeroes:
$\tau_{\pm c}^+$ and $\tau_{\pm c}^-$. Now we must have 
\be
{\tilde {\cal M}}(\tau, x) = {\tilde M}_-
{\tilde M}_+ \;\;\;,\;\;\;  {\tilde M}_+ (0) = W_+^{-1} (0) \, \mathbb{I} \;,
\ee
and hence the two columns of ${\tilde M}_+^{-1}$ are given by the solutions of 
\be
{\tilde {\cal M}}(\tau, x) \, \Phi_+ (\tau) = \Phi_- (\tau) \;,
\ee
where $ \Phi_+ (\tau)$ and $\Phi_- (\tau)$ are analytic and bounded inside and outside the unit
disk, respectively, and satisfying $ (\Phi_+)_1 (0) = W_+ (0)$ and $(\Phi_+)_2 (0) = 0$
for the first column, and $ (\Phi_+)_1 (0) = 0, \; (\Phi_+)_2 (0) = W_+(0)$ for the second
column. Considering the latter and writing $(\Phi_+)_1$ as $\tau \, ({\tilde \Phi}_+)_1$
to account for the zero at $\tau = 0$, it follows that 
\begin{eqnarray}
\frac{P}{\tau^2} \, \tau ({\tilde \Phi}_+ )_1 + 2 a m (\Phi_+)_2 &=& (\Phi_-)_1 = \frac{A \tau + B}{\tau}
\;, \nonumber\\
2 a m \tau ({\tilde \Phi}_+ )_1 + \frac{\tilde P}{\tau^2} \, (\Phi_+)_2 &=& (\Phi_-)_2 = 
\frac{C \tau^2 + D \tau + E}{\tau^2} \;,
\end{eqnarray}
where $A, B, C, D, E$ are constants. From this we obtain
\begin{eqnarray}
 ({\tilde \Phi}_+ )_1 &=& \frac{ A \tau + B - 2 a m \tau (\Phi_+)_2 }{P} \;, \\
 ({\Phi}_+ )_2 &=& \frac{P (C \tau^2 + D \tau + E) - 2 a m \tau^3 (A \tau + B)}{P {\tilde P}
 - 4 a^2 m^2 \tau^4 } \;.
 \label{relPhi2}
 \end{eqnarray}
Since $(\Phi_+)_2 (0) = W_+ (0)$, we see that 
 \be
 E = {\tilde P} (0) W_+ (0) = \tau_c^+ \, \tau_{-c}^+  \;.
 \ee
   The other
 constants $A, B, C, D$ are obtained by imposing two double zeroes, for $\tau = \tau_c^+$
 and $\tau = \tau_{-c}^+$, on the numerator on the right hand side of \eqref{relPhi2}, in order
 to guarantee the analyticity of $(\Phi_+)_2$ in the unit disc.
 
 The resulting system of equations reads 
 \begin{eqnarray}
 T \begin{pmatrix}
A \\
B\\
C\\
D
\end{pmatrix} = - E 
 \begin{pmatrix}
(m - c)^2 + a^2 \\
\rho (m-c) (1 + \frac{1}{(\tau_c^+)^2}) \\
(m+c)^2 + a^2 \\
 \rho (m+c) (1 + \frac{1}{(\tau_{-c}^+)^2})
\end{pmatrix} \;,
\label{systemT}
\end{eqnarray}
where 
\begin{eqnarray}
T_{11} &=& - 2 a m \, (\tau^+_c)^2 \;\;\;,\;\;\; T_{12} = - 2 a m \, \tau_c^+ \;\;\;,\;\;\; T_{13} = 
\left((m - c)^2 + a^2 \right)  (\tau_c^+)^2 \;, \nonumber\\
T_{14} &=& 
\left((m - c)^2 + a^2 \right)  \tau_c^+ \;\;\;,\;\;\; T_{21} = - 4 a m \, \tau_c^+ \;\;\;,\;\;\; T_{22} = - 2 a m 
\;, \nonumber\\
T_{23} &=& \rho (m-c) \left(1 + \frac{1}{(\tau_c^+)^2}\right) (\tau_c^+)^2 + 2 \left((m - c)^2 + a^2 \right)  \tau_c^+ \;,
\nonumber\\
T_{24} &=&  \rho (m-c) \left(1 + \frac{1}{(\tau_c^+)^2}\right) \tau_c^+ + (m - c)^2 + a^2 \;,
\end{eqnarray}
and where the elements in the third and fourth rows of the matrix $T$ can be obtained from the first
and second rows, respectively, by replacing $c$ by $-c$.

We note that the system \eqref{systemT} has a unique solution iff $\det T \neq 0$, 
which yields a condition on $x=(\rho,v)$ involving the parameters $m, c, a$.

Assuming that $\det T \neq 0$, and
solving the system \eqref{systemT} for $A$ and $C$, we obtain $\lim_{\tau \rightarrow \infty} (\Phi_-)_1
= A$, $\lim_{\tau \rightarrow \infty} (\Phi_-)_2 = C$, i.e., we obtain the second column of the matrix
$M_0 (x) = \lim_{\tau \rightarrow \infty} {\tilde M}_- (\tau, x)$. 
In addition, from \eqref{decompWpm}, we infer  $\lim_{\tau \rightarrow \infty} W_- = 1$.
Thus, the factorization of \eqref{Monokerr} results in ${\cal M}(\tau, x) = M_- (\tau, x) M_+ (\tau, x)$, with
$\lim_{\tau \rightarrow \infty} {M}_- (\tau, x) = M_0 (x) $.
Then, if \eqref{Monokerr}, obtained
from the Kerr solution $M(u,y)$ by applying the substitution rule to the latter, was indeed 
the monodromy matrix corresponding to $M(u,y)$, we should find
\be
M_0 (\rho, v)\vert_{\rho = \sqrt{(u^2 - c^2) (1-y^2)}\;, v = u y} = M(u,y) \;.
\ee
And indeed, it can be verified that for $v > c$ this relation holds. For instance,
take $y=1/2$ and $u=20 c$, in which case $v = 10 c$ and $\rho = \sqrt{1197/4} \, c$.
Furthermore, we choose $m=1, a=1/2$, so that $c = \sqrt{3/4}$. 
We obtain 
\be
A/E = 0.00501567 \;,\; B/E = - 0.00306683 \;,\; C/E =  3.36756 \;,\; D/E= 3.66873
\label{valuesKerrabcd}
\ee
as well as $E = 1/3$. 
This results in $A= 0.00167189$ and $C=1.12252$, in 
agreement with the values of the first and third entries of the second column of $M(u,y=\frac12)$
(we recall that $M_0(x)$ is symmetric with $\det M_0=1$, and hence, it suffices to verify the matching
with the second column of $M(u,y=\frac12)$).

Next, let us discuss the second way, which consists in setting $y = - 1$, which results in
$v=-u<-c$. Substituting $y=-1$ and $u = - \omega$ in 
\eqref{Mxkerr}, we obtain a candidate monodromy matrix 
that is the inverse of \eqref{Monokerr} (and is the one considered in \cite{Katsimpouri:2012ky}, up to
a sign change of $a$).  This inverse matrix
equals \eqref{Monokerr} with $m$ replaced by $-m$. We can thus use the factorization results
described above, with $m$ replaced by $-m$, and verify the matching of the resulting matrix $M_0$
with $M(u,y)$. To this end, let 
us take $y=-1/2$ and $u=20 c$, in which case $v = -10 c$ and $\rho = \sqrt{1197/4} \, c$.
Choosing $m=1, a=1/2$, so that $c = \sqrt{3/4}$, we obtain
\be
A/E = -0.00501567 \;,\; B/E = 0.00306683 \;,\; C/E =  3.36756 \;,\; D/E= -3.66873
\label{valuesKerrabcd2}
\ee
as well as $E = 1/3$. This results in $A= -0.00167189$ and $C=1.12252$, in 
agreement with the values 
of the first entry of the second column of $M(u,y=-\frac12)$.

Finally, let us discuss the third way, which consists in setting $u=c$, which results in 
$-c < v= c \, y < c$. Substituting $u=c$ and $v = c \, y  =  \omega$ in 
\eqref{Mxkerr}, we obtain the following candidate monodromy matrix,
\begin{equation}
{\cal M}(\omega)= \frac{1}{a^2( \omega^2 - c^2) } \begin{pmatrix}
 c^2 (c-m)^2 + a^2 \, \omega^2 & \quad 2 a m c \, \omega   \\
2 a m c \, \omega  & \quad c^2 (c+m)^2 + a^2 \, \omega^2  
\end{pmatrix} \;\;\;,\;\;\; \det {\cal M} =1 \;.
\label{Monokerrinterm}
\end{equation}
We now  factorize \eqref{Monokerrinterm}. Proceeding in a 
manner analogous to the one described above, and using the 
same notation as above, we obtain
\begin{eqnarray}
 ({\tilde \Phi}_+ )_1 &=& \frac{ A \tau + B - 2 a m c \, Q \,  (\Phi_+)_2 }{P} \;, \\
 ({\Phi}_+ )_2 &=& \frac{P (C \tau^2 + D \tau + E) - 2 a m c \, Q\,   \tau^2 (A \tau + B)}{P {\tilde P}
 - 4 a^2 m^2 c^2 \, Q^2\,  \tau^2 } \;,
\label{relPhi3}
 \end{eqnarray}
where $Q(\tau) = \tau v - \rho (\tau^2 -1)/2 $, and  $E = \tau_c^+ \, \tau_{-c}^+$.
 As before,  the
 constants $A, B, C, D$ are obtained by imposing two double zeroes, for $\tau = \tau_c^+$
 and $\tau = \tau_{-c}^+$, on the numerator on the right hand side of \eqref{relPhi3}, in order
 to guarantee the analyticity of $(\Phi_+)_2$ in the unit disc. This results in a system of equations
 similar to \eqref{systemT}, which we refrain from giving here.
 The factorization of \eqref{Monokerrinterm} yields ${\cal M}(\tau, x) = M_- (\tau, x) M_+ (\tau, x)$, with
$\lim_{\tau \rightarrow \infty} {M}_- (\tau, x) = M_0 (x) $. We then verify whether
\be
M_0 (\rho, v)\vert_{\rho = \sqrt{(u^2 - c^2) (1-y^2)}\;, v = u y} = M(u,y) 
\ee
holds for $-c < v < c$. And indeed, it holds. For instance, set  $y=0$, in which case
$v=0$. Inspection of \eqref{Mxkerr} then shows that the entries of 
the second column of $M(u,y)$ are $0$ and 
\be
\frac{u + m}{u-m} \;,
\ee
respectively.  These we compare with the entries of the second column of $M_0 (\rho, v)$.
For concreteness, we pick
$m=1, a=1/2$, so that $c = \sqrt{3/4}$, and we pick $\rho = 1$, so that $u = \sqrt{7/4}$.
We obtain 
\be
A = 0  \;,\; B= 0.758265 \;,\; C= 7.19434 \;,\; D= 0 \;,
\ee
in agreement with the second column of  \eqref{Mxkerr}.

Summarizing, we find that the factorization of the candidate monodromy matrix for each of the regions,
$v>c, \, -c < v < c, \, v < -c$, correctly yields $M(u,y)$ given in \eqref{Mxkerr}. 
Thus, the substitution rule works in all three regions for the non-extremal Kerr black hole.
Note, in particular, that we have obtained the monodromy matrix for the intermediate
region $-c < v < c$.  This is in stark contrast with what happens for the Schwarzschild black hole.
In the latter case, the substitution rule fails to assign a monodromy matrix in the intermediate
region $-c < v < c$, as we will discuss in subsection \ref{sec:schwarz}. 
Observe that \eqref{Monokerrinterm} does not possess a finite limit when switching off
the rotation parameter $a$, i.e. when taking
$a \rightarrow 0$. Thus, we conclude that the non-extremal Kerr black hole and the Schwarzschild
black hole have a distinct behaviour from the point of view of the substitution rule.

We also note the following curiosity.  Inspection of \eqref{Monokerrinterm} shows that $\omega$ always comes multiplied
by $a$. We may thus introduce the rescaled variable $\tilde \omega = a \, \omega$, and consider a double scaling
limit which consists in sending $a \rightarrow 0$ while keeping $\tilde \omega$ finite.  In this double scaling limit,
we obtain
\begin{equation}
{\cal M}(\tilde \omega)= \frac{1}{{\tilde \omega}^2 } \begin{pmatrix}
{\tilde \omega}^2 & \quad 2  m^2 \, {\tilde \omega }  \\
2  m^2 \, {\tilde \omega } & \quad 4 m^4  +  {\tilde \omega}^2  
\end{pmatrix} \;\;\;,\;\;\; \det {\cal M} =1 \;.
\label{Monokerrds}
\end{equation}
Factorization of this non-diagonal monodromy matrix will result in a new solution to Einstein's field equations.
We will, however, not pursue this here.


We conclude with the following observation.
For $c \neq 0$, the solution \eqref{Mxkerr} describes an interpolating non-extremal Kerr solution.
In subsection \ref{subsecsl3} we will consider the extremal Kerr solution in the near-horizon limit.
The extremal Kerr solution has $c=0$.
We will
show that the substitution rule also applies in this case. We note that the near-horizon  extremal Kerr solution is written
in terms of coordinates $(t,\phi)$ that differ from the usual Boyer-Lindquist coordinates $(t,\phi)$ used to describe the non-extremal Kerr black hole \cite{Bardeen:1999px}.


\section{Canonical factorization yields solution of equations of motion }

Now we show, building upon the ideas in \cite{Breitenlohner:1986um}, that the converse to the results of the previous section also holds, i.e., that a canonical factorization \eqref{factorcan2} of ${\cal M} (\omega)$, with $\omega=\frac{1}{\tau}\,[v \, \tau  + \tfrac12 \, \rho \,  (1 - \tau^2) ]$ and $\cal M^\natural=\cal M$,  gives rise to a solution
to the two-dimensional equations of motion \eqref{eqrA} with $A=M^{-1}dM$, where $M$ is the middle factor on the right hand side of \eqref{factorcan2}.
Here, we again denote the Weyl coordinates $\rho$ and $v$ collectively by $x$.

To this end, we will assume that the factor $X(\tau, x)$ of the canonical factorization \eqref{factorcan2}
is not only analytic in the region $D_+$, but that its analyticity properties extend slightly beyond the region
$D_+$ delimited by the curve $\Gamma$ ($\partial D_+ = \Gamma$). In particular, we demand $X(\tau, x)$ to be analytic in $\tau = \pm i \in \Gamma$, since the proof will make use of this property. Then,
$X^{\natural}(-1/\tau, x)$ will not only be analytic in $D_-$, but its domain of analyticity will also extend partially
into $D_+$, and, in particular, $X^{\natural}(-1/\tau, x)$ will be analytic in $\tau = \pm i$.\footnote{We note that demanding analyticity in $\tau = \pm i$ does not constitute a significant restriction, in general.}

Given $X(\tau, x)$, let $X(x)$ denote its composition with $\tau( x) = 
\frac{1}{\rho} \left( v -\omega  \pm \sqrt{\rho^2 + (v - \omega)^2} \right) $, where $\omega$ is a complex parameter. Define also
\be
S (\tau, x) = \frac{(1 + \tau^2)}{\tau} \, d X \, X^{-1}
\ee
and let $S(x)$ denote  its composition with $\tau( x)=\frac{1}{\rho} \left( v -\omega  \pm \sqrt{\rho^2 + (v - \omega)^2} \right)$.

First, using the algebraic curve relation \eqref{wtau3}, we infer
\be
( 1 + \tau^2) \, d \tau = \frac{\tau}{\rho} \Big( (1-\tau^2) \, d\rho + 2 \, \tau \, dv \Big) \;
\ee
and hence we obtain
\begin{eqnarray}
(1+\tau^2)dX=(1+\tau^2)\,dx \, \partial_x X(\tau,x)_{\big|_{\tau}} 
+ \frac{\tau}{\rho}  \Big( (1-\tau^2) \, d\rho + 2 \, \tau \,  dv   \Big)   \partial_{\tau}  X(\tau,x)_{\big|_x}\,.
\end{eqnarray}
It follows that
\be
\label {Sx}
S (x) = \left[   \left( \frac{(1 + \tau^2)}{\tau} \, dx \, \partial_x X(\tau,x)
+   \frac{\Big( (1-\tau^2) \, d\rho + 2 \, \tau \,  dv  \Big) }{\rho}  \partial_{\tau}  X(\tau,x)   \right)   X^{-1}(\tau, x)  \right]_{\big|_{\tau=\tau(x)}} .
\ee

Next, using that $\natural$ acts as an anti-homomorphism at the group level,
we compute $S^{(\natural)}(x)=S^{\natural} (-1/\tau, x)_{|_{\tau=\tau(x)}}$:
\begin{eqnarray}
S^{\natural} (-1/\tau, x)  &=& -  \frac{(1 + \tau^2)}{\tau} \, (X^{ -1})^\natural (-1/\tau,x) \, 
( d X )^{\natural} (-1/\tau, x) \nonumber\\
&=& -  \frac{(1 + \tau^2)}{\tau} \, (X^{ -1})^\natural (-1/\tau,x) \, 
d \Big( X ^{\natural} (-1/\tau, x) \Big) \;.
\end{eqnarray}

Now from the canonical factorization \eqref{factorcan2},
\be
{\cal M}(\tau,x)_{|_{\tau=\tau(x)}}=X^{(\natural)}(x)\, M(x)\, X(x) \;,
\ee
where $X^{(\natural)}(x)=X ^{\natural} (-1/\tau, x)_{|_{\tau=\tau(x)}}\,,\,X(x)=X(\tau,x)_{|_{\tau=\tau(x)}}.$
Taking into account that $ {\cal M}(\tau,x)_{|_{\tau=\tau(x)}}$ does not depend on $x$ (but only on the complex parameter $\omega$), we obtain
\begin{eqnarray}
0 = d {\cal M}(\tau,x)_{|_{\tau=\tau(x)}} &=& d \Big( X ^{(\natural)} (x) \Big)  \, M(x) \, X( x) + 
X^{(\natural)} (x) \, dM (x) \, X(x) \nonumber\\
&& + X^{(\natural)}(x) \, M (x) \, d X  (x) \;.
\end{eqnarray}
Multiplying by $(X^{\natural})^{-1}$ and $X^{-1}$ from the left and from the right, respectively, we obtain
\be
- S^{\natural} (-1/\tau, x)_{|_{\tau=\tau(x)}} \, M(x) + \Big(\frac{(1 + \tau^2)}{\tau}\Big)_{|_{\tau=\tau(x)}} \, d M(x) + M(x) \, S(\tau, x)_{|_{\tau=\tau(x)}} = 0 \;.
\ee
This we write as
\be
S^{(\natural)} ( x) \, M(x) \,-\, \frac{1}{\tau(x)} \, d M(x) = M(x) \, S(x) \,+ \,\tau(x) \, d M(x)\,=:\,G(x).
\ee
 Thus,
\begin{eqnarray}
S(x) &=& M^{-1} (x) \Big( G(x) - \tau(x) \, d M(x) \Big) \;, \nonumber\\
S^{(\natural)} (x) &=&  \left( G(x) + \frac{1}{\tau(x)} \, d M(x) \right) M^{-1} (x) \;.
\label{SGM}
\end{eqnarray}
Moreover, since $M^{\natural} = M$, we see from the definition of $G(x)$ that
\be
G^{(\natural)} = G \;.
\label{GnatG}
\ee
Now we note that $\tau (x)=\tau(\rho,v)$ depends on $\omega$ according to \eqref {tauw2}, i.e., for each $\omega$ we have $\tau_\omega(x)$, and $\tau (x)=\pm i$ if and only if $v=\Re (\omega)\,,\,\rho=\mp \Im (\omega)$ by \eqref{wtau3}. Let $x_\omega=(\rho_\omega,v_\omega)$ with $v_\omega=\Re (\omega)\,,\,\rho_\omega= \mp \Im (\omega)$.

Then, from \eqref{Sx} and \eqref{SGM} we have
\begin{eqnarray}\label{Si}
S(\tau = \pm i,x_\omega) &=& \frac{2}{\rho_\omega} \left( d\rho \pm i \, d v \right)  \left( \partial_{\tau}  X(\tau,x) \right) X^{-1}(\tau,x)_{|_{\tau = \pm i\,,\,x=x_\omega}}\nonumber\\
&=&M^{-1} (x_\omega) \Big( G(x_\omega) \mp i \, d M(x_\omega) \Big) \;.
\end{eqnarray}
Taking the Hodge dual $\star$ of this equation and using $\star d \rho = - dv, \, \star dv = d \rho$,
we get
\be
\star S(\tau = \pm i,x_\omega) = \pm i \, S (\tau = \pm i,x_\omega) \;,
\ee
and thus, from \eqref{Si}, 
\be
M^{-1} (x_\omega) \Big( \star G(x_\omega) \mp i \, \star d M(x_\omega) \Big) =\pm i M^{-1} (x_\omega) \Big( G(x_\omega) \mp i \, d M(x_\omega) \Big) \;.
\ee
Hence $
\pm i 
 \Big( G(x_\omega) \mp i \, d M(x_\omega) \Big) = 
\star G(x_\omega) \mp i \, \star d M(x_\omega)  \;$
and, since this holds for all $\omega$, we conclude that 
\be
\pm i 
 \Big( G(x) \mp i \, d M(x) \Big) = 
\star G(x) \mp i \, \star d M(x)
\ee
 for all $x$.
This results in
\be
\Big( \star \mp i \Big) \Big( G(x) + \star d M (x) \Big) = 0 \;,
\ee
where we used $\star^2 = - {\rm id}$. Thus we obtain
\be
 \star d M(x) = -G(x)\:,
\label{GM}
\ee
which satisfies \eqref{GnatG}.  
Inserting \eqref{GM} into \eqref{SGM} yields
\be
S(x) = - \Big( \star + \tau (x) \Big) A \;,
\ee
with $A = M^{-1} d M$. This is equivalent to \eqref{Xeq}, and hence we have shown that 
a canonical factorization \eqref{factorcan2} gives rise to a solution
to the two-dimensional equations of motion \eqref{eqrA}.

\section{Integrability and extended $G$-symmetries \label{sec:intGsymm}}

In this section we briefly review the construction of conserved currents for two-dimensional
$G/H$ sigma
models both with and without gravity \cite{Breitenlohner:1986um,Nicolai:1991tt,Schwarz:1995af,Lu:2007jc}.

\subsection{Conserved currents in two-dimensional $G/H$ sigma
models}

We have seen that the equations of motion for a 
sigma model (\ref{ActionSigma3d}) are conservation 
equations (\ref{ConserveM}). As a preparation for the
two-dimensional sigma model describing the stationary 
axisymmetric sector of four- or higher dimensional gravity,
let us first consider the two-dimensional version of
a $G/H$ sigma model (\ref{ActionSigma3d}). 
In two dimensions there are further, hidden symmetries. Let us write
the equation of motion as $d(\star A) =0$, where $A=A_m dx^m$, and
$\star$ is the Hodge dual. The dual one-form $\star A$ is closed,
and therefore has, locally, a potential $X_1$:
\[
d (\star A) = 0 \Rightarrow A = \star d X_1 \;.
\]
Now, define a new current (one-form) by taking the covariant
exterior derivative of the potential $X_1$:
\[
J_1 = (d+A) X_1 \;.
\]
This is a new conserved current, because
\[
d(\star J_1) = d \star [(d+A) X_1] = d \star d X_1 + d(\star A) X_1 -
\star A \wedge d X_1  \;.
\]
Using that $\star A$ is conserved, the definition $\star d X_1 = A$,
and that $\alpha \wedge \beta
=\star \alpha \wedge \star \beta$ for one-forms in two Euclidean
dimensions we find
\[
d (\star J_1) =  d A + A \wedge A = 0 \;,
\]
where we used that $A=M^{-1}dM$ is a flat connection. Iterating this
construction by defining new potentials $X_{k+1}$ by
$J_k= (d + A) X_k = \star dX_{k+1}$,
where $J_0=A$, $X_0=\mathbb{I}$, and $k=0,1,2\ldots$ we obtain an infinite hierarchy
of conserved currents $J_k$ and potentials $X_{k+1}$.
Introducing an auxiliary variable $\tau$, which at present
is independent of $x$, we can collect the currents
$J_k$ and potentials $X_k$ into a master current and master potential:
\[
J = \sum_{k=0}^\infty J_k \tau^k \;,\;\;\;
X = \mathbb{I} + \sum_{k=1}^\infty X_k \tau^k \;.
\]
The master potential satisfies the linear equation
\begin{equation}
\label{linearEq}
\tau J = \tau (d+A) X = \star d X \;,
\end{equation}
which combines the relations $J_k = (d+A) X_k = \star dX_{k+1}$.
The integrability condition of the linear system
(\ref{linearEq}) is the conservation equation for the master current $J$:
\[
0 = d \star \star d X = \tau d\star [(d+A) X] = \tau d\star J \;.
\]
Expanding in $\tau$ we obtain as lowest order term
$d\star J_0 = d\star A=0$, which is the equation of motion 
of our sigma model. Thus, the sigma model equation of motion 
is an integrability condition for the linear system (\ref{linearEq}).

By construction the current $J=\sum_{k=0}^\infty J_k \tau^k$  is
$\mathfrak{g}$-valued and can therefore be interpreted as an 
element of the loop algebra $\tilde{\mathfrak{g}}$. This indicates
that the rigid $G$-invariance of the $G/H$ sigma model extends
to a rigid invariance under the infinite-dimensional loop group
$\tilde{G}$ upon reduction to two dimensions. If $G$ is a real
simple Lie group, then the associated loop group $\tilde{G}$ 
is the group of maps $S^1 \rightarrow G$.\footnote{If 
we take $\tau$ to be complex valued (without restriction to the 
unit circle), we obtain a group of meromorphic functions valued
in the complexification of $G$. We will not distinguish by notation
between this group and the loop group. See \cite{Breitenlohner:1986um} for
a discussion of regularity assumptions.}
If $T_a$ are generators of $\mathfrak{g}$, then $T_a^m=T_a\tau^m$ 
are generators of $\tilde{\mathfrak{g}}$, with commutation relations
\[
[T^m_a, T^n_b]=f_{ab}^{\;\;\;\;c} T_c^{m+n} \;.
\]
Since $J$ is obtained from $X$ by differentiation, the 
master potential $X(\tau,x)$ is $G$-valued, and can for
fixed $x$ be interpreted as an element of the group $\tilde{G}$.
The same applies to ${\cal P}(\tau,x) = V(x) X(\tau,x)$. 
To summarize, for a standard two-dimensional $G/H$ sigma model, the 
rigid $G$-symmetry is enhanced to an infinite-dimensional 
rigid $\tilde{G}$-symmetry.

\subsection{Ernst sigma models and the spectral curve}

The $G/H$ sigma models arising from dimensional reduction of 
four-dimensional gravity to two dimensions contain an additional
factor of $\rho$ in the action and equations of motion. Such
generalized sigma models are sometimes called Ernst sigma models.
The equation of motion is still a conservation equation related
to rigid $G$-symmetry, but the conservation equation is modified
from $d(\star A)=0$ to 
\be
d(\rho \star A) = 0 \Rightarrow d\rho \wedge \star A +
\rho d(\star A) =0 \;.
\label{eomwrho}
\ee
In order to still be able to obtain this equation as the
integrability condition of the linear system
\begin{equation}
\label{LinSys}
\tau (d+A) X = \star dX \;,
\end{equation}
one allows the spectral parameter $\tau$ to become spacetime
dependent, $\tau=\tau(x)$, in order to compensate for the
spacetime dependent modification $\rho=\rho(x)$ of the conservation
equation. Based on the above discussion of standard sigma
models, the natural candidate for the conserved master current is
$J =  \tau (d+A)X$, because its conservation, $d(\star J)=0$, 
is the integrability condition of the linear system
(\ref{LinSys}). Based on our previous discussion of this linear
system, we expect that the spectral curve arises as a condition 
involving terms proportional to $d\tau$ in $d(\star J)=0$. 
Indeed, after reorganising terms and using the flatness of
the connection $A$ we find 
\[
d(\star J) = d\tau \wedge (2\tau (d+A)X + \star A X)
+ \tau d(\star A) X \;.
\]
For constant $\tau$ conservation of $J$ is equivalent to conservation 
of $A$. For non-constant $\tau$ we obtain, using \eqref{eomwrho},
\[
d(\star J)X^{-1}= d\tau \wedge (2 \tau dX X^{-1} + 2\tau A + \star A)
-\frac{\tau}{\rho} d\rho \wedge \star A \;.
\]
Here we have used that $X(\tau,x)$ is a $G$-valued function 
(or, for fixed $x$, and element of $\tilde{G}$), which implies
the existence of $X^{-1}$. Using \eqref{Xeq}, we can now eliminate
$dXX^{-1}$ in terms of $A$ and $\star A$ to obtain
\[
d(\star J)X^{-1} = d\tau \wedge \left( \frac{1-\tau^2}{1+\tau^2} \star A 
+ \frac{2\tau}{1+\tau^2} A\right) -\frac{\tau}{\rho} d\rho \wedge
\star A \;.
\]
As we have shown previously (c.f. \eqref{B.10}), the vanishing of the right hand side, 
which then implies $d(\star J)=0$, determines $\tau(x)$ through the
spectral curve. 
Note that the master current $J=\tau (d+A)X$ is different from the
`Kac-Moody current' of \cite{Nicolai:1991tt}, which
in our conventions takes the form $J^{(KM)}=\rho K(\rho,\tau) X^{-1} A
X= \frac{1}{2} \rho K(\rho,\tau) {\cal P}^{-1} P {\cal P}$.

We now turn to the group theoretical interpretation of the hidden
symmetry. The following paragraphs follow closely 
Section 4 of \cite{Nicolai:1991tt}, which we translate into
our notation.\footnote{Note in particular that in \cite{Nicolai:1991tt} the
reduced two-dimensional spacetime has Minkowski signature, and
that compared to our conventions left and right group actions
have been exchanged.} 
The manifest symmetries acting on coset representatives
$V(x)$ are
\[
V(x) \mapsto h(x) V(x) g^{-1} \;,\;\;\;h(x)\in H\;,\;\;\;g \in G \;.
\]
The natural generalisation of $V(x)$ is ${\cal P}(\tau,x) = V(x)
X(\tau, x)$. We therefore expect that the loop group $\tilde{G}$
acts on ${\cal P}(\tau,x)$ from the right, while some
subgroup $\tilde{H}\subset \tilde{G}$ acts from the left. 
Moreover, $\tilde{G}$ should act as a rigid symmetry group, while
$\tilde{H}$ should act as a gauge group. 

Starting with the rigid symmetry group 
$\tilde{G}$, its action should be independent of $x$,
and therefore $\tilde{G}$ should act by $G$-valued functions of
the constant spectral parameter $\omega$:
\[
\tilde{G} \;: {\cal P}(\tau,x) \mapsto {\cal P}(\tau,x) g^{-1}(\omega) \;.
\]
Such a transformation will of course in general not respect 
any gauge conditions we have imposed, and therefore will need
to be accompanied by a compensation $\tilde{H}$ gauge transformation.
We will come back to this, but first turn our attention to infinitesimal
$\tilde{G}$-transformations. Elements of the 
loop algebra $\tilde{\mathfrak{g}}$
are Laurent series in a complex variable, which for the $\tilde{G}$-action is $\omega$:
\[
\tilde{\mathfrak{g}} \ni \delta g(\omega) = 
\sum_{n=-\infty}^\infty \delta g_n \omega^n  \;,\;\;\;
\delta g_n \in \mathfrak{g} \;.
\]
Thus the infinitesimal $\tilde{G}$-action is
\[
\tilde{\mathfrak{g}} \;: 
{\cal P}(\tau,x) \mapsto - {\cal P}(\tau,x) \delta g(\omega) \;.
\]
Next, one needs to identify a subgroup $\tilde{H}\subset \tilde{G}$
which acts as a gauge group. The subgroup $H\subset G$ is characterised
by its invariance under the involutive automorphism $\Theta$, which 
is associated with the symmetric decomposition:
\[
H = \{ h \in G | \Theta(h) = h \} \;.
\]
The natural candidate for extending this to an involutive 
automorphism $\tilde{\Theta}$ of $\tilde{G}$ is the combination
of $\Theta$ with the involutive map
\[
\tau \rightarrow - \frac{1}{\tau}\;,
\]
which leaves $\omega$ invariant, but inverts the spacetime dependent
spectral parameter $\tau$ and exchanges the two Riemann sheets of
the spectral curve. This is closely related to the 
operation ${\cal M}(\tau) \mapsto {\cal M}^\natural(-1/\tau)$ 
under which the monodromy matrix is invariant (c.f. \eqref{calMnatM}). Remember that 
quantities representing the coset $G/H$ are invariant 
under $\natural$-transposition, while `anti-invariant'
under the automorphisms $\Theta$ and $\theta$.\footnote{
`Anti-invariant' means that group elements are inverted while
Lie algebra elements are multiplied by $-1$.}
Conversely elements of $H$ 
are invariant under the automorphisms $\Theta$ and $\theta$,
but `anti-invariant' under $\natural$-transposition. Since
we are looking for a generalization of $H$,  we 
define 
\[
\tilde{\Theta}\;: g(\tau) \mapsto \tilde{\Theta}(g(\tau)) =
(\Theta g)(-1/\tau) \;,
\]
and 
$\tilde{H}\subset \tilde{G}$ is defined by  
\[
\tilde{H} =  \{ h(\tau) \in \tilde{G} | (\Theta h)(\tau) = 
h(-1/\tau) \} \;.
\]
For reference, the following conditions are equivalent: 
\[
\tilde{\Theta}(h(\tau)) = h(\tau) \Leftrightarrow
(\Theta h)(\tau) = h(-1/\tau) \Leftrightarrow
h(\tau)^{-1} = h(-1/\tau)^\natural \;.
\]
We denote the Lie algebra of $\tilde{H}$ by $\tilde{\mathfrak{h}}$.
Elements $\delta h(\tau) \in \tilde{\mathfrak{h}}$ are characterised
by 
\[
\tilde{\theta} (\delta h(\tau)) = \delta h(\tau) 
\Leftrightarrow (\Theta h)(-1/\tau) = h(\tau) \;.
\]
Therefore:
\[
\delta h(\tau) = \delta h_0 +\sum_{n=1}^\infty \delta h_n 
\left( \tau^n +(-1)^n \frac{1}{\tau^n} \right) 
+ \sum_{n=1}^\infty \delta p_n \left( \tau^n -(-1)^n \frac{1}{\tau^n}\right)
\;,\;\;\;
\delta h_n \in \mathfrak{h} \;,\;\;\;
\delta p_n \in \mathfrak{p} \;.
\]
Since $\tilde{H}$ should act as a gauge group, we should
also allow that $h$ and $\delta h$ depend explicitly on $x$, so we
take  $\delta h_n$, $\delta p_n$ to be $x$-dependent. Thus the 
$H\times G$ action on $G/H$ generalises to the following
$\tilde{H}\times \tilde{G}$ action on ${\cal P}(\tau,x)$:
\[
{\cal P}(\tau , x) \mapsto h(\tau, x) {\cal P}(\tau,x) g^{-1}(\omega) \;,
\]
or, infinitesimally
\[
{\cal P}(\tau,x) \mapsto \delta h(\tau, x) {\cal P}(\tau, x) -
{\cal P}(\tau,x) \delta g(\omega) \;.
\]
For fixed $x$, ${\cal P}(\tau,x)$ is interpreted as an element
of an infinite-dimensional generalisation $\tilde{G}/\tilde{H}$ 
of $G/H$. This space is defined by the equivalence relation
\[
{\cal P}(\tau,x) \simeq h(\tau,x) {\cal P}(\tau,x) \;.
\]
The space $\tilde{G}/\tilde{H}$ admits a right action of $\tilde{G}$,
which in general requires a compensating $\tilde{H}$ transformation
to maintain any gauge we have imposed.

According to \cite{Breitenlohner:1986um}, solutions to
the sigma model equation of motion $d(\rho \star A)=0$ correspond
to elements of $\tilde{G}/\tilde{H}$. 
The following explanation is taken from \cite{Nicolai:1991tt}.
If ${\cal P}(\tau, x)$ is a solution to the linear system,
then 
\[
d{\cal P}{\cal P}^{-1} = Q + \frac{1-\tau^2}{1+\tau^2} P 
- \frac{2\tau}{1+\tau} \star P \in \tilde{h} \;,
\]
since under the extended involution $Q \mapsto Q$, $P\mapsto -P$
and $\tau \mapsto -1/\tau$. Since ${\cal P}(\tau,x) \in \tilde{G}$, this
suggests that it must have the form
\[
{\cal P}(\tau,x) = h(\tau,x) g^{-1}(\omega) \;.
\]
But not every expression of the form $h(\tau,x) g^{-1}(\omega)$
will have the $\tau$-dependence required by the linear system.
Moreover, to obtain a solution $V(x)$ of the sigma model
equations we must impose that ${\cal P}$ is regular at $\tau=0$,
so that ${\cal P}(0,x)=V(x)$ exists. Given its existence,
the integrability condition of the linear system guarantees that
it is a solution. Writing ${\cal P}(\tau,x) = V(x) X(\tau,x)$, 
the regularity condition is $X(0,x)=\mathbb{I}$. To see under which 
conditions a ${\cal P}(\tau,x)$ with the required properties
exists, consider the associated monodromy matrix
\[
{\cal P}^\natural(-1/\tau,x) {\cal P}(\tau,x) =
g^{-1,\natural}(\omega) g^{-1}(\omega) = {\cal M}(\omega) \;.
\]
If the canonical factorization problem for ${\cal M}(\omega)$
has a solution 
\[
{\cal M}(\omega(\tau,x)) = {\cal P}^\natural(-1/\tau, x) {\cal P}(\tau,x)
\;,\;\;\;{\cal P}(0,x) = V(x) \Leftrightarrow X(0,x)=\mathbb{I} \;,
\]
then we obtain a solution ${\cal P}(\tau,x)$ of the linear system
which has the required regularity properties and yields a 
solution of the sigma model equations of motion. 
This regular solution takes the form 
\[
{\cal P}(\tau,x) = h(\tau,x,g) g^{-1}(\omega) \;,
\]
where the dependence of $h$ on $g$ indicates that for a given
$g^{-1}(\omega)$ we must determine a suitable, i.e. regular
coset representative. This solution is of course 
$\tilde{H}$-gauge equivalent to other solutions
$h(\tau,x) {\cal P}(\tau,x)$, 
but due to the presence of negative powers of $\tau$ in 
$\delta h(\tau,x)$ any such gauge equivalent solution of the
linear system is not regular at $\tau=0$, except if we take
$\delta h = \delta h_0 \in \mathfrak{h}$. However, if we have 
already chosen a triangular gauge (or any other gauge) for 
$V(x)$, then $\delta h_0=0$ and $h(\tau,x)=\mathbb{I}$.
Thus solutions
of the sigma model correspond to elements of $\tilde{G}/\tilde{H}$ and
the Riemann-Hilbert problem picks a canonical, regular 
coset representative, which upon setting $\tau=0$ provides a 
solution to the sigma model equations of motion.

By acting with $g(\omega) \in \tilde{G}$ one can obtain new 
sigma model solutions from known solutions.
For pure gravity the action of $\tilde{G}$
is transitive, so that all stationary axisymmetric solutions of 
pure gravity can be related this way \cite{Nicolai:1991tt}. 
For general theories this 
question seems to be open. 
We remark that even in pure gravity there are further hidden
symmetries, which centrally extend the loop group $\tilde{G}$
to the associated `Kac-Moody group'  \cite{Breitenlohner:1986um}.
In supergravity, the presence of fermions leads to further extensions.
For example, in maximal supergravity, the affine Kac Moody group
$E_9$, is extended to a hyperbolic Kac Moody group
\cite{Nicolai:1991tt}. The search for further extensions has 
led to $E_{11}$
\cite{West:2001as}, and,
most recently, infinite-dimensional tensor hierarchy algebras
\cite{Bossard:2017wxl} as candidates 
for the full symmetry group underlying 
eleven-dimensional supergravity and its UV-completion, M-theory.

\section{Explicit canonical factorization examples}
In this section we consider three different gravitational theories in four dimensions that, upon dimensional reduction to
two dimensions, give scalar field models based on cosets $G/H = SL(2, \mathbb{R})/SO(2)$, $SU(2,1)/(SL(2, \mathbb{R}) \times U(1))$
and $SL(3, \mathbb{R})/SO(2,1)$. We focus on black hole solutions in these models.
Assuming validity of the substitution rule mentioned in section \ref{sec:monM}, we associate monodromy matrices
to these solutions. These monodromy matrices possess canonical factorizations, whose 
explicit
factorization yields the black hole solutions.

We begin by considering the Schwarzschild solution in the model based $G/H = SL(2, \mathbb{R})/SO(2)$. Subsequently, we turn
to the near-horizon limit of dyonic extremal black hole solutions in the models based on $G/H = SU(2,1)/(SL(2, \mathbb{R}) \times U(1))$
and $SL(3, \mathbb{R})/SO(2,1)$.

\subsection{$G/H = SL(2, \mathbb{R})/SO(2)$: the Schwarzschild solution \label{sec:schwarz}}

We consider the exterior region of the Schwarzschild solution in four dimensions, 
\be 
ds_4^2 = - \Delta \, dt^2 + \Delta^{-1} \, dr^2 + r^2 \left( d \theta^2 + \sin^2 \theta \, d \phi^2 \right) \;,
\label{lineschw}
\ee
with $\Delta = 1 - \frac{2m}{r} > 0$. Here, $m$ denotes the mass of the black hole, and the event horizon is at $r = 2m$.
In Weyl coordinates $(\rho, v)$, the line element is given by (see \eqref{line32} and \eqref{weylrv})
\begin{equation}
ds_4^2 = - \Delta \, dt^2 + \Delta^{-1} \, \rho^2 \, d\phi^2 + \Delta^{-1} \, e^{\psi} \, \left(d\rho^2 + d v^2 \right)  \;,
\label{4dline}
\end{equation}
with $\Delta = \Delta(\rho, v)$ and $\psi = \psi (\rho, v)$. 
Upon dimensional reduction over $(t, \phi)$, the resulting 
representative $M(\rho,v)$ is given by \eqref{Msl2} with $B=0$.

The coordinate change from spherical coordinates $(r, \theta)$ with $ r > 2m$ and 
$0 < \theta < \pi$ 
to Weyl coordinates 
$(\rho, v)$ is given by
\begin{eqnarray}
\rho &=& \sqrt{r^2 - 2 m r} \, \sin \theta \;, \nonumber\\
v &=& (r - m ) \cos \theta \;,
\label{rvrt}
\end{eqnarray}
where $\rho$ and $v$ are harmonic conjugates, c.f. \eqref{funcv}.
Note that $\rho > 0$.

Introducing the abbreviations
\begin{eqnarray}
l_{\pm} &=& \sqrt{\rho^2 + (v \pm m)^2} \;,
\label{deflpm}
\end{eqnarray}
we compute
\begin{eqnarray}
l^2_+ &=& \left(r - m + m \cos \theta \right)^2 \;, \nonumber\\
l^2_- &=& \left(r - m - m \cos \theta \right)^2 \;.
\end{eqnarray}
Since, by definition, $l_{\pm}$ are positive, we obtain
\begin{eqnarray}
l_+ &=& r - m + m \cos \theta  \;, \nonumber\\
l_- &=& r - m - m \cos \theta  \;, 
\end{eqnarray}
which indeed satisfy $l_{\pm} > 0$ in the patch $ r > 2m$ and $0 < \theta <
 \pi$.
Next, we compute
\begin{eqnarray}
v - m + l_- &=& (r-2m) (1 + \cos \theta) \;, \nonumber\\
v - m - l_- &=& r (\cos \theta -1) \;, \nonumber\\
v + m + l_+ &=& r (1 + \cos \theta) \;, \nonumber\\
v + m - l_+ &=& (r-2m) (\cos \theta -1) \;,
\end{eqnarray}
and hence we may express $\Delta =  1 - \frac{2m}{r} >0$ in terms of $(\rho, v)$ in various ways, as follows,
\begin{eqnarray}
\Delta (\rho, v) &=& 
\frac{v - m + l_-}{v + m + l_+} = \frac{v + m - l_+}{v - m - l_-}  \nonumber\\
&=&- \frac{(v - m + l_-)(v + m - l_+)}{ \rho^2} =
- \frac{\rho^2}{(v - m - l_-)(v + m + l_+)}
\;.
\label{Drvsch}
\end{eqnarray}

Now let us consider the substitution rule \eqref{substruleMM}. 
We want to study the behaviour of \eqref{Drvsch} in the limit $\rho \rightarrow 0^+$.
To discuss this limit, we have to consider the following three cases: $v > m$, $- m < v < m$ and $v < -m$. 
In the case $v >m$, and using L'Hopital's rule, we find that the limit exists for all four expressions in \eqref{Drvsch}, and that it is given by
 \begin{equation}
\Delta(\rho =0, v) =
\frac{v - m }{v + m } > 0 \;,
\end{equation}
so that, when $ v>m$,
\begin{eqnarray}
{M} (\rho =0 ,v) = \begin{pmatrix}
\frac{v - m}{v+m} & 0\\
0 & \frac{v+m}{v-m}  
\end{pmatrix} \;.
\label{Mschwarz1}
\end{eqnarray}
When $- m < v < m$, we obtain $\Delta =0$ in the limit  $\rho \rightarrow 0^+$, and hence $M(x)$
in \eqref{Msl2} diverges in this limit.
This means that the
substitution rule is not applicable in this case, because $M(x)$ does not have a continuous
extension to $\rho =0$.
And finally, when  $v < -m$, the limit exists for all four expressions in \eqref{Drvsch} and is given by
 \begin{equation}
\Delta(\rho =0, v) =
\frac{v + m }{v - m } > 0 \;,
\end{equation}
so that, when $v < -m$, 
\begin{eqnarray}
{M} (\rho =0 ,v) = \begin{pmatrix}
\frac{v + m}{v-m} & 0\\
0 & \frac{v-m}{v+m}  
\end{pmatrix} \;.
\label{Mschwarz2}
\end{eqnarray}
Then, the substitution rule \eqref{substruleMM} associates the monodromy matrix
\begin{eqnarray}
{\cal M} (\omega) = \begin{pmatrix}
\frac{\omega - m}{\omega+m} & 0\\
0 & \frac{\omega+m}{\omega-m}  
\end{pmatrix} \;\;\;,\;\;\; \det {\cal M} = 1 \;,
\label{monodromyschwarz}
\end{eqnarray}
to \eqref{Mschwarz1}, the monodromy matrix
\begin{eqnarray}
{\cal M} (\omega) = \begin{pmatrix}
\frac{\omega + m}{\omega-m} & 0\\
0 & \frac{\omega-m}{\omega+m}  
\end{pmatrix} \;\;\;,\;\;\; \det {\cal M} = 1 \;,
\label{monodromyschwarz2}
\end{eqnarray}
to \eqref{Mschwarz2}, and fails to assign a monodromy matrix when $-m < v<m$.

The two monodromy matrices \eqref{monodromyschwarz} and \eqref{monodromyschwarz2}
are the inverse of one another, 
and being diagonal matrices, it suffices to factorize one of them. This will be done below.

We may also ask a reversed question: namely, starting from a monodromy matrix such as \eqref{monodromyschwarz}
and factorizing it in the three regions $v>m, -m<v<m, v< -m$, what are the resulting solutions $M(x)$ and how do they compare with the Schwarzschild solution \eqref{lineschw}?
As will be see below, in region $v>m$ one obtains the Schwarzschild solution $M(r)$ as
expected, while for $v < -m$ one obtains its inverse $M^{-1}(r)$,
which describes a Schwarzschild solution with parameter $-m$. For $-m < v < m$, however, one
obtains $M(\theta)$, with diagonal entries that are negative for $\theta >0$.

Let us now consider the factorization of \eqref{monodromyschwarz}.
This monodromy matrix is a $2 \times 2$ matrix that depends on a single parameter $m$. Its entries are rational
functions of $\omega$, with two simple poles at $\omega = \pm m$ \cite{Maison:1988zx}.
This monodromy matrix is bounded, with ${\cal M} (\infty) = \mathbb{I}$.

We introduce the short hand notation $x=(v, \rho)$.
Using \eqref{tauw2}, we evaluate $\tau(\omega, x)$ on the poles $\omega = \pm m$, 
\begin{eqnarray}
\tau_1 &=& \tau(m, x) = \frac{1}{\rho} \left( v -m  \pm \sqrt{\rho^2 + (v - m)^2} \right) \;, \nonumber\\
\tau_2 &=& \tau(-m, x) = \frac{1}{\rho} \left( v +m  \pm \sqrt{\rho^2 + (v + m)^2} \right) \;.
\label{tauw12}
\end{eqnarray}
Note that there are two possible choices for both $\tau_1$ and $\tau_2$. 
We also recall here that $\rho > 0$.

Next, we compute the combination
\begin{eqnarray}
(\tau - \tau_1) ( \tau^{-1} + \tau_1) = (1- \tau^2_1) - \frac{\tau_1}{\tau} (1 - \tau^2) \;,
\end{eqnarray}
and using \eqref{wtau2}, we infer
\begin{eqnarray}
(\tau - \tau_1) ( \tau^{-1} + \tau_1) = \frac{2}{\rho} \tau_1 (m - \omega) \;,
\end{eqnarray}
and hence,
\begin{eqnarray}
\omega-m &=&  - \frac{\rho}{2 \tau_1} (\tau - \tau_1) ( \tau^{-1} + \tau_1) \;,
\nonumber\\
\omega+ m &=& - \frac{\rho}{2 \tau_2} (\tau - \tau_2) ( \tau^{-1} + \tau_2) \;.
\end{eqnarray}
Thus we obtain
\begin{equation}
\frac{\omega - m}{\omega +m} = \frac{\tau_2}{\tau_1} \frac{(\tau - \tau_1)}{(\tau - \tau_2)} \frac{ ( \tau^{-1} + \tau_1)}{
 ( \tau^{-1} + \tau_2)} \;.
 \label{wt1t2gen}
\end{equation}
Note that \eqref{wt1t2gen} holds in general, irrespective of the particular choices for $\tau_{1,2}$.

Next, we factorize \eqref{monodromyschwarz}. Here, we have to make a choice of $\tau_{1,2}$.
We may choose $\tau_{1,2}$ to both lie outside (inside) the unit circle, or one may 
lie inside and the other one outside of the unit circle.  In view of the fact that
\eqref{wt1t2gen} holds for any of the possible choices for $\tau_{1,2}$, it suffices to consider one such choice.
Thus, in the following, we will pick
\begin{eqnarray}
\tau_1 &=& \tau(m, x) = \frac{1}{\rho} \left( v -m  + \sqrt{\rho^2 + (v - m)^2} \right) \;, \nonumber\\
\tau_2 &=& \tau(-m, x) = \frac{1}{\rho} \left( v +m  + \sqrt{\rho^2 + (v + m)^2} \right) \;.
\label{tauw12choice}
\end{eqnarray}

We note that, to perform the factorization of the monodromy matrix, $\tau_{1,2}$ are not allowed to be on the unit circle.
When do $\tau_{1,2}$ lie on the unit circle? This happens when $v = \pm m$, respectively. This means
that we have to exclude these spacetime points from our analysis.
We will thus consider three regions in the following, namely $v>m$, $-m < v < m$ and $v < -m$.

First, let us consider the factorization of \eqref{wt1t2gen} in region $v>m$, in which case
$\tau_{1,2} >1$. We obtain
\begin{equation}
\frac{\omega - m}{\omega +m} = m_- \, m_+ \;,
\end{equation}
where
\begin{eqnarray}
m_+ &=& 
\frac{\tau_2}{\tau_1} \frac{(\tau - \tau_1)}{(\tau - \tau_2)} \;, \nonumber\\
 m_- &=& \frac{\tau_1}{\tau_2} \frac{(\tau +1/\tau_1)}{(\tau + 1/\tau_2)} \;,
 \end{eqnarray}
with $m_+$ and $m_+^{-1}$ analytic and bounded inside the unit circle, and with $m_-$ and $m_-^{-1}$ analytic and bounded outside the unit circle.
Note that $m_+$ is normalized, i.e. $m_+ (\tau =0, x) = 1$. We thus obtain
\begin{equation}
{\cal M} (\omega) = M_- (\tau, x) \, M_+ (\tau, x) 
\end{equation}
with
\begin{eqnarray}
M_+ (\tau, x) = \begin{pmatrix}
\frac{\tau_2}{\tau_1} \frac{(\tau - \tau_1)}{(\tau - \tau_2)} & 0\\
0 & {\frac{\tau_1}{\tau_2}} \frac{(\tau - \tau_2)}{(\tau - \tau_1)}
\end{pmatrix} \;\;\;,\;\;\;
M_-(\tau,x)  =  \begin{pmatrix}
\frac{\tau_1}{\tau_2} \frac{(\tau +1/\tau_1)}{(\tau + 1/\tau_2)} & 0\\
0 & \frac{\tau_2}{\tau_1} \frac{(\tau +1/\tau_2)}{(\tau + 1/\tau_1)} \end{pmatrix} \;,
\end{eqnarray}
with $M_+$ and $M_+^{-1}$ analytic and bounded inside the unit circle, and with $M_-$
and $M_-^{-1}$ analytic and bounded outside the unit circle.
The spacetime solution $M(x)$ is read off from
\begin{eqnarray}
\label{Mt12}
M(x) = M_-(\tau = \infty,x)  =  \begin{pmatrix}
\frac{\tau_1}{\tau_2}  & 0\\
0 & \frac{\tau_2}{\tau_1} \end{pmatrix} \;.
\end{eqnarray}
Using \eqref{Drvsch}, we infer
\be
\frac{\tau_1}{\tau_2} = \frac{v -m + l_-}{v + m + l_+} = \Delta (\rho, v) \;,
\ee
and hence, when $v >m$, $M(x)$ describes the outside region of the Schwarzschild black hole.

Next, we 
consider the factorization of \eqref{wt1t2gen} in region $-m < v<m$. In this case $\tau_1 <1, \tau_2 >1$. We obtain
\begin{eqnarray}
M_+ (\tau, x) = \begin{pmatrix}
- \tau_1 \tau_2 \,  \frac{(\tau +1/\tau_1)}{(\tau - \tau_2)} & 0\\
0 & -\frac{1}{\tau_1 \tau_2}  \frac{(\tau -\tau_2)}{(\tau +1/ \tau_1)}
\end{pmatrix} \;\;\;,\;\;\;
M_- (\tau, x) =  \begin{pmatrix}
- \frac{1}{\tau_1 \tau_2} \frac{(\tau - \tau_1)}{(\tau +1/\tau_2)} & 0\\
0 & - \tau_1 \tau_2 \frac{(\tau +1/\tau_2)}{(\tau - \tau_1)}
 \end{pmatrix} \;, \nonumber\\
\end{eqnarray}
with $M_+$ and $M_+^{-1}$ analytic and bounded inside the unit circle, and with $M_-$
and $M_-^{-1}$ analytic and bounded  outside the unit circle.
The spacetime solution $M(x)$ is read off from
\begin{eqnarray}
M(x) = M_-(\tau = \infty,x)  =  \begin{pmatrix}
-\frac{1}{ \tau_1 \tau_2}  & 0\\
0 & - \tau_1 \tau_2 \end{pmatrix} \;.
\label{t12M}
\end{eqnarray}
Using \eqref{Drvsch}, we infer
\be
- \frac{1}{\tau_1 \tau_2} = - \frac{\rho^2}{(v -m + l_-)(v + m + l_+)} = \frac{\cos \theta -1}{\cos \theta +1}  \;,
\ee
which is negative for $\theta >0$ and solely depends on $\theta$. Therefore, this does not
describe the outside region of the Schwarzschild black hole.

Finally, we 
consider the factorization of \eqref{wt1t2gen} in region $v< -m$. In this case $\tau_{1,2} <1$. We obtain
\begin{eqnarray}
M_+ (\tau, x) = \begin{pmatrix}
\frac{\tau_1}{\tau_2} \frac{(\tau +1/\tau_1)}{(\tau +1/ \tau_2)} & 0\\
0 & {\frac{\tau_2}{\tau_1}} \frac{(\tau +1/\tau_2)}{(\tau +1/ \tau_1)}
\end{pmatrix} \;\;\;,\;\;\;
M_-(\tau,x)  =  \begin{pmatrix}
\frac{\tau_2}{\tau_1} \frac{(\tau -\tau_1)}{(\tau - \tau_2)} & 0\\
0 & \frac{\tau_1}{\tau_2} \frac{(\tau - \tau_2)}{(\tau - \tau_1)} \end{pmatrix} \;,
\end{eqnarray}
with $M_+$ and $M_+^{-1}$ analytic and bounded inside the unit circle, and with $M_-$
and $M_-^{-1}$ analytic and bounded outside the unit circle.
The spacetime solution $M(x)$ is read off from
\begin{eqnarray}
M(x) = M_-(\tau = \infty,x)  =  \begin{pmatrix}
\frac{\tau_2}{\tau_1}  & 0\\
0 & \frac{\tau_1}{\tau_2} \end{pmatrix} \;,
\end{eqnarray}
which is the inverse of \eqref{Mt12}. It describes a Schwarzschild solution
with  parameter $-m$,
as can be easily verified by performing a change of the radial coordinate $r$ and replacing
$m$ by $-m$.

Finally, when evaluating $M(x)$ at $\rho \rightarrow 0^+$, we obtain in all
three regions
$M(\rho =0, v) = {\cal M} (\omega =v)$, in agreement with the substitution rule \eqref{substruleMM}.


\subsection{Attractors: near horizon solutions \label{rotexam}}

Next, we turn to extremal single-centre static and rotating black hole solutions. These exhibit the attractor mechanism \cite{Ferrara:1996dd,Astefanesei:2006dd}.
Due to this 
attractor mechanism, the near-horizon solution is, by itself, an exact solution to the equations
of motion in four dimensions.

In the following, we focus on these near-horizon solutions and derive them by canonical factorization
of associated monodromy matrices.  These monodromy matrices will be rational matrices with a {\sl double
pole} at $\omega =0$. We will make use of the values of $\tau(\omega, x)$ at the pole $\omega =0$, which we will
denote by $\tau_{0 \pm}$. Using
\eqref{tauw2}, we obtain
\begin{eqnarray}
\tau_{0 \, \pm} = \frac{1}{\rho} \left( v  \pm \sqrt{\rho^2 +  v^2} \right) \;.
\label{tauw0}
\end{eqnarray}
Recall that  $\rho >0$. We also take $v\neq 0$, so that 
$\tau_{0 \pm}$ are never on the unit circle.
Note that $\tau_{0 +} \tau_{0 -} = -1$, so that one of them lies inside and the other one outside
of the unit circle.  In the following, we will denote the one inside the unit circle by $\tau_0^+$, and the one
outside of the unit circle by $\tau_0^-$.
Depending on the region in the $(\rho, v)$ plane, $\tau_0^+$ will equal either  $\tau_{0 +}$ or $ \tau_{0 -}$.
When $v> 0$, we have
\be
\tau_0^+ = \frac{v - \sqrt{v^2 + \rho^2}}{\rho} \;\;\;,\;\;\; \tau_0^- = \frac{v + \sqrt{v^2 + \rho^2}}{\rho} \;,
\label{tvpos}
\ee
while when $v < 0$ we have
\be
\tau_0^+ = \frac{v + \sqrt{v^2 + \rho^2}}{\rho} \;\;\;,\;\;\; \tau_0^- = \frac{v - \sqrt{v^2 + \rho^2}}{\rho} \;.
\label{tvneg}
\ee
We also note the useful relation
\begin{equation}
\omega = - \frac{\rho}{2} \frac{(\tau - \tau_0^+)(\tau - \tau_0^-)}{\tau} \;,
\label{wtaupm0}
\end{equation}
which we will be using repeatedly below.

\subsubsection{$G/H = SU(2,1)/(SL(2, \mathbb{R}) \times U(1))$: static attractors}

In this subsection we consider the near-horizon limit of a static extremal black hole supported by one
electric charge $q$ and one magnetic charge $p$. In this limit, the solution describes an $AdS_2 \times S^2$ background, with a line element 
\begin{equation}
ds^2_4 = - e^{- \varphi} \, dt^2 + e^{\varphi} \, ds_3^2 \;,
\label{lineads2su21}
\end{equation}
where in  spherical coordinates 
$ds_3^2 =  dr^2 + r^2 \, \left(d \theta^2 + \sin^2 \theta \, d \phi^2 \right)$ with
\be
e^{- \varphi (r)} = |Q|^{-2} \, r^2 \;\;\;, \;\;\; Q = q + i p  \;\;\;,\;\;\; q, p \in \mathbb{R}\;.
\ee
We take the gauge potentials that support the solution to be (c.f. \eqref{sigmZ})
\begin{equation}
 \chi_e (r) = \frac{q}{|Q|^2} \, r \;\;\;,\;\;\; 
\chi_m (r) = \frac{p}{|Q|^2} \, r \;\;\;,\;\;\;
\sqrt{2} \lambda = 2 \chi_e \chi_m \;,
\label{backg}
\end{equation}
in which case the $\natural$-symmetric coset representative \eqref{cosetrepMsu21} is given by
\begin{eqnarray}
M = 
\begin{pmatrix}
e^{\varphi} & \quad  \sqrt{2} \,e^{\varphi} \, Z & \quad 1 \,  \\
- \sqrt{2} \,e^{\varphi} \,  {\bar Z}  & \quad - 1  & \quad    0 
  \\
1 \,   & \quad  0    &  \quad 0 
\end{pmatrix} \;,
\label{Mads2}
\end{eqnarray}
with 
$Z = \frac{Q}{|Q|^2} \, r$ and $\Sigma = |Z|^2 = e^{- \varphi}$.  The gauge potentials $\chi_e$ and $\chi_m$ can always be shifted by constants.
Note that with the above choice \eqref{backg} of the gauge potentials, the matrix $M$
exhibits a triangular structure that will be inherited by the monodromy matrix ${\cal M}$, making it simpler to factorize it. Adding constants to
the gauge potentials will result in more complicated monodromy matrices. These can be dealt
with in a manner described below.

In Weyl coordinates $(\rho,v)$, given by
\begin{eqnarray}
\rho &=& r\, \sin \theta \;, \nonumber\\
v &=& r \cos \theta \;,
\label{weylads2}
\end{eqnarray}
with $r> 0 $ and $0 < \theta < \pi$. Hence $\rho > 0$. The line element \eqref{lineads2su21} takes the form \eqref{line32} and \eqref{weylrv}
with $e^{\psi} = 1$, and we obtain for $M$,
\begin{eqnarray}
M(\rho, v) = \frac{1}{\rho^2 + v^2}
\begin{pmatrix}
\tfrac12 |a|^2 & \quad  a \, \sqrt{\rho^2 + v^2} & \quad  \rho^2 + v^2 \\
- {\bar a} \, \sqrt{\rho^2 + v^2} 
 & \quad - (\rho^2 + v^2)  & \quad    0 
  \\
 \rho^2 + v^2   & \quad  0    &  \quad 0 
\end{pmatrix} \;,
\label{outbhsol}
\end{eqnarray}
where $a = \sqrt{2} \, Q$.

Now we consider the substitution rule \eqref{substruleMM}. In the limit $\rho \rightarrow 0^+$ we obtain
\begin{eqnarray}
M(\rho=0, v) = \frac{1}{v^2}
\begin{pmatrix}
\tfrac12 |a|^2 & \quad  a \, |v| & \quad  v^2 \\
- {\bar a} \, |v|
 & \quad - v^2  & \quad    0 
  \\
 v^2   & \quad  0    &  \quad 0 
\end{pmatrix} \;.
\end{eqnarray}
For $v > 0$,  the substitution rule \eqref{substruleMM} associates the monodromy matrix
\begin{eqnarray}
{\cal M}(\omega) = \frac{1}{\omega^2}
\begin{pmatrix}
\tfrac12 |a|^2 & \quad  a \, \omega & \quad  \omega^2 \\
- {\bar a} \, \omega
 & \quad - \omega^2  & \quad    0 
  \\
 \omega^2   & \quad  0    &  \quad 0 
\end{pmatrix}  \;\;\;,\;\; \det {\cal M} = 1 \;, 
\label{monads2}
\end{eqnarray}
to $M(\rho, v)$, while for $v< 0$ it associates the monodromy matrix 
\begin{eqnarray}
{\cal M}(\omega) = \frac{1}{\omega^2}
\begin{pmatrix}
\tfrac12 |a|^2 & \quad  - a \, \omega & \quad  \omega^2 \\
{\bar a} \, \omega
 & \quad - \omega^2  & \quad    0 
  \\
 \omega^2   & \quad  0    &  \quad 0 
\end{pmatrix}  \;\;\;,\;\; \det {\cal M} = 1  \;.
\label{outbhsolvneg}
\end{eqnarray}
Both monodromy matrices are related by the replacement $a \rightarrow - a$.
 Thus, it suffices to analyze
the factorization of \eqref{monads2}.

Let us return to the choice of gauge potentials \eqref{backg}. Shifting these gauge potentials
by constants,
\be
\chi_e \rightarrow \chi_e + {\rm Re} \, \alpha \;\;\;,\;\; \chi_m \rightarrow \chi_m + {\rm Im} \,  \alpha \;\;\;,\;\;\; \alpha \in \mathbb{C} \;,
\label{shiftgaugep}
\ee
results in a  $\natural$-symmetric coset representative given by
\begin{eqnarray}
{\tilde M} = 
\begin{pmatrix}
e^{\varphi} & \quad  \sqrt{2} \,e^{\varphi} \, \left( Z + \alpha \right) & \quad e^{\varphi} \tilde{ \Sigma} \,  \\
- \sqrt{2} \,e^{\varphi} \, \left(  {\bar Z} + \bar \alpha \right)  & \quad 1 - 2 e^{\varphi} | Z + \alpha|^2   & \quad    \sqrt{2} \left( \bar Z + \bar \alpha \right) 
\left(1 - e^{\varphi} \, {\tilde \Sigma} \right)
  \\
e^{\varphi} \bar{ \tilde{\Sigma}} \, \,   & \quad  -  \sqrt{2} \left(  Z + \alpha \right) 
\left(1 - e^{\varphi} \, \bar{\tilde \Sigma} \right)
    &  \quad e^{\varphi} |\tilde \Sigma|^2 - 2 | Z + \alpha |^2 + e^{- \varphi}
\end{pmatrix} 
\end{eqnarray}
with 
\be
\tilde \Sigma = |Z + \alpha|^2 - \alpha \, {\bar Z} + \bar{\alpha} \, Z \;.
\ee
This coset representative does not any longer exhibit a triangular structure. However, it is related to $M$ given in \eqref{Mads2}
by group multiplication, 
\be
{\tilde M} = g^{\natural} (\alpha) \, M \, g(\alpha) \;,
\label{shiftgpg}
\ee
with 
\begin{equation}
g (\alpha) =  \left( \begin{array}{ccc}
1 & \quad  \sqrt{2} \alpha  & \quad  |\alpha|^2 \\
0  & \quad 1 & \quad \sqrt{2} \bar{\alpha} \\
0 & \quad 0 & \quad 1 \end{array} \right) 
\end{equation}
and
\begin{equation}
g^{\natural} (\alpha) =  \left( \begin{array}{ccc}
1 & \quad 0 & \quad 0 \\
-\sqrt{2} \bar{\alpha} & \quad 1 & \quad 0 \\
|\alpha|^2 & \quad -\sqrt{2} \alpha & \quad 1 \end{array} \right) \;,
\end{equation}
which satisfies the group relation \eqref{gkgk}.  The associated monodromy matrix is then given by 
\be
{\tilde {\cal M}} = g^{\natural} (\alpha) \, {\cal M} \, g(\alpha) \;,
\ee
with ${\cal M}$ given by either \eqref{monads2} or \eqref{outbhsolvneg}.

Let us now consider the factorization of ${\cal M} (\omega)$
given in
\eqref{monads2}. Note that
${\cal M} (\omega)$
 is bounded and
asymptotes to
\begin{eqnarray}
{\cal M} (\omega = \infty) = 
\begin{pmatrix}
0 & \quad 0 & \quad 1  \\
0  & \quad -1 &  \quad  0 
  \\
1 \,   & \quad 0
   & \quad  0 
\end{pmatrix} \;.
\end{eqnarray}

To obtain the canonical factorization of \eqref{monads2}, we use the factorization technique
described in section \ref{intromath}.
We set up the factorization problem in the form
\be
{\cal M} (\omega) \,M_+^{-1}  = M_- \;\;\;,\;\; \tau \in \Gamma \;,
\ee
and we consider the vectorial factorization problem
\begin{eqnarray}
\label{matrvec}
\frac{1}{\omega^2}
\begin{pmatrix}
\frac12 |a|^2 &  \quad a \, \omega \,  &\quad  \omega^2 \,  \\
- {\bar a} \, \omega  & \quad - \omega^2  & \quad   0 
  \\
\omega^2 \,   & \quad 0
   &  \quad 0 
\end{pmatrix} 
\begin{pmatrix}
\phi_{1+} \\
\phi_{2+} \\
\phi_{3+}  
\end{pmatrix} = 
\begin{pmatrix}
\phi_{1-} \\
\phi_{2-} \\
\phi_{3-}  
\end{pmatrix} \;,
\end{eqnarray}
where the $\pm$-columns denote the first columns of $M_+^{-1}$ and of $M_-$, respectively.
Similar equations hold for the
second and third columns of $M_+^{-1}$ and of $M_-$, respectively.

Since $M_+^{-1}(\tau=0) = \mathbb{I}$, we have the normalization conditions 
\be
\phi_{1+} (0) = 1 \;\;\;,\;\;\; \phi_{2+} (0) = 0 \;\;\;,\;\;\; \phi_{3+} (0) = 0 \;.
\label{normcolumn1}
\ee
Now we spell out the three relations that we get from \eqref{matrvec}. 
The last one gives
\be
\phi_{1+} = \phi_{3-} \;\;\;,\;\;\;  \tau \in \Gamma \;.
\ee
Since $\phi_{1+} $ is analytic and bounded in the interior $D_+$ of the unit disc, and since $\phi_{3-}$ is analytic and bounded in the exterior region $D_-$, and since both are equal to one another on the unit circle $\Gamma$,
they together define an entire function $f(\tau)$ given by
\begin{equation}
f(\tau) = \left\{
\begin{matrix}
\phi_{1+}  \;\;\;&,&\;\;\; \tau \in D_+\\
\phi_{1+} = \phi_{3-}  \;\;\;&,&\;\;\; \tau \in \Gamma \\
\phi_{3-}  \;\;\;&,&\;\;\;  \tau \in D_-
\end{matrix} 
\right.
\end{equation}
Hence, by Liouville's theorem, $f(\tau)$ is constant, 
\be
\phi_{1+} = \phi_{3-} = K \;,
\ee
with $K$ a constant. Imposing the normalization condition \eqref{normcolumn1} we get
\be
\phi_{1+} = \phi_{3-} = 1 \;.
\ee

The second relation resulting from \eqref{matrvec} yields
\be
- \frac{\bar a}{\omega} \, \phi_{1+} - \phi_{2+} = \phi_{2-} \;\;\;,\;\;\; \tau \in \Gamma\;,
\ee
and hence
\be
- \frac{\bar a}{\omega}  - \phi_{2+} = \phi_{2-} \;\;\;,\;\;\; \tau \in \Gamma \;.
\ee
Using \eqref{wtaupm0} this yields
\be
 \frac{2\bar a}{\rho} \, \frac{\tau}{(\tau - \tau_{0}^+) (\tau - \tau_{0}^-)} - \phi_{2+} = \phi_{2-} \;.
\label{secrel}
\ee
We decompose
\be
 \frac{\tau}{(\tau - \tau_{0}^+) (\tau - \tau_{0}^-)} =  \frac{A}{(\tau - \tau_{0}^+) } + \frac{B}{(\tau - \tau_{0}^-)}
 \;,
\ee
where
\be
A = \frac{\tau_0^+}{\tau_0^+ - \tau_0^-} \;\;\;,\;\; B = - \frac{\tau_0^-}{\tau_0^+ - \tau_0^-} \;.
\ee
Then, \eqref{secrel} results in
\be
 \frac{2\bar a}{\rho} \, \frac{B}{ (\tau - \tau_{0}^-)} - \phi_{2+} = \phi_{2-} 
 - \frac{2\bar a}{\rho} \, \frac{A}{(\tau - \tau_{0}^+) }\;,
\ee
for $\tau$ on the unit circle $\Gamma$.
The terms on the left hand side are analytic and bounded in $D_+$, while the terms on the right hand side are
analytic and bounded in $D_-$. Since they are equal to one another on the unit circle $\Gamma$, when taken together, they
define an entire function. Hence, by Liouville's theorem,
\be
 \frac{2\bar a}{\rho} \, \frac{B}{ (\tau - \tau_{0}^-)} - \phi_{2+} = \phi_{2-} 
 - \frac{2\bar a}{\rho} \, \frac{A}{(\tau - \tau_{0}^+) } = k \;,
\ee
with $k$ a constant. Then, at $\tau =0$ we obtain
\be
k =  - \frac{2\bar a}{\rho} \, \frac{B}{  \tau_{0}^-} \;,
\ee
and hence
\be
\phi_{2+} = \frac{2\bar a}{\rho} \,B \,  \frac{\tau}{ (\tau - \tau_{0}^-)} = 
\frac{-2 \bar a}{\rho \, (\tau_0^+ - \tau_0^-) } \,\frac{\tau}{(\tau - \tau_0^-)}\;,
\ee
as well as
\be
\phi_{2-} =  
\frac{2 \bar a}{\rho \, (\tau_0^+ - \tau_0^-) } \,\frac{\tau}{(\tau - \tau_0^+)}\;.
\ee

Finally, the first relation in \eqref{matrvec} reads
\be
\tfrac12 \frac{|a|^2}{\omega^2} \, \phi_{1+} + \frac{|a|^2}{\omega} \, \phi_{2+} + \phi_{3+} = \phi_{1-} \;.
\label{aaom}
\ee
Using \eqref{wtaupm0}, we decompose
\be
\frac{1}{\omega^2} = \frac{4}{\rho^2} \left[ \frac{C_1}{(\tau - \tau_0^+)^2} + \frac{C_2}{\tau - \tau_0^+}
+ \frac{ C_3 }{  (\tau - \tau_0^- )^2 } 
+ \frac{C_4}{\tau - \tau_0^-}  
\right] \;,
\label{omdecomp}
\ee
where
\be
C_1 = \frac{(\tau_0^+)^2}{(\tau_0^+ - \tau_0^-)^2} \;\;\;,\;\;\; C_2 = - \frac{2 \tau_0^+ \tau_0^-}{(\tau_0^+ - \tau_0^-)^3} = - C_4 \;\;\;,\;\;\; C_3 = \frac{(\tau_0^-)^2}{(\tau_0^+ - \tau_0^-)^2} \;.
\label{valuesC}
\ee
We will also need 
\be
\frac{\tau^2}{(\tau - \tau_0^+)(\tau - \tau_0^-)^2} = \frac{D_1}{\tau - \tau_0^+} + \frac{D_2}{(\tau - \tau_0^-)^2}
+ \frac{D_3}{\tau - \tau_0^-} \;,
\label{decomptt}
\ee
where
\be
D_1 = \frac{(\tau_0^+)^2}{(\tau_0^+ - \tau_0^-)^2} \;\;\;,\;\;\; D_2 = - \frac{(\tau_0^-)^2}{\tau_0^+ - \tau_0^-} 
\;\;\;,\;\;\; D_3 = \frac{- 2 \tau_0^+ \tau_0^- + (\tau_0^-)^2}{(\tau_0^+ - \tau_0^-)^2} \;.
\label{valuesD}
\ee
Making use of both \eqref{omdecomp} and \eqref{decomptt} in \eqref{aaom} we obtain
\begin{eqnarray}
 \frac{2|a|^2}{\rho^2} \left[ \frac{C_3}{  (\tau - \tau_0^- )^2 } 
+ \frac{C_4}{ \tau - \tau_0^- } \right] + \frac{4 |a|^2}{\rho^2 \, (\tau_0^+ - \tau_0^-) } \left[ \frac{D_2}{  (\tau - \tau_0^- )^2 } 
+ \frac{D_3}{ \tau - \tau_0^- } \right] + \phi_{3+} = \nonumber\\
\phi_{1-} - 
 \frac{2|a|^2}{\rho^2} \left[ \frac{C_1}{  (\tau - \tau_0^+ )^2 } 
+ \frac{C_2}{ \tau - \tau_0^+ } \right] - \frac{4 |a|^2}{\rho^2 \, (\tau_0^+ - \tau_0^-) }  \frac{D_1}{ \tau - \tau_0^+ } \;.
\label{firrelpp}
\end{eqnarray}
The terms on the left hand side are analytic and bounded in $D_+$, while the terms on the right hand side are
analytic and bounded in $D_-$. Since they are equal to one another on the unit circle $\Gamma$, 
when taken together, they
define an entire function, and hence, by Liouville's theorem, both sides of \eqref{firrelpp} have be constant.
Denoting this constant by $K$, its value can be inferred by evaluating the left hand side at $\tau =0$,
\be
K = \frac{2|a|^2}{\rho^2} \left[ \frac{C_3}{  (\tau_0^- )^2 } 
- \frac{C_4}{ \tau_0^- } \right] + \frac{4 |a|^2}{\rho^2 \, (\tau_0^+ - \tau_0^-) } \left[ \frac{D_2}{  (\tau_0^- )^2 } 
- \frac{D_3}{ \tau_0^- } \right] = \frac{2 |a|^2}{\rho^2 \, (\tau_0^+ - \tau_0^-)^2} \;,
\ee
where we used \eqref{normcolumn1}. Then, we obtain from \eqref{firrelpp}
\begin{eqnarray}
\phi_{1-} &=&
\frac{2 |a|^2}{\rho^2 \, (\tau_0^+ - \tau_0^-)^2} \frac{\tau^2}{(\tau - \tau_0^+)^2} \;, \nonumber\\
\phi_{3+} &=& \frac{2 |a|^2}{\rho^2 \, (\tau_0^+ - \tau_0^-)^2} \frac{\tau^2}{(\tau - \tau_0^-)^2} \;.
\end{eqnarray}

Proceeding in a similar manner for the vectorial factorization problem based on the second and third columns of 
$M_+^{-1}$ and of $M_-$, respectively, and imposing the normalization condition $M_+^{-1}(\tau =0, x) = \mathbb{I}$,
we determine the explicit form of $M_+^{-1}$ and of $M_-$,
\begin{eqnarray}
{M}_+^{-1} (\tau, x) &=& 
\begin{pmatrix}
1  & \quad 0 & \quad 0  \\
\frac{-2 \bar a}{\rho \, (\tau_0^+ - \tau_0^-) } \,\frac{\tau}{\tau - \tau_0^-}  & \quad 1  &  \quad  0 
  \\
\frac{2 |a|^2}{\rho^2 \, (\tau_0^+ - \tau_0^-)^2} \,  \frac{\tau^2}{(\tau - \tau_0^-)^2} \qquad  & \quad 
\frac{- 2 a }{\rho \, (\tau_0^+ - \tau_0^-) } \,\frac{\tau}{\tau - \tau_0^-} 
   & \quad  1 
\end{pmatrix} \;
\label{facMpunder}
\end{eqnarray}
as well as 
\begin{eqnarray}
{M}_- (\tau, x) &=& 
\begin{pmatrix}
\frac{2 |a|^2}{\rho^2 \, (\tau_0^+ - \tau_0^-)^2} \,  \frac{\tau^2}{(\tau - \tau_0^+)^2} \qquad  
& \quad \frac{-2 a}{\rho \, (\tau_0^+ - \tau_0^-) } \,\frac{\tau}{\tau - \tau_0^+} 
& \quad 1 \,  \qquad \\
 \frac{2 \bar a }{\rho \, (\tau_0^+ - \tau_0^-) } \,\frac{\tau}{\tau - \tau_0^+}
  & \quad -1  &   \quad 0 \qquad 
  \\
1 \,   & \quad 0
   &  \quad 0 
\end{pmatrix} \;.
\end{eqnarray}
$M_+$ is analytic inside the unit circle and satisfies $M_+ (\tau =0, x) = \mathbb{I}$, 
while $M_-$ is analytic outside the unit circle. The spacetime solution $M(x)$ is given by
\begin{eqnarray}
M(x) = {M}_- (\tau = \infty, x) = 
\begin{pmatrix}
\frac{2 |a|^2}{\rho^2 \, (\tau_0^+ - \tau_0^-)^2}  \qquad  
&\quad  \frac{-2 a}{\rho \, (\tau_0^+ - \tau_0^-) } 
& \quad 1 \,  \qquad \\
 \frac{2 \bar a }{\rho \, (\tau_0^+ - \tau_0^-) } 
  &\quad  -1  & \quad   0 \qquad 
  \\
1 \,   & \quad 0
   & \quad  0 
\end{pmatrix} \;.
\end{eqnarray}

Having determined $M(x)$ from factorization, we now verify that it satisfies the substitution rule
\eqref{substruleMM}. When $v > 0$, we have \eqref{tvpos}, and hence
\begin{eqnarray}
M(\rho, v) = 
\begin{pmatrix}
\frac{ |a|^2}{2(\rho^2 + v^2) }  \qquad  
& \quad \frac{ a}{ \sqrt{\rho^2 + v^2} } 
& \quad 1 \,  \qquad \\
\frac{- \bar a }{ \sqrt{\rho^2 + v^2} } 
  & \quad -1  &  \quad  0 \qquad 
  \\
1 \,   & \quad 0
   & \quad 0 
\end{pmatrix} \;.
\end{eqnarray}
This precisely equals  \eqref{outbhsol}, and hence we have recovered the solution describing the near-horizon regime of 
a static extremal black hole with parameter $a$.
Setting $\rho=0$, we obtain
\begin{eqnarray}
M(\rho=0, v) = \frac{1}{v^2} 
\begin{pmatrix}
\tfrac12 |a|^2  \qquad  
& \quad a \,  v 
& \quad v^2 \,  \qquad \\
- {\bar a} \, v 
  &\quad  -v^2  & \quad   0 \qquad 
  \\
v^2\,   & \quad 0
   &  \quad 0 
\end{pmatrix} \;,
\end{eqnarray}
and hence $M(\rho=0, v) = {\cal M} (\omega=v)$, in accordance with the substitution rule \eqref{substruleMM}. 

On the other hand, when $v < 0$, we have \eqref{tvneg}, and hence
\begin{eqnarray}
M(\rho, v) = 
\begin{pmatrix}
\frac{ |a|^2}{2(\rho^2 + v^2) }  \qquad  
& \quad  \frac{- a}{ \sqrt{\rho^2 + v^2} } 
& \quad 1 \,  \qquad \\
\frac{\bar a }{ \sqrt{\rho^2 + v^2} } 
  &\quad  -1  & \quad   0 \qquad 
  \\
1 \,   &\quad  0
   &  \quad 0 
\end{pmatrix} \;.
\label{mvs0ma}
\end{eqnarray}
Note that this describes the near-horizon regime of 
a static extremal black hole with parameter $-a$. 
Let us compare \eqref{mvs0ma} with the solution, for $v<0$,
that we would obtain by factorizing the monodromy matrix \eqref{outbhsolvneg}.
Both solutions are related by $a \rightarrow -a$, since the associated monodromy
matrices \eqref{monads2} and \eqref{outbhsolvneg} are related in that manner.

In  \eqref{mvs0ma}, 
setting $\rho =0$, we get
\begin{eqnarray}
M(\rho=0, v) = \frac{1}{v^2} 
\begin{pmatrix}
\tfrac12 |a|^2  \qquad  
& \quad -a \,  |v| 
& \quad v^2 \,  \qquad \\
{\bar a} \, |v| 
  & \quad -v^2  & \quad   0 \qquad 
  \\
v^2\,   &\quad  0
   &\quad   0 
\end{pmatrix} \;,
\end{eqnarray}
and hence $M(\rho=0, v) = {\cal M} (\omega=v)$, again in accordance with the substitution rule \eqref{substruleMM}.

\subsubsection{$G/H=SL(3, \mathbb{R})/SO(2,1)$: rotating attractors \label{subsecsl3}}

In this section we consider 
the near-horizon region of extremal rotating black holes in a 
four-dimensional Einstein-Maxwell-dilaton theory that arises by dimensional reduction on a circle of a five-dimensional
Einstein theory. These solutions, which may be underrotating or overrotating 
\cite{Astefanesei:2006dd},
are  supported
by one electric charge $Q$ and one magnetic charge $P$, and they carry angular momentum $J$. 
The underrotating black holes have a static limit, and the associated monodromy matrix is
bounded, while in the overrotating case it is unbounded.

The associated five-dimensional line elements are of the form \cite{Chakrabarty:2014ora}
\begin{equation}
ds^2_5 = e^{\frac{\phi_1}{\sqrt{3}} + \phi_2} ds^2_3
- e^{\frac{\phi_1}{\sqrt{3}} - \phi_2}
 \left( dt + {\cal A}_2 \right)^2 + e^{\frac{-2\phi_1}{\sqrt{3}}} \left(d {\tilde \psi} + \chi_1 dt + {\cal A}_1\right)^2
 \;.
 \end{equation}
Reducing to three dimensions along the two isometry directions $(t, {\tilde \psi})$, 
the one-forms ${\cal A}_{1,2}$ can be dualized into scalars $\chi_2, \chi_3$ in three dimensions: introducing
the field strengths ${\cal F}_1 = d  {\cal A}_1 + {\cal A}_2 \wedge d \chi_1$ and 
${\cal F}_2 = d  {\cal A}_2$, the dualization is performed by means of the relations
\cite{Cortes:2014jha,Chakrabarty:2014ora}
\be
e^{- \sqrt{3} \phi_1 - \phi_2} * {\cal F}_1 = d \chi_2 \;\;\;,\;\;\; 
- e^{ -2 \phi_2} * {\cal F}_2 = d \chi_3 - \chi_1 \, d \chi_2 \;,
\label{f1f2}
\ee
where the dual $*$ is with respect to the three-dimensional metric.

Let us first consider the near-horizon regime of an {\sl underrotating} extremal black hole.
In spherical coordinates, the line element describing the near-horizon region is given by \cite{Astefanesei:2006dd}
\begin{equation}
ds^2_4 = - \frac{r^2}{v_1 (\theta)} \left( dt - \frac{J \, \sin^2 \theta}{8 \pi \, r} \, d\phi \right)^2 + 
v_1(\theta) \left(\frac{dr^2}{r^2} + d \theta^2 + \sin^2 \theta \, d\phi^2 \right)
\end{equation}
with 
\begin{equation}
v_1 (\theta) = \frac{1}{8 \pi} \sqrt{P^2 Q^2 - J^2 \, \cos^2 \theta} \;.
\end{equation}
Written as
\begin{eqnarray}
ds^2_4 = - e^{-\phi_2} \left(dt + B_{\phi} \, d \phi \right)^2 + e^{\phi_2} \, \left[dr^2 + r^2 \left( d \theta^2 + \sin^2 \theta \, d\phi^2 \right) \right]  \;,
\label{lineund}
\end{eqnarray}
we infer
\begin{eqnarray}
e^{\phi_2} &=& \frac{v_1(\theta)}{r^2} \;, \nonumber\\
B_{\phi}&=& \frac{b(\theta)}{r} \;\;\;,\;\;\; b (\theta) = - \frac{J \, \sin^2 \theta}{8 \pi } \;.
\label{B-phi}
\end{eqnarray}
The fields entering in the $\natural$-symmetric representative \eqref{cosetrepM} and in \eqref{sigphi} are
\begin{eqnarray}
e^{2 \phi_1/\sqrt{3}} &=& \left( \frac{P}{Q} \right)^{2/3} \, \frac{PQ - J \cos \theta}{PQ + J \cos \theta} \;\;\;,\;\;\;
e^{\phi_2} = \frac{v_1(\theta)}{r^2} \;,
\nonumber\\
\chi_1 
&=& r \, g_1 (\theta) \;\;\;,\;\;\; g_1(\theta) =  - \frac{4 \sqrt{\pi}}{Q} \frac{1}{1 + \mu \, \cos \theta} \;, \nonumber\\
\chi_2 &=& r \, g_2(\theta) \;\;\;,\;\;\; g_2(\theta) = \left( \frac{F_{\theta \phi} + b(\theta) \, \partial_{\theta} g_1 (\theta)}{v_1 (\theta) \, \sin \theta} \right) e^{-\sqrt{3} \, \phi_1}
\;, \nonumber\\
\chi_3 &=& \tilde{\phi} + \tfrac12 \, \chi_1 \chi_2 = \tfrac12 \, r^2 \,
g_3 (\theta) 
 \;\;\;,\;\;\; \tilde{\phi} = \tfrac12 r^2 \, \frac{ \partial_{\theta} b (\theta)}{v_1^2 (\theta) \, \sin \theta} \;, \nonumber\\
\end{eqnarray}
where
\begin{eqnarray}
v_1 (\theta) &=& \frac{1}{8 \pi} \sqrt{P^2 Q^2 - J^2 \, \cos^2 \theta} \;, \nonumber\\
 b (\theta) &=&- \frac{J \, \sin^2 \theta}{8 \pi } \;, \nonumber\\
 F_{\theta \phi} &=& \frac{1}{2 \sqrt{\pi}} \, P \, \sin \theta \frac{(1- \mu^2)}{(1 + \mu \, \cos \theta)^2} \;, \nonumber\\
 \mu &=& \frac{J}{PQ} \;.
 \label{fivbF}
 \end{eqnarray}
 We compute
 \begin{eqnarray}
  g_1(\theta) &=&-  \frac{4 \sqrt{\pi}  P}{\lambda_+} \;, \nonumber\\
 g_2(\theta) &=& \frac{4 \sqrt{\pi}  Q}{\lambda_-} \;, \nonumber\\
g_3 (\theta) &=& - \frac{16 \pi}{\lambda_-} \;,
\end{eqnarray}
where
\begin{equation}
\lambda_{\pm} = PQ \pm J \cos \theta \;.
\end{equation}
Thus, we obtain for the coset representative $M(r, \theta)$ evaluated on the solution,
 \begin{eqnarray}
{M}(r, \theta) = \begin{pmatrix}
\frac{1}{8 \pi} \left(\frac{P}{Q} \right)^{1/3} \frac{(PQ - J \cos \theta)}{r^2} & \frac{P^{1/3} Q^{2/3} }{2 \sqrt{\pi}\, r}   &- \left( \frac{P}{Q} \right)^{1/3}
 \\
- \frac{P^{1/3} Q^{2/3} }{2 \sqrt{\pi} \,r}   & - \left( \frac{Q}{P} \right)^{2/3}
 &  0
  \\
  - \left( \frac{P}{Q} \right)^{1/3} &
0
   & 0
\end{pmatrix} \;.
\label{rotatattr}
\end{eqnarray}
Note that $M(r, \theta)$ is bounded at $r = \infty$ and that 
$\det M(r, \theta) = 1 $.
Also
note that while the expressions in \eqref{fivbF} exhibit a rather complicated dependence on $\theta$, 
the $\natural$-symmetric coset representative $M(r,\theta)$ only depends on $\cos \theta$ in a simple way.

We now associate a monodromy matrix ${\cal M} (\omega)$ to $M$.
To this end, we convert to Weyl coordinates $(\rho, v)$. Using  \eqref{line32} and \eqref{weylrv}, we
infer
\begin{equation}
e^{\psi} = 1 \;\;\;,\;\;\; \rho = r \, \sin \theta \;\;\;,\;\;\; v = r \, \cos \theta \;,
\label{valuepsiweyl}
\end{equation}
with $r> 0 $ and $0 < \theta < \pi$, and hence $\rho > 0$. Then, the coset representative $M$ becomes
 \begin{eqnarray}
{M}(\rho, v) = \frac{1}{\rho^2 + v^2} \begin{pmatrix}
- \frac{B^2}{2D} + \frac{(2AD + B^2)}{2D} \frac{v}{ \sqrt{\rho^2 + v^2}  } & \quad B \sqrt{\rho^2 + v^2}   &\quad C (\rho^2 + v^2)
 \\
- B \sqrt{\rho^2 + v^2}  & D (\rho^2 + v^2)
 &  0
  \\
  C (\rho^2 + v^2) &
0
   & 0
\end{pmatrix} \;,
\label{rotatattrrv}
\end{eqnarray}
where
\begin{eqnarray}
\label{valuesABCD}
B= \frac{1}{2 \sqrt{\pi}} \, P^{1/3} \, Q^{2/3} \;\;,\;\; C = - \left(\frac{P}{Q} \right)^{1/3}
 \;\;,\;\; D = - \left(\frac{Q}{P} \right)^{2/3} \;\;,\;\; 2 AD + B^2 = \frac{1}{4 \pi}  \left(\frac{Q}{P} \right)^{1/3} \, J  \;. \nonumber\\
\end{eqnarray}
Note that $- C^2 D =1$.
Static black holes correspond to
\begin{equation}
 2 AD + B^2 = 0 \;.
 \end{equation}

Now we consider the substitution rule \eqref{substruleMM}. In the limit $\rho \rightarrow 0^+$ we obtain
 \begin{eqnarray}
{M}(\rho=0, v) = \frac{1}{v^2} \begin{pmatrix}
- \frac{B^2}{2D} + \frac{(2AD + B^2)}{2D} \frac{v}{ |v|  } & \quad B |v|   &\quad C  v^2
 \\
- B |v|  & D v^2
 &  0
  \\
  C v^2 &
0
   & 0
\end{pmatrix} \;.
\label{rotatattrrv2}
\end{eqnarray}
 For $v > 0$,  the substitution rule \eqref{substruleMM} associates the monodromy matrix
\begin{eqnarray}
{\cal M} (\omega) = 
 \frac{1}{\omega^2} \begin{pmatrix}
A  & \quad B \omega   &\quad C  \omega^2
 \\
- B \omega  & D  \omega^2
 &  0
  \\
  C \omega^2 &
0
   & 0
\end{pmatrix} 
\label{underrotmonvg0}
\end{eqnarray}
to $M(\rho, v)$, while for $v < 0$ it associates 
 the monodromy matrix
\begin{eqnarray}
{\cal M} (\omega) = 
 \frac{1}{\omega^2} \begin{pmatrix}
 - \frac{(AD + B^2)}{D}  & \quad - B \omega   &\quad C  \omega^2
 \\
 B \omega  & D  \omega^2
 &  0
  \\
  C \omega^2 &
0
   & 0
\end{pmatrix} \;.
\label{underrotmonvl0}
\end{eqnarray}
Both monodromy matrices are related by 
$B \rightarrow -B, \; J \rightarrow -J$ and hence, it suffices to study the factorization of
\eqref{underrotmonvg0} .

Note that we may shift the gauge potentials $\chi_1$ and $\chi_2$ by constants. For instance, shifting $\chi_2$ by 
a constant can be effected by the transformation $g^{\natural} (\alpha) {\cal M} g(\alpha)$ with
\begin{equation}
g(\alpha) =  \begin{pmatrix}
1 &\quad  \alpha & \quad 0 \\
0& 
\quad 1 & \quad 0\\
0& \quad 0 & \quad 1
\end{pmatrix} 
\end{equation}
which, when applied to \eqref{underrotmonvg0}, results in 
\begin{equation}
{\tilde {\cal M}} = g^{\natural} (\alpha) \, {\cal M} \, g(\alpha)=  \begin{pmatrix}
\frac{A}{\omega^2} & \quad \frac{\alpha A}{\omega^2}+\frac{B}{\omega}& \quad C\\
- \frac{\alpha A}{\omega^2}-\frac{B}{\omega} &\quad  \frac{-\alpha ^2 A}{\omega^2}-\frac{2 \alpha B}{\omega} +D & \quad - \alpha C\\
C  &\quad  \alpha C & \quad 0
\end{pmatrix} \;.
\end{equation}

Let us now factorize the monodromy matrix 
\eqref{underrotmonvg0} associated to an underrotating black hole. Note that 
$\det {\cal M}= 1$, since $- C^2 D = 1$.
 Then, factorizing  ${\cal M} (\omega)$ into
${\cal M} (\omega) = M_- (\tau, x) M_+ (\tau,x)$ using the vectorial factorization technique described in the previous subsection and the relation \eqref{wtaupm0}, yields
\begin{eqnarray}
{M}_- (\tau, x) = 
\begin{pmatrix}
m_{11} & \quad - \frac{2 B}{\rho (\tau_0^+ - \tau_0^-)} \, \frac{\tau}{(\tau - \tau_0^+)} &\quad  C \\
 \frac{2 B}{\rho (\tau_0^+ - \tau_0^-)} \, \frac{\tau}{(\tau - \tau_0^+)} & \quad D &\quad    0 
  \\
C  & \quad 0
   & \quad 0 
\end{pmatrix} \;,\nonumber\\
{M}^{-1}_+ (\tau, x) = 
\begin{pmatrix}
1 & \quad 0 &\quad  0 \\
\frac{2 B}{D \, \rho (\tau_0^+ - \tau_0^-)} \, \frac{\tau}{(\tau - \tau_0^-)}
 & \quad 1 &   \quad 0 
  \\
m_{31}  &\quad  - \frac{2 B}{C \, \rho (\tau_0^+ - \tau_0^-)} \, \frac{\tau}{\tau -\tau_0^-)}
   &  \quad 1
\end{pmatrix} \;.
\label{undefMM}
\end{eqnarray}
The entries $m_{11}$ and $m_{31}$ are given by
\begin{eqnarray}
m_{11} &=& \frac{4A}{\rho^2} \left(\frac{c_1}{(\tau - \tau_0^+)^2} + \frac{c_2}{(\tau - \tau_0^+)} \right) - \frac{4 B^2}{D \rho^2  (\tau_0^+ - \tau_0^-)} \frac{d_1}{(\tau - \tau_0^+)}
\nonumber\\
&&+ \frac{4}{ \rho^2  (\tau_0^+ - \tau_0^-)^2 } \left(A - \frac{(2 AD + B^2)}{D} \, \frac{\tau_0^+}{ (\tau_0^+ - \tau_0^-)} \right) \;, \nonumber\\
m_{31} &=& \frac{4}{ C \, \rho^2  (\tau_0^+ - \tau_0^-)^2 } \left(A - \frac{(2 AD + B^2)}{D} \, \frac{\tau_0^+}{ (\tau_0^+ - \tau_0^-)} \right) - 
\frac{4 A}{C \, \rho^2} \left( \frac{c_3}{(\tau - \tau_0^-)^2} + \frac{c_4}{(\tau- \tau_0^-)} \right) \nonumber\\
&& + \frac{4 B^2}{CD \, \rho^2 \, (\tau_0^+ - \tau_0^-)} \left( \frac{d_2}{(\tau - \tau_0^-)^2} + \frac{d_3}{(\tau - \tau_0^-)} \right) \;,
\label{m11m31}
\end{eqnarray}
with constants $c_i, d_i$ that equal those given in \eqref{valuesC}
and in \eqref{valuesD}, respectively.
Given these expressions, it can be checked in a direct manner that the above matrix expressions satisfy ${\cal M} \, M_+^{-1} = M_-$, and that $M_+^{-1} (\tau=0, x) = \mathbb{I}$.

The spacetime solution $M(x)$ is given by
\begin{eqnarray}
M(x) = {M}_- (\infty, x) = 
\begin{pmatrix}
\frac{4}{\rho^2 (\tau_0^+ - \tau_0^-)^2}\left( A - \frac{(2AD + B^2)}{D} \frac{\tau_0^+}{\tau_0^+ - \tau_0^-} \right) & \quad - \frac{2B}{\rho (\tau_0^+ - \tau_0^-)} & \quad C \,  \\
\frac{2B}{\rho (\tau_0^+ - \tau_0^-)}  & \quad D &  \quad  0 
  \\
C  & \quad 0
   &  \quad 0 
\end{pmatrix} \;.\nonumber\\
\label{minftm}
\end{eqnarray}

Having determined $M(x)$ from factorization, we now verify that it satisfies the substitution rule
\eqref{substruleMM}. When $v > 0$, we have \eqref{tvpos}, and hence
\begin{eqnarray}
M(\rho,v) = \frac{1}{\rho^2 + v^2}
\begin{pmatrix}
A + \frac{(2AD + B^2)}{2D}\left( \frac{v}{\sqrt{\rho^2 + v^2}}-1\right)  & \quad B \, \sqrt{\rho^2 +v^2}  & \quad  C \,  (\rho^2 + v^2) \\
- B \, \sqrt{\rho^2 +v^2}  & \quad D \,  (\rho^2 + v^2)&   \quad 0 
  \\
C  \,  (\rho^2 + v^2) & \quad 0
   &  \quad 0 
\end{pmatrix} \;.
\nonumber\\
\label{minftm2}
\end{eqnarray}
This precisely equals \eqref{rotatattrrv}, and hence describes the near-horizon region of the underrotating extremal black hole.
Setting $\rho=0$, we obtain
\begin{eqnarray}
M(\rho=0,v) = \frac{1}{v^2}
\begin{pmatrix}
A  & \quad B \, v  & \quad  C \,  v^2 \\
- B \, v  & \quad D \,  v^2&  \quad  0 
  \\
C  \,  v^2 &\quad  0
   & \quad  0 
\end{pmatrix} ,
\label{minftm3}
\end{eqnarray}
and hence $M(\rho=0, v) = {\cal M} (\omega=v)$, in accordance with the substitution rule \eqref{substruleMM}. 

On the other hand, when $v < 0$, we have \eqref{tvneg}, and hence
\begin{eqnarray}
M(\rho,v) = \frac{1}{\rho^2 + v^2}
\begin{pmatrix}
A - \frac{(2AD + B^2)}{2D}\left( \frac{v}{\sqrt{\rho^2 + v^2}}+1\right)  & \quad - B \, \sqrt{\rho^2 +v^2}  & \quad  C \,  (\rho^2 + v^2) \\
B \, \sqrt{\rho^2 +v^2}  & \quad D \,  (\rho^2 + v^2)&   \quad 0 
  \\
C  \,  (\rho^2 + v^2) & \quad 0
   &  \quad 0 
\end{pmatrix} \;. \nonumber\\
\label{minftm4}
\end{eqnarray}
Note that this is related to \eqref{minftm2} by $B \rightarrow -B, \; 2 AD + B^2 \rightarrow  - (2 AD + B^2)$.
Setting $\rho=0$, we obtain
\begin{eqnarray}
M(\rho=0,v) = \frac{1}{v^2}
\begin{pmatrix}
A  & \quad B \, v  & \quad  C \,  v^2 \\
- B \, v  & \quad D \,  v^2&  \quad  0 
  \\
C  \,  v^2 &\quad  0
   &  \quad 0 
\end{pmatrix} ,
\label{minftm3}
\end{eqnarray}
and hence $M(\rho=0, v) = {\cal M} (\omega=v)$, again  in accordance with the substitution rule \eqref{substruleMM}. 


Now, let us discuss the {\sl overrotating case}.  To keep the expressions as simple as possible, we will only consider the case
of vanishing charges $Q = P =0$. The resulting spacetime solution will
describe
the near-horizon geometry of an extremal Kerr black hole, supported by
a constant scalar field.  
We use the results of \cite{Astefanesei:2006dd} to write down the associated 
near-horizon solution, as follows. 
We begin by setting ${\tilde q} = {\tilde p} = 2 M_K$ in eq. (5.51) of \cite{Astefanesei:2006dd}, obtaining
\begin{equation}
Q = P = 0 \;\;\;,\;\;\; M^2_K = \frac{J}{16 \pi} > 0\;.
\end{equation}
Inspection of (5.66) and (5.67) in \cite{Astefanesei:2006dd} yields
\begin{equation}
f_p = f_q = M^2_K \, (1 + \cos^2 \theta) \;.
\end{equation}
The scalar field is thus constant,
\begin{equation}
e^{- 4 \Phi/\sqrt{3}} = \frac{f_p}{f_q} = 1 \longrightarrow \Phi = 0 \;.
\end{equation}
Then, inspection of (5.61) and of (5.64) in \cite{Astefanesei:2006dd} gives
\begin{equation}
\alpha = 1 \;\;\;,\;\;\; \Omega = \frac{J}{8 \pi} \, \sin \theta \;\;\;,\;\;\;
e^{- 2 \psi} = \frac{J^2 \, \sin^2 \theta }{64 \pi^2 \, v_1 (\theta)}
\;\;\;,\;\;\; v_1 (\theta) = \frac{J}{16 \pi} \, (1 + \cos^2 \theta) \;,
\end{equation}
and the line element becomes
\begin{equation}
ds^2_4 = \frac{J}{16 \pi} \left( 1 + \cos^2 \theta \right) \left( - r^2 \, dt^2 + \frac{dr^2}{r^2} + d \theta^2 \right)
+ \frac{J}{4 \pi} \, \frac{\sin^2 \theta}{(1 + \cos^2 \theta)} \left(d \phi - r \, dt \right)^2 \;.
\label{lineextkerr}
\end{equation}
We now bring the line element in the form
\begin{equation}
ds^2_4 = - \Delta \, (dt + B)^2 + \Delta^{-1} \, ds^2_3 
\end{equation}
with
\begin{equation}
ds^2_3 = e^{\lambda} \left(d\rho^2 + dv^2 \right) + \rho^2 \, d \phi^2 \;.
\label{3dmetric}
\end{equation}
We obtain
\begin{eqnarray}
\Delta &=& \frac{J}{16 \pi} \, r^2 \, \left(1 + \cos^2 \theta -  \frac{4 \sin^2 \theta}{(1 + \cos^2 \theta)}
\right)  = \frac{J}{16 \pi} \, r^2 \, \frac{ \left( \cos^4 \theta + 6 \cos^2 \theta -3 \right)}{1 + \cos^2 \theta}
 \;, \nonumber\\
B &=& B_{\phi} \, d\phi \;\;\;\;,\;\;\; B_{\phi} = \frac{J}{4 \pi} \, \frac{r}{\Delta} \,
\frac{\sin^2 \theta}{(1 + \cos^2 \theta)} = \frac{b(\theta)}{r} \;\;\;,\;\;\;
b(\theta) = \frac{4 \sin^2 \theta}{(1 + \cos^2 \theta)^2 - 4 \sin^2 \theta} \;, \nonumber\\
e^{\lambda} &=& \frac{1}{4} \, \left[(1 + \cos^2 \theta)^2  -  4 \sin^2 \theta
\right] \;\;\;,\;\;\;
\rho = \frac{J}{8 \pi} \, r \, \sin \theta \;\;\;,\;\;\; v =  \frac{J}{8 \pi} \, r \, \cos \theta \;.
\label{fieldsweylover}
\end{eqnarray}
Observe that the requirement for a space-like reduction, $\Delta > 0$ and $e^{\lambda}  > 0$, impose the following
restriction on the range of $\theta$,
\begin{equation}
(1 + \cos^2 \theta)^2  -  4 \sin^2 \theta > 0 \;.
\end{equation}
For values of $\theta$ outside this range we have a time-like reduction, which can be dealt with
along similar lines:  in \eqref{Xeq} one has to change $\tau^2$ to $-\tau^2$ and $\tau$ into
$- \tau$, see \cite{Lu:2007jc}.
Since when factorizing we never use the explicit form of \eqref{Xeq}, but only the spectral curve, whose
form doesn't change, the results we will obtain are valid for 
any $\theta \in (0,  \pi)$ 
as long as
\begin{equation}
(1 + \cos^2 \theta)^2  -  4 \sin^2 \theta \neq 0 \;.
\end{equation}
Observe that this sign change of $\Delta$ is due to the presence of an ergo region, which distinguishes the
overrotating from the underrotating case. This feature is not present in the underrotating case.

To obtain the coset representative  \eqref{cosetrepM}, we need to dualize $B_{\phi}$ into $\chi_3$.
The dualization is with respect to the three-dimensional metric \eqref{3dmetric}.
Using
\begin{equation}
\sqrt{g_3} \, g^{rr} = \frac{J}{8 \pi} \, r^2 \, \sin \theta \;,
\end{equation}
we obtain after dualization, 
\begin{equation}
\partial_r \chi_3 = \frac{8 \pi}{J} \, \frac{\Delta^2}{r^3} \frac{\partial_{\theta} b(\theta)}{\sin \theta} \;,
\end{equation}
which integrates to 
\begin{eqnarray}
\chi_3 = \frac12 \, r^2 \, \frac{J}{32 \pi} \, \frac{\partial_{\theta} b(\theta)}{\sin \theta}
 \,  \left(1 + \cos^2 \theta -  \frac{4 \sin^2 \theta}{(1 + \cos^2 \theta)}
\right)^2 
=  \frac{J}{8\pi} \, r^2 \, \cos \theta \,  \frac{ \left( 3 -  \cos^2 \theta  \right)}{
1 + \cos^2 \theta} \;.
\end{eqnarray}
Thus, the fields appearing in the $\natural$-symmetric coset representative \eqref{cosetrepM} are given by
\begin{eqnarray}
e^{2 \Sigma_1} = e^{\phi_2} = \frac{1}{\Delta} \;\;\;,\;\; e^{2 \Sigma_2} = 1 \;\;\;,\;\;\; 
e^{2 \Sigma_3} = e^{- \phi_2}  = \Delta \;\;\;,\;\;\; \chi_1 = \chi_2 = 0 \;,
\end{eqnarray}
and $\chi_3$, where we used $\phi_1 = - 2 \Phi =0$.
The coset representative \eqref{cosetrepM} is thus given by 
\begin{eqnarray}
M(r, \theta) = \begin{pmatrix}
\Delta^{-1}   &\quad  0  & \quad  \Delta^{-1} \, \chi_3 \\
0 & \quad 1
& 0\\
 \Delta^{-1} \, \chi_3 & \quad 0
 & \quad   \Delta^{-1} \, \chi_3^2 +  \Delta
\end{pmatrix} \;.
\label{Mstmat}
\end{eqnarray}
We evaluate
\begin{eqnarray}
\Delta^{-1} \, \chi_3 &=& 2 \, \frac{ \cos \theta \, (3 - \cos^2 \theta)}{ \cos^4 \theta + 6 \cos^2 \theta -3} \;, \nonumber\\
 \Delta^{-1} \, \chi_3^2 + \Delta &=& \frac{J}{16 \pi} \, r^2 \, \frac{(\cos^6 \theta + 15 \, \cos^4 \theta - 9 \, \cos^2 \theta + 9 )}{\cos^4 \theta + 6 \cos^2 \theta
- 3} \;.
\label{Moverr}
\end{eqnarray}
Note that the coset representative $M(r, \theta)$ has a complicated dependence on $\cos \theta$ in contrast
with the coset representative \eqref{rotatattr} for the underrotating case.

We now associate a monodromy matrix ${\cal M} (\omega)$ to $M$.
To this end, we convert to Weyl coordinates $(\rho, v)$ given in \eqref{fieldsweylover}.
Then, $M(\rho, v)$ is obtained from $M(r, \theta)$ by the substitutions $r = \sqrt{\rho^2 + v^2}$
and $\cos \theta = v/ \sqrt{\rho^2 + v^2}$.
Now we consider the substitution rule \eqref{substruleMM}. In the limit $\rho \rightarrow 0^+$ we obtain
\begin{eqnarray}
M(\rho=0, v) = \begin{pmatrix}
\beta \, \frac{1}{v^2}  &\quad  0  & \quad   \frac{v}{|v|} \\
0 & \quad 1
& 0\\
 \frac{v}{|v|}  & \quad 0
 & \quad   \frac{2}{\beta} v^2
\end{pmatrix} \;\;\;,\;\; \beta = \frac{8 \pi}{J } \;.
\label{Mstmat2}
\end{eqnarray}
For $v > 0$,  the substitution rule \eqref{substruleMM} associates the monodromy matrix
\begin{eqnarray}
{\cal M}(\omega) = \frac{1}{\omega^2} \begin{pmatrix}
\beta  &\quad  0  & \quad   \omega^2 \\
0 & \quad \omega^2
& 0\\
\omega^2  & \quad 0
 & \quad   \frac{2}{\beta} \, \omega^4
\end{pmatrix} \;\;\;,\;\; \beta = \frac{8 \pi}{J } \;.
\label{Mmono1}
\end{eqnarray}
to $M(\rho, v)$, while for $v < 0$ it associates
\begin{eqnarray}
{\cal M}(\omega) = \frac{1}{\omega^2} \begin{pmatrix}
\beta  &\quad  0  & \quad  - \omega^2 \\
0 & \quad \omega^2
& 0\\
-\omega^2  & \quad 0
 & \quad   \frac{2}{\beta} \, \omega^4
\end{pmatrix} \;\;\;,\;\; \beta = \frac{8 \pi}{J } \;.
\label{Mmono2}
\end{eqnarray}
Note that both \eqref{Mmono1} and \eqref{Mmono2} are {\sl unbounded} at $\omega = \infty$,
in stark contrast with the monodromy matrices \eqref{underrotmonvg0} and \eqref{underrotmonvl0} in the underrotating case.

We may combine both monodromy matrices \eqref{Mmono1} and \eqref{Mmono2} into 
\begin{eqnarray}
{\cal M} (\omega) = \frac{1}{\omega^2} \begin{pmatrix}
\beta  & 0  & s \, \omega^2\\
0 & \omega^2
& 0\\
s \,  \omega^2 & 0
 &  \frac{2 }{\beta} \, \omega^4 
\end{pmatrix} \;\;\;,\;\;\; s = \pm 1 \;\;\;,\;\;\; 
\det {\cal M} (\omega) = 1 \;.
\label{ompl}
\end{eqnarray}
The non-trivial part of the factorization pertains to the factorization of the 
$2 \times 2$ monodromy matrix 
\begin{eqnarray}
{\cal M} (\omega) = \frac{1}{\omega^2} \begin{pmatrix}
\beta    & s \, \omega^2\\
s \,  \omega^2
 &  \frac{2 }{\beta} \, \omega^4 
\end{pmatrix} \;.
\label{effectmon}
\end{eqnarray}
Note that when $s = -1$, we can pull out an overall sign if we perform the replacement $\beta \rightarrow
- \beta$, in which case we obtain
\begin{eqnarray}
{\cal M} (\omega) = - \frac{1}{\omega^2} \begin{pmatrix}
\beta    &  \omega^2\\
  \omega^2
 &  \frac{2 }{\beta} \, \omega^4 
\end{pmatrix} \;.
\label{effectmon2}
\end{eqnarray}
Thus, the factorization of the $2 \times 2$  monodromy matrix with $s =-1$ can be easily related to the factorization 
of the $2 \times 2$ monodromy matrix with $s=1$. It therefore suffices to focus on the case $s=1$.

We proceed to factorize the  monodromy matrix \eqref{effectmon} with $s =1$. To obtain
${\cal M} (\omega) = M_- (\tau, x) M_+ (\tau,x)$,  we use the vectorial factorization technique described in the previous subsection and the relation \eqref{wtaupm0}. We require $M_- (M_+)$ to be bounded and analytic in the respective domain,
and we impose the 
normalization condition $M_+(\tau=0,x) = \mathbb{I}$.
We obtain
\begin{eqnarray}
{M}_- (\tau, x) &=& 
\begin{pmatrix}
\frac{A \tau^2 + B \tau + C}{(\tau - \tau_0^+)^2}  &\quad  \frac{{\tilde A} \tau^2 + {\tilde B} \tau + {\tilde C}}{(\tau - \tau_0^+)^2}   \\
\\
 D + \frac{E}{\tau } &\quad  {\tilde D} + \frac{\tilde E}{\tau } + \frac{\tilde F}{\tau^2}
\end{pmatrix} \;, \nonumber\\
{M}^{-1}_+ (\tau, x) &=& 
\begin{pmatrix}
\frac{\rho^2}{2 \beta} (\tau - 2 \tau_0^- ) B + 
\frac{\rho^2}{2 \beta} (\tau- \tau_0^-)^2 A - D   & \quad  \frac{\tilde \Delta}{\tau}  \\
\\
 \frac{\Sigma}{\rho^2 \, (\tau-\tau_0^+)^2 \, (\tau-\tau_0^-)^2}  & \quad  \frac{\tilde \Sigma}{\rho^2 \, (\tau-\tau_0^+)^2 \, (\tau-\tau_0^-)^2} 
\end{pmatrix} \;.
\label{canonfactover}
\end{eqnarray}
where the 
quantities $\Sigma, {\tilde \Sigma}$ and ${\tilde \Delta}$ are expressed in terms of the constants $A, B, C, D, E$
and ${\tilde A}, {\tilde B}, {\tilde C}, {\tilde D}, {\tilde E}, {\tilde F}$ as
\begin{eqnarray}
\Sigma &=& - \left( A \tau^2 + B \tau + C \right) (\tau- \tau_0^-)^2 \, \rho^2 + 4 \beta \, \left(D \tau^2 + E \tau \right) \;, \nonumber\\
{\tilde \Sigma} &=& - \left( {\tilde A} \tau^2 + {\tilde B} \tau + {\tilde C} \right) (\tau- \tau_0^-)^2 \, \rho^2 + 4 \beta \, \left({\tilde D} \tau^2 + {\tilde E} \tau + {\tilde F} \right) \;, \nonumber\\
{\tilde \Delta} &=& \frac{\rho^2}{2 \beta} \, (\tau - \tau_0^-)^2 ({\tilde A}\tau + {\tilde B}) + \frac{\rho^2}{2 \beta} (\tau - 2 \tau_0^-) 
{\tilde C} - {\tilde D} \tau - {\tilde E} \;.
\end{eqnarray}
Analyticity of $M_+$
requires $\Sigma$ and ${\tilde \Sigma}$ to have a double zero at $\tau = \tau_0^+$, i.e. 
$\Sigma (\tau_0^+) = \Sigma^\prime (\tau_0^+) = {\tilde \Sigma} (\tau_0^+) = {\tilde \Sigma}^\prime (\tau_0^+) = 0$.
Also, analyticity of $M_+$ together with the normalization condition $M_+(\tau = 0, x) = \mathbb{I}$ requires
${\tilde \Delta}$ to have a double zero at $\tau =0$, i.e. $\Delta (0) = \Delta^\prime (0) = 0$.
Imposing these various conditions results in (we recall $\tau_0^+ \tau_0^- = -1$)
\begin{eqnarray}
A&=& 
 \frac{4 \beta}{\rho^2} \, \frac{(\tau_0^+)^2 + (\tau_0^-)^2}{
\left[ (\tau_0^+ + \tau_0^-)^4 - 12 \right]  } 
 \;\;\;, \;\;\;
 B = 
- \frac{ 8 \beta}{\rho^2} \, \frac{(\tau_0^+)^2 \, (\tau_0^+ - \tau_0^-)}{ \left[
(\tau_0^+ + \tau_0^-)^4 - 12 \right] } \;\;\;,\;\;\; C = 0 \;, \nonumber\\
D &=& {\tilde A}= 
- \frac{(\tau_0^+ - \tau_0^-) (\tau_0^+ + \tau_0^-) \left[ (\tau_0^+)^2  +( \tau_0^-)^2 + 4 \right]  }{
(\tau_0^+ + \tau_0^-)^4 - 12 } \;\;\;,\;\;\;
E = 
- 4   \,  \frac{ (\tau_0^+ - \tau_0^-)}{
(\tau_0^+ + \tau_0^-)^4 - 12  } 
\;, \nonumber\\
{\tilde B} &=& 2 \,  \frac{\tau_0^+ (\tau_0^+ - \tau_0^-) \left[ (\tau_0^+)^3  + 3 \tau_0^+ -  3 \tau_0^- 
+( \tau_0^-)^3 \right]  }{
(\tau_0^+ + \tau_0^-)^4 - 12 } \;\;\;,\;\;\;
{\tilde C} = (\tau_0^+)^2 \;,
\nonumber\\
{\tilde D} &=&  \frac{\rho^2}{2 \beta} \,\frac{ \left[(\tau_0^+)^6 + 9 (\tau_0^+)^2 + 16 + 9 (\tau_0^-)^2 + (\tau_0^-)^6
\right] }{
(\tau_0^+ + \tau_0^-)^4 - 12 } \;,
\nonumber\\
{\tilde E} &=&   \frac{\rho^2}{\beta} \,  \frac{\left[ (\tau_0^+)^5  - 3 (\tau_0^+)^3 - 2 \tau_0^+ -  2 \tau_0^- - 3 (\tau_0^-)^3
+( \tau_0^-)^5 \right]  }{
(\tau_0^+ + \tau_0^-)^4 - 12 } \;\;\;,\;\;\; 
{\tilde F} =
\frac{\rho^2}{2 \beta} \;.
\end{eqnarray}
It can be checked that ${\cal M} (\omega) \, M^{-1}_+ (\tau,x)= M_- (\tau, x)$.

Evaluating $M_-(\tau,x)$ at $\tau = \infty$ results in 
\begin{eqnarray}
{M}_- (\infty, x) = 
\begin{pmatrix}
A  & \quad \tilde A   \\
\\
 D & \quad {\tilde D}
\end{pmatrix} \;.
\label{mminusfact}
\end{eqnarray}
Having determined $M(x)$ from factorization, we now verify that it satisfies the substitution rule
\eqref{substruleMM}. When $v > 0$, we have \eqref{tvpos}, and setting $\rho=0$, we obtain
\begin{eqnarray}
M(\rho=0,v) =  \begin{pmatrix}
\frac{\beta}{v^2}   & \quad  1 \\
1
 & \quad   \frac{2}{\beta} \, v^2
\end{pmatrix} \;,
\end{eqnarray}
and hence $M(\rho=0, v) = {\cal M} (\omega=v)$, in accordance with the substitution rule \eqref{substruleMM}. 
On the other hand, when $v < 0$, we have \eqref{tvneg}, and we obtain
\begin{eqnarray}
M(\rho=0,v) =  \begin{pmatrix}
\frac{\beta}{v^2}   & \quad  - \frac{|v|}{v} \\
 - \frac{|v|}{v}
 & \quad   \frac{2}{\beta} \, v^2
\end{pmatrix} \;,
\end{eqnarray}
and hence $M(\rho=0, v) = {\cal M} (\omega=v)$, in accordance with the substitution rule \eqref{substruleMM}. 
Finally, we note that when  $v > 0$,
we precisely recover the near-horizon solution
\eqref{Mstmat} and \eqref{Moverr}.

\section{Factorization by group transformations }

In this section we consider the action of a group element $g \in G$ in the Lie group $G$
on a seed monodromy matrix. Thus, we take $g$ to be a matrix that is independent
of $\omega$.  In the next section we will consider the action of $\omega$-dependent 
matrices $g(\omega) \in {\tilde G}$ on a seed monodromy matrix.

We may generate new monodromy matrices by the group action on a seed monodromy matrix,
\begin{equation}
{\cal M}_{\rm seed} (\omega) \rightarrow {\cal M} (\omega)  = g^{\natural} \, {\cal M}_{\rm seed} (\omega) \, g \;,
\label{MseedMtran}
\end{equation}
where $g \in G$ is a constant matrix.
Hence, we may decompose ${\cal M} (\omega)$ in a straightforward manner,
as follows.
Decomposing ${\cal M}_{\rm seed} = M_-^{\rm seed} \, 
M_+^{\rm seed} $ we obtain
 \begin{eqnarray}
 {\cal M} (\omega) = M_- \, M_+ \;\;\;,\;\;\; M_- =  g^{\natural} \, M_-^{\rm seed} \, g \;\;\;,\;\;\;
 M_+ =  g^{-1} \, M_+^{\rm seed} \, g \;.
 \label{factMMM}
\end{eqnarray}
Note that $M_+$ satisfies the normalization condition $M_+ (\tau=0, x) = \mathbb{I}$. Since $g$ is independent of $\omega$,
we immediately obtain a new spacetime solution given by
\begin{equation}
M(x) = M_- (\tau= \infty, x)  =  g^{\natural} \, M^{\rm seed}(x) \,  g \;.
\label{MgMseedg}
\end{equation}
There is thus no need to perform a laborious explicit factorization of ${\cal M} (\omega)$: it suffices to 
know the factorization of ${\cal M}_{\rm seed} (\omega)$.

In particular, some of these transformations 
will generate so-called Harrison transformations \cite{Kinnersley:1977pg,Kinnersley:1977ph,Richterek:2004bb,Cvetic:2012tr}
in spacetime that transform near-horizon
solutions into interpolating solutions. We will address this in the context of the 
scalar field models based on the cosets $G/H = 
SU(2,1)/(SL(2, \mathbb{R}) \times U(1))$
and $SL(3, \mathbb{R})/SO(2,1)$, respectively.
In subsection \ref{subsecsu21} we generate monodromy matrices by group actions that will correspond
to static non-extremal/extremal black holes in flat spacetime and to 
non-extremal black hole solutions in $AdS_2$. In subsection \ref{harrsl3} we generate monodromy
matrices that correspond to interpolating extremal 
black hole solutions.

\subsection{$G/H=SU(2,1)/(SL(2,\mathbb{R})\times U(1)$ \label{subsecsu21}}

We consider the coset representative 
\eqref{cosetrepMsu21} with $\Sigma = |Z|^2 + i \Psi$, where $\Psi = \sqrt{2} \lambda - 2 \chi_e \chi_m$ (c.f. \eqref{sigmZ}). We set $\Psi =0$
and define $\epsilon = e^{-\varphi} - \Sigma$. 
Then the coset representative \eqref{cosetrepMsu21} can be re-written as 
\begin{equation}
M= \frac{1}{\tfrac12 (\epsilon + \bar \epsilon) + |Z|^2}\left( \begin{array}{ccc}
1 & \quad  \sqrt{2}Z & \quad  |Z|^2 - \tfrac12( \epsilon - \bar \epsilon) \\
-\sqrt{2}
{\bar Z}  & \quad \tfrac12 (\epsilon + \bar \epsilon) - |Z|^2 & \quad  \sqrt{2} {\bar Z}\epsilon \\
|Z|^2 + \tfrac12 (\epsilon - \bar \epsilon) &\quad  - \sqrt{2} Z \bar \epsilon & \quad |\epsilon|^2 \end{array} \right) \;.
\label{Meps}
\end{equation}
Now consider the group element $g= e^N$, with $N$ a nilpotent matrix $N^3=0$, as follows, 
\begin{equation}
g (c) =  \left( \begin{array}{ccc}
1 & \quad 0 & \quad 0 \\
-\sqrt{2} \bar{c} & \quad 1 &\quad 0 \\
|c|^2 &\quad  -\sqrt{2} c & \quad 1 \end{array} \right) \;,
\label{groupgc}
\end{equation}
which satisfies the group relation \eqref{gkgk}.
 Using $g^{\natural}(c) = \eta g^{\dagger}(c)
\eta^{-1}$ with $\eta = {\rm diag} (1,-1,1)$ we obtain,
\begin{equation}
g^{\natural} (c) =  \left( \begin{array}{ccc}
1 &  \quad \sqrt{2} c  &\quad  |c|^2 \\
0  & \quad 1 & \quad \sqrt{2} \bar{c} \\
0 &\quad  0 & \quad 1 \end{array} \right) \;.
\end{equation}
Next consider $g^{\natural} (c)  M g(c)$, as in \eqref{MgMseedg}. 
We obtain
\begin{equation}
g^{\natural} (c) M g(c) =  \frac{1}{\tfrac12 (\epsilon + \bar \epsilon) + |Z|^2} \left( \begin{array}{ccc}
m_1 & \quad  m_2 & \quad m_3 \\
- {\bar m}_2 &\quad 
m_4
 & \quad  \sqrt{2} \left( \epsilon \bar Z + \bar c | \epsilon|^2 \right) \\
{\bar m}_3 & \quad 
 - \sqrt{2} \left( {\bar \epsilon}  Z + c | \epsilon|^2 \right)
 & \quad | \epsilon|^2 \end{array} \right) \;,
 \label{matrixgMg}
\end{equation}
with
\begin{eqnarray}
m_1 &=& \left\vert 1 - 2 {\bar c} Z -  \epsilon | c|^2 \right\vert^2 \;,
\nonumber\\
m_2 &=& \sqrt{2} \left( Z - 2 c |Z|^2 + c \epsilon - 2c^2  \epsilon {\bar Z} - |c|^2  \bar \epsilon Z  - c |c \epsilon|^2
\right)
\;,\nonumber\\
m_3 &=& |Z|^2 - \tfrac12 (\epsilon - \bar \epsilon) + 2 c \epsilon {\bar Z} + | c \epsilon|^2 \;, \nonumber\\
m_4 &=& \tfrac12 (\epsilon + \bar \epsilon)- |Z|^2 - 2 (\epsilon  c {\bar Z} + {\bar \epsilon} {\bar c} Z)
- 2 | c \epsilon|^2 \;.
\end{eqnarray}
Now observe that \eqref{matrixgMg} can be obtained by applying the following transformation to the elements of 
\eqref{Meps},
\begin{eqnarray}
\epsilon &\rightarrow& \frac{\epsilon}{1- 2 {\bar c} Z- |c|^2 \epsilon}  \;, \nonumber\\
Z &\rightarrow & \frac{Z + c \epsilon}{1- 2 \bar{c} Z- |c|^2 \epsilon} \;.
\label{epsZtranf}
\end{eqnarray}
This transformation is called Harrison transformation, c.f. page 4 of \cite{Richterek:2004bb}.
Observe that the condition $\Psi =0$
is preserved by the transformation \eqref{epsZtranf}, and so is the condition $\epsilon =0$.

Let us first consider the case $\epsilon =0$. An example with 
$\epsilon =0$ is provided by \eqref{Mads2}, which describes the near-horizon limit of a 
static extremal black hole with electric charge $q$ and magnetic charge $p$, with a particular choice of the gauge potentials, c.f. 
\eqref{backg}. Then, under the transformation \eqref{epsZtranf}, $e^{\varphi}$ and $Z$ transform into
\begin{eqnarray}
e^{\varphi} &\rightarrow& e^{\tilde \varphi} = \left\vert 2 c \right\vert^2  \left\vert 1 - \frac{1}{ 2 \bar{c} Z } \right\vert^2 \;,
\nonumber\\
Z &\rightarrow& {\tilde Z} = \frac{Z}{1 - 2 {\bar c} Z} \;.
\end{eqnarray}
Imposing the asymptotic normalization condition $2 |c| =1$ as well as the condition
\be 
{\rm Im} ( \bar c \, Q) = 0 \;,
\label{cQcond}
\ee
we obtain
\be
2 {\bar c} \, Q =2  c \, {\bar Q} = \pm \sqrt{q^2 + p^2} \;,
\ee
where we recall $Q = q + ip$. Choosing $2 {\bar c} \, Q = - \sqrt{q^2 + p^2}$, we obtain
\begin{eqnarray}
e^{\tilde \varphi} &=& H^2 \;\;\;, \;\;\; H = 1 + \frac{|Q|}{r} \;,
\nonumber\\
{\tilde \chi}_e &=& \frac{q}{|Q|} \frac{r}{(r + |Q|)} \;\;\;,\;\;\; {\tilde \chi}_m = \frac{p}{|Q|} \frac{r}{(r + |Q|)} \;.
\label{interpolsolu}
\end{eqnarray}
This describes an interpolating extremal static black hole solution
\cite{Behrndt:1997ny}.

This is the expected feature of a Harrison transformation: it transforms
a near-horizon black solution into an interpolating black solution. Here, we have implemented the Harrison
transformation by operating on the seed monodromy matrix \eqref{monads2} with the group element $g(c)$, as described
by \eqref{MseedMtran}. The monodromy matrix associated with the interpolating solution \eqref{interpolsolu} is therefore
given by
\begin{eqnarray}
{\cal M} (\omega) = g^{\natural} (c) \, {\cal M}_{\rm seed} (\omega) \, g(c) = 
\begin{pmatrix}
\frac{(\omega + |Q|)^2}{\omega^2} & \quad  \sqrt{2} \frac{(q + ip)}{|Q|} \frac{(\omega + |Q|)}{\omega} \,  & \quad 1  \\
- \sqrt{2} \frac{(q - ip)}{|Q|} \frac{(\omega + |Q|)}{\omega}  &\quad  -1 & \quad   0 
  \\
1 \,   &\quad  0
   & \quad  0 
\end{pmatrix} \;.
\end{eqnarray}

Note that the gauge potentials of the interpolating solution \eqref{interpolsolu} vanish at the horizon $r =0$. 
These gauge potentials may, of course, be shifted
by constants, as described in \eqref{shiftgaugep} and \eqref{shiftgpg}. Performing this shift will result in a coset representative
\eqref{Meps} with $\epsilon \neq 0$.

Now let us consider an example with $\epsilon \neq 0$.
We consider the exterior region of Schwarzschild solution, with line element \eqref{lineads2su21} given by
\be
e^{-\varphi} = \frac{r -m}{r +m} \;\;\;,\;\;\; r > m \;,
\ee
and 
\be
ds_3^2 = dr^2 + (r^2 -m^2) (d \theta^2 + \sin^2 \theta \, d\phi^2) \;.
\ee
The associated coset representative \eqref{Meps} reads, with $Z=0$ and $\epsilon = \bar \epsilon = e^{-\varphi}$,
\begin{eqnarray}
M = 
\begin{pmatrix}
e^{\varphi} & 0 & \quad 0  \\
0 & \quad - 1  & \quad    0 
  \\
0 & \quad  0    &  \quad e^{- \varphi} 
\end{pmatrix} \;.
\label{Mschwtr}
\end{eqnarray}
Now consider the transformation ${\tilde M} = g^{\natural} (c) \, M \, g(c) $
with $g(c)$ given in \eqref{groupgc} and $|c| = 1$. The resulting coset representative ${\tilde M}$ is of the form \eqref{cosetrepMsu21} with 
\begin{eqnarray}
e^{- \tilde \varphi} = \frac{r^2 - m^2}{4 m^2} \;\;\;,\;\;\; 
\tilde{Z} = {\tilde \chi}_e + i {\tilde \chi}_m = \frac{c}{2m} \, (r - m) \;\;\;,\;\;\; {\tilde \Sigma} = |{\tilde Z}|^2 \;.
\end{eqnarray}
This describes a black hole in $AdS_2$ \cite{Sen:2008yk}, with line element \cite{Cvetic:2012tr}
\begin{equation}
ds^2_4 = 4m^2  \left( -  (r^2 - m^2 )\, dt^2 + \frac{dr^2}{r^2 - m^2} \right) + 4 m^2 \, \left(d \theta^2 + \sin^2 \theta \, d \phi^2 \right)\;,
\end{equation}
and with vanishing gauge potentials ar $r = m$.
On the other hand, if we consider the same transformation but with $|c| \neq 1$, we obtain
\be
e^{\tilde \varphi} = \left( e^{\varphi/2} - |c|^2 \, e^{-\varphi/2} \right)^2 \;.
\ee
Asymptotically, $e^{\tilde \varphi} $ tends to $(1 - |c|^2)^2 \neq 0$. We take $|c| <1$, in which case
the associated four-dimensional line element describes a non-extremal Reissner-Nordstrom black hole in flat
spacetime, with outer and inner horizons at $r = \pm m$.
In both cases ($|c| \leq 1$), the associated monodromy matrix is given by 
${\tilde {\cal M}} = g^{\natural} (c) \, {\cal M} \, g(c) 
$,
with ${\cal M}$ the monodromy matrix associated to the Schwarzschild solution, given by
\begin{equation}
{\cal M}(\omega)= \begin{pmatrix}
\frac{\omega + m}{\omega - m} & 0 & 0\\
0 & 1 &0 \\
0 & 0 & \frac{\omega - m}{\omega + m} 
\end{pmatrix} \;.
\end{equation}

\subsection{$G/H = SL(3,\mathbb{R})/SO(2,1)$ \label{harrsl3}}

We pick as seed monodromy matrix ${\cal M}_{\rm seed}  (\omega)$ the matrix
\eqref{underrotmonvg0}
 that describes the near-horizon solution of an 
underrotating black hole.  
Now consider acting with a group element $g = e^N$ on ${\cal M}_{\rm seed}  (\omega)$, as in \eqref{MseedMtran}, with $N$ a 
nilpotent lower triangular matrix given by
\begin{eqnarray}
N =
\begin{pmatrix}
0 & \quad 0 & \quad 0    \\
\beta & \quad 0 & \quad 0  \\
 \mu & \quad \gamma  &\quad  0
\end{pmatrix} \;.
\label{Nnilsl3}
\end{eqnarray}
It depends on 3 real parameters, and can be written as  $N = \beta \, F_2 + \gamma F_1 + \mu \, F_3$, 
with the $F_i$ defined in appendix \ref{backgroundLie}.
The matrix $g^{\natural}$ is an upper triangular matrix,
constructed as follows.
Given an element $Z$ of the Lie algebra, $Z^{\natural}$ is defined by
\begin{equation}
Z^{\natural} = \eta \, Z^T \, \eta \;,
\end{equation}
where $\eta = {\rm diag} (1, \epsilon_2, \epsilon_1)$, with $\epsilon_1 = \pm 1 , \epsilon_2 = \mp 1 $ .
We pick $\epsilon_1 = 1, \epsilon_2 = -1$, and get
\begin{eqnarray}
N^{\natural} = - \beta \, E_2 - \gamma E_1 + \mu \, E_3 = 
\begin{pmatrix}
0 & \quad - \beta & \quad \mu    \\
0  & \quad 0 & \quad - \gamma  \\
0 & \quad 0 & \quad 0
\end{pmatrix} \;.
\end{eqnarray}
Then, $g^{\natural}$ is defined by $g^{\natural} = e^{N^{\natural}}$. We obtain
\begin{eqnarray}
g &=& e^N = \mathbb{I} + N + \tfrac12 \, N^2 = 
\begin{pmatrix}
1 & \quad 0 &\quad 0    \\
\beta & \quad 1 & \quad 0  \\
 {\tilde \mu}  & \quad \gamma  & \quad 1
\end{pmatrix} \;\;\;,\;\; {\tilde \mu} = \mu + \tfrac12 \, \beta \, \gamma \:, \nonumber\\
g^{\natural} &=& e^{N^{\natural}} = \mathbb{I} + N^{\natural} + \tfrac12 \, \left( N^{\natural} \right)^2 = 
\begin{pmatrix}
1 & \quad - \beta &\quad  {\tilde \mu}    \\
0 & \quad 1 & \quad - \gamma  \\
0 & \quad 0  & \quad 1
\end{pmatrix} \;.
\label{ggnatsl3}
\end{eqnarray}
Using \eqref{MseedMtran}, the resulting class of matrices ${\cal M} (\omega)$ 
is given by
\begin{eqnarray}
{\cal M} (\omega) =  \begin{pmatrix}
\frac{A}{\omega^2}  + \frac{A_1}{\omega} + A_2 & \quad \frac{B}{\omega} + B_2 & \quad  C 
 \\
- \frac{B}{\omega} - B_2  & \quad D 
 & \quad  0
  \\
  C  & \quad
0
   & \quad  0
\end{pmatrix} \;,
\label{classmontransf}
\end{eqnarray}
with 
\begin{eqnarray}
A_1 = 2 \, \beta \, B \;\;\;,\;\; A_2 = 2 \, {\tilde \mu} \, C - \beta^2 \, D \;\;\;,\;\;\; 
B_2 = \gamma \, C - \beta \, D \;\;\;,\;\;\; {\tilde \mu} = \mu + \tfrac12 \beta \gamma \;.
\label{valuesbgm}
\end{eqnarray}
Taking any element ${\cal M} (\omega) $ of this class of monodromy matrices as a representative, 
we can write any other element of this class as
$g^{\natural} \, {\cal M} (\omega) \, g $.

Using \eqref{factMMM}, 
any matrix ${\cal M}(\omega)$
of the form \eqref{classmontransf} can be factorized in a straightforward way.  Using as seed matrix the matrix
$M(x)$ given in \eqref{minftm2}, valid for $v >0$, we obtain
\begin{eqnarray}
g^{\natural} \, M(\rho,v) \, g = 
\begin{pmatrix}
\frac{A}{\rho^2 + v^2} + \frac{A_1}{\sqrt{\rho^2 +v^2} } + A_2 + 
 \frac{(2AD + B^2)}{2D}\left( \frac{v}{\sqrt{\rho^2 + v^2}}-1\right) \frac{1}{\rho^2 + v^2} & \quad 
\frac{B}{\sqrt{\rho^2 +v^2} } + B_2 & \quad C \\
- \frac{B}{\sqrt{\rho^2 +v^2}} - B_2 & \quad D &  \quad 0 
  \\
C  & \quad 0
   & \quad  0 
\end{pmatrix} . \nonumber\\
\end{eqnarray}
We proceed with the spacetime interpretation of this solution.  
To this end, let us first consider the case when $2AD + B^2=0$,
thus restricting to the case of static seed solution. 
Then, $g^{\natural} \, M(\rho,v) \, g$
is  simply given by
 \begin{eqnarray}
g^{\natural} \, M(r) \, g = 
\begin{pmatrix}
\frac{\alpha(r) }{r^2}   & \quad \frac{\beta(r)}{r}  & \quad C 
 \\
- \frac{\beta(r)}{r}   & \quad D 
 & \quad  0
  \\
  C  & \quad 
0
   & \quad 0
\end{pmatrix} \;,
\label{transfgmg}
\end{eqnarray}
where
\begin{eqnarray}
 \alpha(r) &=& A_2 r^2 + A_1 r + A \;,\nonumber\\
\beta(r) &=& B_2 r + B \;.
\label{valab}
\end{eqnarray}
Comparing with the coset representative \eqref{cosetrepM} we obtain,
\begin{eqnarray} 
\chi_1 &=& \frac{C r \beta(r)}{\alpha(r) D + \beta^2(r)} \;\;\;,\;\; 
\chi_2 = \frac{r \beta(r)}{\alpha(r)} \;\;\;,\;\;\;
\chi_3 =\frac{C r^2}{\alpha(r)} \;, \nonumber\\
e^{2 \Sigma_1}&=& \frac{\alpha(r)}{r^2} \;\;\;,\;\; 
e^{2 \Sigma_2}= D + \frac{\beta^2(r)}{\alpha(r)} \;\;\;,\;\;\;
e^{2\Sigma_3}=\frac{r^2}{\alpha(r) D + \beta^2(r)} \;.
\end{eqnarray}
The associated four-dimensional spacetime line element is of the form \eqref{lineund}
with $e^{2 \phi_2} = e^{2(\Sigma_1-\Sigma_3)}$.
To obtain a static line element, we require (c.f. \eqref{f1f2})
\be
d\chi_3 - \chi_1 d \chi_2 = 0 \;,
\label{vadchi3}
\ee
which results in the condition
\be
 r \, \left( 2 \alpha(r) D + \beta^2(r) \right)' = 2  \left( 2 \alpha(r) D + \beta^2(r) \right) \;.
 \ee
This results in 
\be
 2 \alpha(r) D + \beta^2(r) = \mu \, r^2 \;,
 \label{alberel}
 \ee
with $\mu \in \mathbb{R}$. Note that we can absorb the term on the right hand side by redefining
the constant $A_2$ in $\alpha(r)$.  We may thus set $\mu =0$ without loss of generality.
Using the explicit expressions \eqref{valab}, we obtain the relations
\be
2 A_2 D + B_2^2 = 0\;\;\;,\;\;\; A_1 D + B_2 B = 0 \;.
\label{relAAB}
\ee
When $A_2 \neq 0$, this describes an extremal Reissner-Nordstrom black hole solution, as follows.
Using \eqref{alberel} with $\mu =0$, we obtain $e^{\phi_2} = \sqrt{- D \, \alpha^2}/{r^2}$.
Inspection of \eqref{valuesABCD} shows that $D <0$. Using the relations 
$2AD + B^2=0$ and \eqref{relAAB}, we infer $A/A_2 = \left(B/B_2 \right)^2 > 0$ and
$A_1/A_2 = 2 B/B_2$, so that $A_1/A_2 = 2 \sqrt{A/A_2}$. This
results in $e^{2 \phi_2} = - D A_2^2 \, H^4(r)$, with $H(r) = 1 +  \left(\sqrt{A/A_2}\right)/r $. 
Choosing the normalization $- D A_2^2 =1$, we obtain $e^{\phi_2} = H^2(r)$, which describes
the line element of an extremal static black hole. Now consider $\chi_1$, which can be written as 
$\chi_1 = - \left( C B_2 \right)/\left( D A_2 \, H(r) \right)$.  Using \eqref{valuesABCD} 
we compute $A/A_2 = (PQ)/(8 \pi) > 0$, to obtain $\chi_1 = - \left(\sqrt{2} P\right)/\left( \sqrt{QP} \, H(r) \right)$.
This describes the electric potential of the extremal  Reissner-Nordstrom black hole (c.f. \eqref{interpolsolu}). Finally, we note  that the solution is supported by a constant scalar field $e^{2 \Sigma_2} = -D$.

Thus, by acting with a group element $g$, given in terms of a nilpotent matrix $N$, 
we have obtained a static interpolating black hole solution $g^{\natural} \, M(r) \, g$ that is supported
by a constant scalar field from a 
near-horizon solution $M(r)$.

Next, let us briefly consider the case when we start from an underrotating seed solution with $2 AD + B^2 \neq 0$. This leads to a change of the function $\alpha$ in the matrix \eqref{transfgmg},
which now becomes
\be
\alpha(r, \theta) = A_2 r^2 + A_1 r  - \frac{B^2}{2D} +  \frac{(2AD + B^2)}{2D} \cos \theta \;.
\ee
In this case,  the  field strength ${\cal F}_2$ in \eqref{f1f2}
is non-trivial,
\be
{\cal F}_2 = - \frac{C}{2} \, * d \left(\frac{2 D \, \alpha + \beta^2}{r^2} \right) \;.
\ee
We obtain ${\cal F}_2 = d {\cal A}$, where ${\cal A} = {\cal A}_{\phi} \, d \phi$ with
\be
{\cal A}_{\phi} = B_{\phi} - C \left( A_1 D + B_2 B \right) \cos \theta \;,
\ee
where $B_{\phi}$ is given in \eqref{B-phi}.  Consider the case when
$ A_1 D + B_2 B =0$, so that ${\cal A}_{\phi} = B_{\phi}$. Furthermore, take
$A_2 \neq 0$.  Then, the solution asymptotes to flat spacetime, while in the limit 
$r\rightarrow 0$ it becomes the 
near-horizon solution of the dyonic underrotating black hole.  Therefore, this interpolating solution describes an 
 underrotating extremal black hole in flat spacetime, supported by a non-constant scalar field.

Interestingly, a subset of the monodromy matrices in the class described by \eqref{classmontransf}
can be generated by
M\"obius type transformation of $\omega$, as follows.  Starting from the seed monodromy matrix  
\eqref{underrotmonvg0},
we consider a 
 fractional linear transformation of
$\omega$ that does not modify the pole structure of ${\cal M}_{\rm seed}(\omega)$, which has a double pole at $\omega =0$.
Hence we only consider a two-parameter
set of transformations
\begin{equation}
\omega \rightarrow \tilde \omega = \frac{\omega}{c \, \omega + d} \;\;\;,\;\; c, d \in \mathbb{R} \;,
\end{equation}
which results in matrices of the form
\begin{eqnarray}
{\cal M} (\omega) = {\cal M}_{\rm seed} (\tilde \omega) = 
 \begin{pmatrix}
\frac{A \, d^2}{\omega^2}  + \frac{2cd}{\omega} + c^2 & \quad \frac{Bd}{\omega} + B c & \quad  C
 \\
- \frac{B d}{\omega} - B c & \quad  D 
 &  \quad 0
  \\
  C  & \quad 
0
   & \quad  0
\end{pmatrix} \;.
\end{eqnarray}
This is only in the class \eqref{classmontransf} provided we set $d=1$, in which case 
\begin{equation}
A_1 = 2 c \;\;\;,\;\;\; A_2 = c^2  \;\;\;,\;\;\; B_2 = B c \;.
\label{valuesbgm2}
\end{equation}
To which transformation $g$ does this correspond to? We equate \eqref{valuesbgm} with \eqref{valuesbgm2}, and infer
\begin{equation}
\beta = \frac{c}{B} \;\;\;,\;\;\; \gamma = c \, \frac{(B^2 + D)}{BC} \;\;\;,\;\;\;
\mu = 0 
 \;.
\end{equation}

\section{Action of $g(\omega)$ on a monodromy matrix}

In this section we will consider the action of a $\omega$-dependent 
matrix $g(\omega) \in {\tilde G}$ on a seed monodromy matrix, thereby generating a new 
solution of the reduced field equations.

We choose as seed matrix the underrotating monodromy matrix 
\eqref{underrotmonvg0}.
We take $g(\omega)$ to be given by 
\be
g(\omega) = e^{N/\omega^2} \;,
\ee
where $N$ is the nilpotent matrix given in \eqref{Nnilsl3} with parameters $\beta = \mu =0$,
while $\gamma$ is taken to be non-vanishing.  Defining $\alpha = \gamma \, C$, we obtain
\begin{eqnarray}
{\cal M}(\omega) = g^{\natural} (\omega) \, {\cal M}_{\rm seed}  (\omega) \, g(\omega)=
\frac{1}{\omega^2} \begin{pmatrix}
A  & \quad B \,  \omega + \alpha & \quad C \, \omega^2
 \\
- B  \, \omega - \alpha & \quad D \,  \omega^2
 &  \quad 0
  \\
  C \, \omega^2 &\quad 
0
   &\quad  0
\end{pmatrix} \;.
\label{Mstmo2alp}
\end{eqnarray}
Since now $g$ depends on $\omega$, we need to explicitly perform the factorization 
${\cal M} (\omega) = M_- (\tau, x) M_+ (\tau,x)$. 
We obtain
\begin{eqnarray}
M_- (\tau, x) &=& M^{(0)}_- + \alpha \, \Delta_1^- + \alpha^2 \, \Delta_2^- \;, \nonumber\\
M_+(\tau, x) &=& M^{(0)}_+  + \alpha \, \Delta_1^+ + \alpha^2 \, \Delta_2^+ \;,
\end{eqnarray}
where $M^{(0)}_{\pm}$ denote the undeformed values computed in \eqref{undefMM}, and where
\begin{eqnarray}
\Delta_1^- &=&  \begin{pmatrix}
\Sigma_1^- & \quad  - \phi_2^-& \quad 0
 \\
 \phi_2^- &\quad  0
 &  \quad 0
  \\
 0 &\quad 
0
   & \quad 0
\end{pmatrix} \;\;\;,\;\;\;
\Delta_2^- = \begin{pmatrix}
\Sigma_2^- &\quad  0 &\quad  0
 \\
 0 & \quad 0
 &  \quad 0
  \\
 0 & \quad 0 & \quad 0
\end{pmatrix} \;, \nonumber\\
\Delta_1^+ &=&  \begin{pmatrix}
0 & \quad 0 &\quad  0
 \\
\frac{ \phi_2^+}{D} & \quad 0
 &  \quad 0
  \\
  \Gamma_1^+ & \quad - \frac{ \phi_2^+}{C}
   & \quad 0
\end{pmatrix} \;\;\;,\;\;\;
\Delta_2^+ = \begin{pmatrix}
0 & \quad 0 & \quad 0
 \\
 0 & \quad 0
 & \quad  0
  \\
  \Gamma_2^+ & \quad 0 & \quad 0
\end{pmatrix} \;,
\end{eqnarray}
with
\begin{eqnarray}
\phi_2^- &=& - \frac{4}{\rho^2} \left[\frac{c_1}{(\tau - \tau_0^+)^2} + \frac{c_2}{\tau - \tau_0^+} + \frac{c_3}{(\tau_0^-)^2} - \frac{c_4}{\tau_0^-}
\right] \;, \nonumber\\
\phi_2^+ &=& \frac{4}{\rho^2} \left[\frac{c_3}{(\tau - \tau_0^-)^2} + \frac{c_4}{\tau - \tau_0^-} - \frac{c_3}{(\tau_0^-)^2} + \frac{c_4}{\tau_0^-}
\right] \;, \nonumber\\
k_1 &=& - \frac{2 \, \tau_0^+ \, ( \tau_0^+ + 2 \tau_0^-)}{ (\tau_0^+ - \tau_0^-)^4} \;\;\;,\;\;\;
k_2 = - \frac{ \tau_0^+ \, (\tau_0^+ + \tau_0^-) \, (\tau_0^+ + 4 \tau_0^-) }{(\tau_0^+ - \tau_0^-)^7} \;, \nonumber\\
\Sigma_1^- &=&  \frac{8 \, B }{D \, \rho^3 \, (\tau_0^+ - \tau_0^-)} \Big[
k_1 + \frac{c_1}{(\tau - \tau_0^+)^2} + \frac{c_2}{\tau- \tau_0^+} + \frac{\tau_0^- \, c_1}{(\tau_0^+ - \tau_0^-) \, (\tau- \tau_0^+)^2} \nonumber\\
&& + \frac{\tau_0^- \, c_2}{(\tau_0^+ - \tau_0^-) \, (\tau - \tau_0^+)} 
- \frac{\tau_0^- \, c_1}{(\tau_0^+ - \tau_0^-)^2 \, (\tau - \tau_0^+)} 
- \frac{\tau_0^+ \, c_4}{(\tau_0^+ - \tau_0^-) \, (\tau - \tau_0^+)} \nonumber\\
&& - \frac{\tau_0^+ \, c_3}{(\tau_0^+ - \tau_0^-)^2 \, (\tau - \tau_0^+)} - \frac{\tau_0^+}{(\tau - \tau_0^+)} \left( \frac{c_4}{\tau_0^-} - 
\frac{c_3}{(\tau_0^-)^2} \right)
\Big] \;, \nonumber\\
\Sigma_2^- &=& \frac{16 }{D \, \rho^4 } \Big[ k_2 + \frac{c_1\, c_3}{(\tau_0^+ - \tau_0^-)^2}
\left( \frac{1}{(\tau - \tau_0^+)^2} - \frac{2}{(\tau_0^+ - \tau_0^-) \, (\tau - \tau_0^+)} \right) \nonumber\\
&&+ \frac{c_1 \, c_4}{(\tau_0^+ - \tau_0^-) \, (\tau - \tau_0^+)^2} - 
\frac{c_1 \, c_4}{(\tau_0^+ - \tau_0^-)^2 \, (\tau - \tau_0^+)} 
+ \frac{c_2 \, c_3}{(\tau_0^+ - \tau_0^-)^2 \, (\tau - \tau_0^+)}\nonumber\\
&& +
\frac{c_2 \, c_4}{(\tau_0^+ - \tau_0^-) \, (\tau - \tau_0^+)} 
 + \left( - \frac{c_3}{(\tau_0^-)^2} + \frac{c_4}{\tau_0^-} \right) \left( \frac{c_1}{(\tau - \tau_0^+)^2} + \frac{c_2}{\tau - \tau_0^+}
\right) \Big] \;, \nonumber\\
\Gamma_1^+ &=&  \frac{8 \, B }{C \, D \, \rho^3 \, (\tau_0^+ - \tau_0^-)} \Big[
k_1 - \frac{c_3}{(\tau - \tau_0^-)^2} - \frac{c_4}{\tau - \tau_0^-} - \frac{\tau_0^- \, c_3}{(\tau - \tau_0^-)^3}
- \frac{\tau_0^- \, c_4}{(\tau - \tau_0^-)^2} \nonumber\\
&&+ \frac{\tau_0^- \, c_2}{(\tau_0^+ - \tau_0^-) \, (\tau - \tau_0^-)} 
 - \frac{\tau_0^- \, c_1}{(\tau_0^+ - \tau_0^-)^2 \, (\tau - \tau_0^-)} - \frac{\tau_0^+ \, c_3}{(\tau_0^+ - \tau_0^-) \, (\tau - \tau_0^-)^2} 
- \frac{\tau_0^+ \, c_3}{(\tau_0^+ - \tau_0^-)^2 \, (\tau - \tau_0^-)} \nonumber\\
&& - \frac{\tau_0^+ \, c_4}{(\tau_0^+ - \tau_0^-) \, (\tau -\tau_0^-)} 
- \frac{\tau_0^-}{(\tau - \tau_0^-)} \left(
\frac{c_3}{(\tau - \tau_0^-)^2} + \frac{c_4}{(\tau -\tau_0^-)} + \frac{c_4}{\tau_0^-} - \frac{c_3}{(\tau_0^-)^2}  \right)
\Big] \;, \nonumber\\
\Gamma_2^+ &=&  \frac{16 }{C \, D \, \rho^4 } \Big[
k_2 - \left(  \frac{c_3}{(\tau - \tau_0^-)^2} +  \frac{c_4}{\tau - \tau_0^-} \right)
\left( - \frac{c_3}{(\tau_0^-)^2} + \frac{c_4}{\tau_0^-} \right) -  \left(  \frac{c_3}{(\tau - \tau_0^-)^2} +  \frac{c_4}{\tau - \tau_0^-} \right)^2 \nonumber\\
&&- \frac{c_1 \, c_3}{(\tau_0^+ - \tau_0^-)^2} \left( \frac{1}{(\tau - \tau_0^-)^2} + \frac{2}{(\tau_0^+ - \tau_0^-) \, (\tau - \tau_0^-)} \right)
- \frac{c_1 \, c_4}{(\tau_0^+ - \tau_0^-)^2 \, (\tau - \tau_0^-)} \nonumber\\
&&+ \frac{c_2 \, c_3}{(\tau_0^+ - \tau_0^-)^2 \, (\tau - \tau_0^-)} 
+ \frac{c_2 \, c_3}{(\tau_0^+ - \tau_0^-) \, (\tau - \tau_0^-)^2} + \frac{c_2 \, c_4}{(\tau_0^+ - \tau_0^-) \, (\tau - \tau_0^-)}
\Big] \;.
\end{eqnarray}
Given these expressions, it can be checked in a direct manner that the above matrix expressions satisfy ${\cal M} \, M_+^{-1} = M_-$, and that $M_+^{-1} (\tau=0, x) = \mathbb{I}$.

We obtain for 
$M_-(\tau, x)$ at $\tau = \infty$,
\begin{eqnarray}
M(x) = M_- (\infty, x) = M^{(0)}_-(\infty, x) + \alpha \, \Delta_1^-(\infty, x) + \alpha^2 \, \Delta_2^-(\infty, x) \;,
\label{defMx}
\end{eqnarray}
where $M^{(0)}_-(\infty, x)$ denotes the undeformed solution \eqref{minftm}, and where
\begin{eqnarray}
\Delta_1^-(\infty, x)  &=&  \begin{pmatrix}
 \frac{8 \, B \, k_1 }{D \, \rho^3 \, (\tau_0^+ - \tau_0^-)} &\quad  - \frac{4}{\rho^2} \frac{(\tau_0^+ + \tau_0^-)}{(\tau_0^+ - \tau_0^-)^3} 
  &\quad  0
 \\
 \frac{4}{\rho^2} \frac{(\tau_0^+ + \tau_0^-)}{(\tau_0^+ - \tau_0^-)^3} 
& \quad 0
 & \quad  0
  \\
 0 &\quad 
0
   & \quad 0
\end{pmatrix} \;,\;
\Delta_2^- (\infty, x)= \begin{pmatrix}
 \frac{16 \, k_2 }{D \, \rho^4 }  &\quad  0 &\quad  0
 \\
 0 & \quad 0
 &  \quad 0
  \\
 0 &\quad  0 & \quad 0
\end{pmatrix} \;.
\nonumber\\
\label{alpha-ma}
\end{eqnarray}
Having determined $M(x)$ from factorization, we now verify that it satisfies the substitution rule
\eqref{substruleMM}. In the limit $\rho \rightarrow 0^+$ (with $v \neq 0$), we have
\begin{eqnarray}
 \frac{8 \, B \, k_1 }{D \, \rho^3 \, (\tau_0^+ - \tau_0^-)} &\longrightarrow &0 \;\;\;,\;\;\;
  \frac{16 \, k_2 }{D \, \rho^4 }\longrightarrow 0 \;\;\;,\;\;\;
\frac{4 }{\rho^2} \frac{(\tau_0^+ + \tau_0^-)}{(\tau_0^+ - \tau_0^-)^3}  \longrightarrow  - \frac{1}{v^2} \;,
\label{defbeh}
\end{eqnarray}
in agreement with \eqref{Mstmo2alp}. The terms proportional to $k_1$ and to $k_2$  are thus projected out
and do not appear in the monodromy matrix ${\cal M}(\omega)$.

The deformed solution exhibits the following interesting features. First note that although the deformation of monodromy
matrix \eqref{Mstmo2alp} is linear in $\alpha$, the spacetime solution \eqref{defMx} receives corrections that are of
order $\alpha^2$.  Second, using \eqref{psi2}, we have verified that $e^{\psi} =1$ continues
to hold in the presence of the deformation $\alpha$. Third, 
the deformation of the spacetime solution disappears in the limit $\rho \rightarrow \infty$, so that
in this limit, the solution describes an attractor background.  
For the case when $2 A D + B^2 =0$, which corresponds to a static attractor, 
we have verified that
the solution is well behaved in the quadrant $\rho > 0, v \geq 0$, and that is remains well
behaved in  the limit $\rho \rightarrow 0^+$, as long as  $v \neq 0$. 
At $\rho = v=0$, the solution breaks down, but so does
the canonical factorization.
This is depicted in the two figures:
figure 1 displays the runaway behaviour of the scalar field $e^{2\Sigma_2}$ near $\rho =0, v=0$,
while figure 2 displays the behaviour of the Ricci scalar $R$, which blows up at near  $\rho =0, v=0$.
Thus, the transformation $g(\omega)$ produces a deformation of the original
solution that generates a flow from the near-horizon solution 
describing an $AdS_2 \times S^2$ background towards a geometry that is singular
at $\rho=0, v=0$, in five-dimensional Einstein-Hilbert gravity.

\begin{figure}
  \caption{Runaway behaviour of the scalar field near $(\rho=0, v=0)$.}
  \centering
    \includegraphics[width=0.5\textwidth]{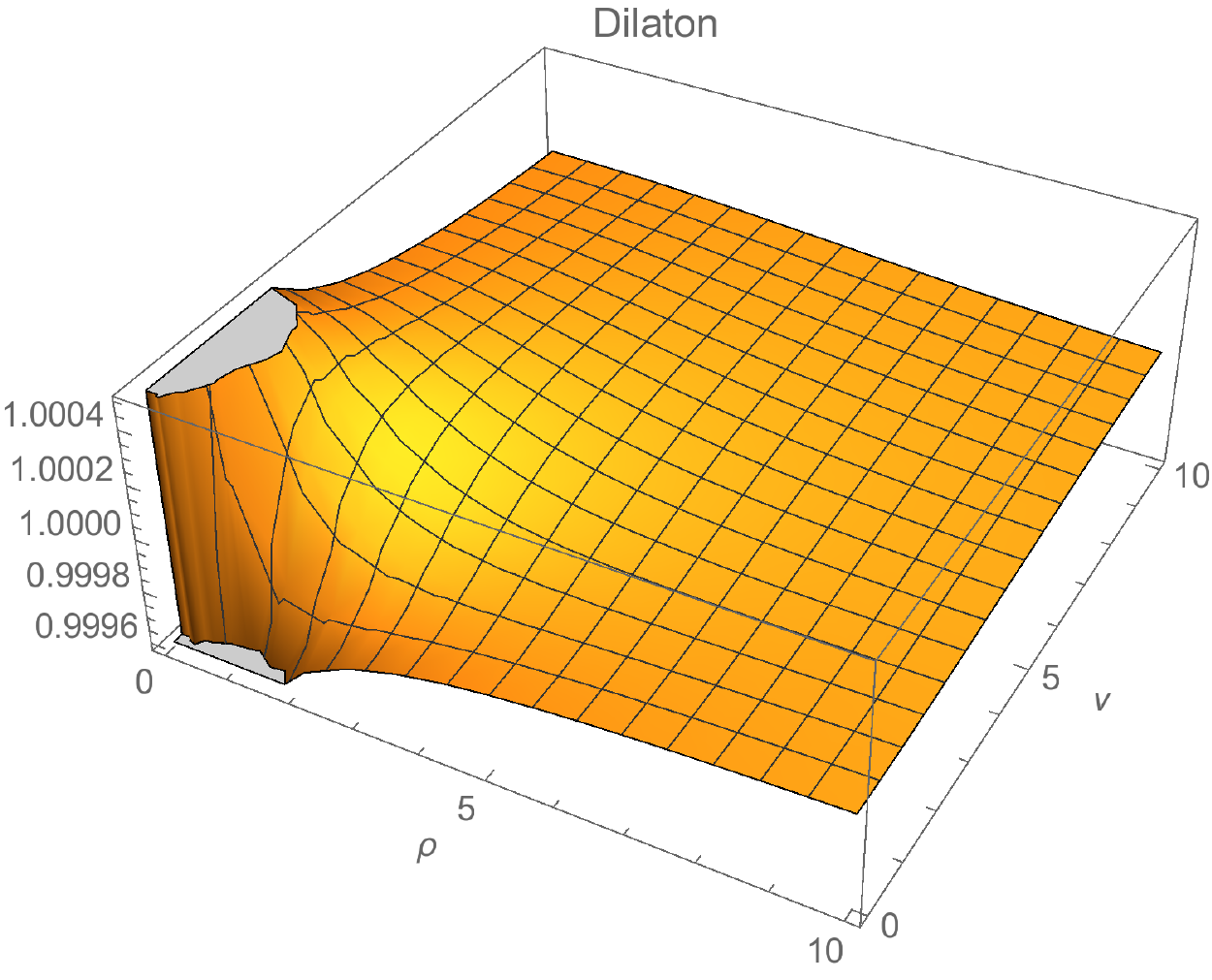}
\end{figure}

\begin{figure}
  \caption{Behaviour of Ricci scalar near $(\rho=0, v=0)$.}
  \centering
    \includegraphics[width=0.5\textwidth]{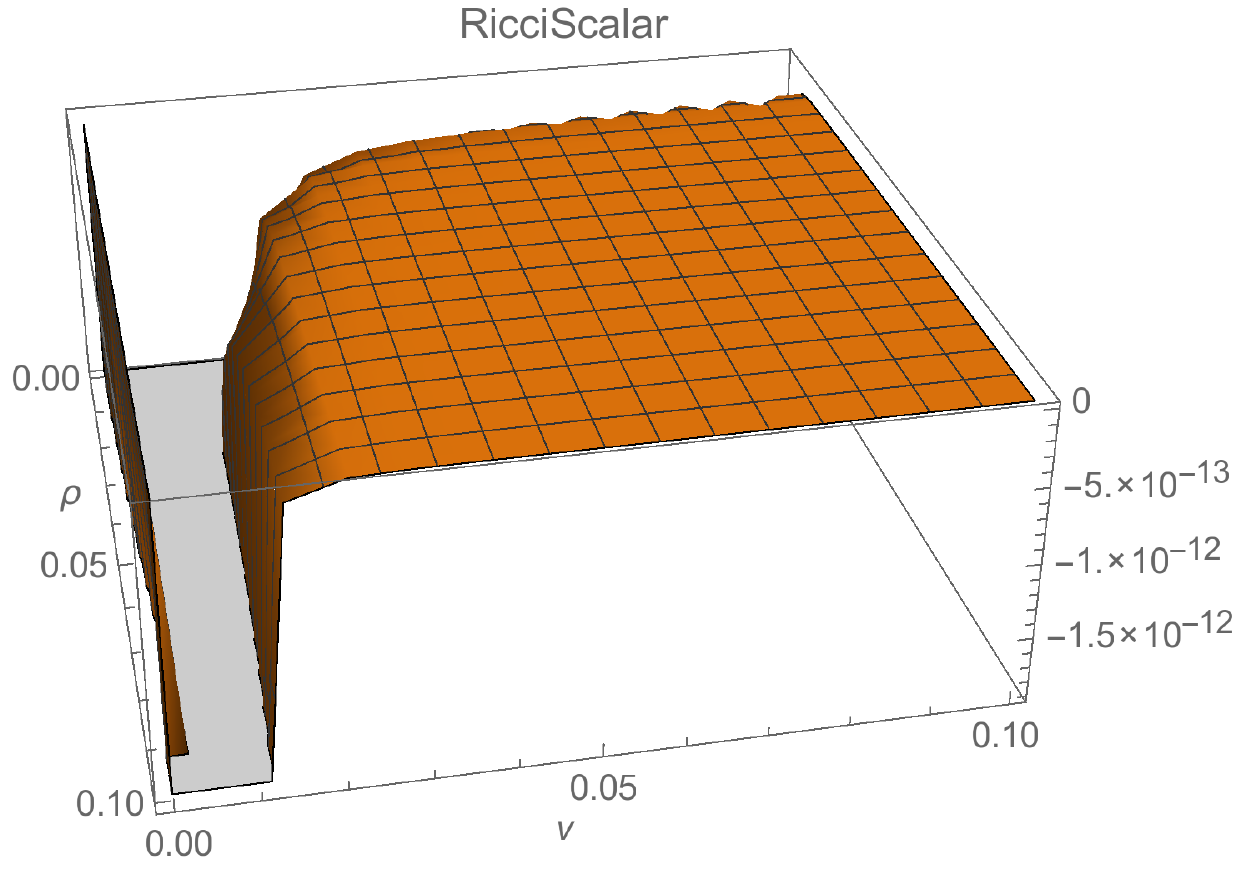}
\end{figure}

Let us further comment on the spacetime interpretation of $M(x)$ given in \eqref{defMx}. We consider the case $v > 0$.
Using \eqref{tvpos}, we obtain
\begin{eqnarray}
\frac{k_1}{\rho^2 }&=& - \frac{(\cos \theta -1) (3 \cos \theta + 1)}{8 \, r^2} \;, \nonumber\\
\frac{k_2}{\rho^4} &=& \frac{\cos \theta \, (\cos \theta -1) \, (5 \, \cos \theta + 3)}{64 \, r^4} \;,
\end{eqnarray}
and hence
\begin{eqnarray}
 \frac{8 \, B \, k_1 }{D \, \rho^3 \, (\tau_0^+ - \tau_0^-)} &=& \frac{B \, (\cos \theta -1) (3 \cos \theta + 1)}{2 \, D  \, r^3} \;,
 \nonumber\\
  \frac{16 \, k_2 }{D \, \rho^4 } &=&  \frac{\cos \theta \, (\cos \theta -1) \, (5 \, \cos \theta + 3)}{4 \, D \,  \, r^4} \;,
  \nonumber\\  
\frac{4 }{\rho^2} \frac{(\tau_0^+ + \tau_0^-)}{(\tau_0^+ - \tau_0^-)^3}  &=& - \frac{\cos \theta}{r^2} \;.
\end{eqnarray}
Introducing the Legendre polynomials $P_0(x) = 1, P_1(x) = x, P_2(x) = \frac12 (3 x^2 -1),  P_3 (x) = \frac12 (5x^3 -3x) $, we obtain
\begin{eqnarray}
 \frac{8 \, B \, k_1 }{D \, \rho^3 \, (\tau_0^+ - \tau_0^-)} &=& \frac{B \left(P_2 (\cos \theta) - P_1 (\cos \theta) \right) }{ D  \, r^3} \;,
 \nonumber\\
  \frac{16 \, k_2 }{D \, \rho^4 } &=&  \frac{  P_3 (\cos \theta) - \tfrac13 (2 P_2 (\cos \theta) + P_0 (\cos \theta)) }{2 \, D \,  \, r^4} \;.
\end{eqnarray}
Collecting all the terms in the first entry of \eqref{defMx} gives
\begin{eqnarray}
&& \frac{1}{r^2} \left[ A + \frac{(2AD + B^2)}{2D}\left(  P_1 (\cos \theta) - P_0 (\cos \theta) \right) \right] \nonumber\\
&&+  \frac{\alpha \, B}{D \, r^3} \, \left[  P_2 (\cos \theta) - P_1 (\cos \theta) \right]
 + \frac{\alpha^2 }{2 D \, r^4} \, \left[  P_3 (\cos \theta) - \tfrac13 (2 P_2 (\cos \theta) + P_0 (\cos \theta)) \right] \nonumber\\
  &&=  \frac{1}{r^2} \left[ - \frac{B^2}{2D} \, P_0 (\cos \theta)
    + \frac{(2AD + B^2)}{2D} \,   P_1 (\cos \theta)  \right]  \nonumber\\
&&+  \frac{\alpha \, B}{D \, r^3} \, \left[  P_2 (\cos \theta) - P_1 (\cos \theta) \right]
 + \frac{\alpha^2 }{2 D \, r^4} \, \left[  P_3 (\cos \theta) - \tfrac13 (2 P_2 (\cos \theta) + P_0 (\cos \theta)) \right] \nonumber\\
&&=   \frac{1}{D \, r^2} \left[ - \frac{B^2}{2} \, P_0 (\cos \theta) -  \frac{\alpha \, B}{ r} P_1 (\cos \theta)  - \frac{\alpha^2 }{3  \, r^2} \, P_2 (\cos \theta) \right] \\
&&+  \frac{1}{D \, r^2} \left[   \frac{(2AD + B^2)}{2D} \,   P_1 (\cos \theta)  
+   \frac{\alpha \, B}{ r} \, P_2 (\cos \theta) + \frac{\alpha^2 }{2 \,  r^2} \,  P_3 (\cos \theta) \right] +  \frac{\alpha^2 }{2  D \, r^4}  P_0 (\cos \theta) \;. \nonumber
   \end{eqnarray}
Curiously, this structure, although somewhat reminiscent of the series expansion (for $|\alpha| /r < 1$),
\be
\frac{1}{\sqrt{r^2 + \alpha^2 - 2 \alpha \, r \, \cos \theta}} = \frac{1}{r} \sum_{k=0}^{\infty} \left(\frac{\alpha}{r}\right)^k \, P_k (\cos \theta) \;,
\ee
associated with two-center solutions, is distinct from the latter.


\subsection*{Acknowledgements}

\noindent
This work was partially supported by FCT/Portugal through
UID/MAT/04459/2013, through 
grant EXCL/MAT-GEO/0222/2012 (G.L. Cardoso) and an associated visiting fellowship (T. Mohaupt), and
through FCT fellowship SFRH/BPD/101955/2014 (S. Nampuri).
This work was also supported by the COST action MP1210
"The String Theory Universe". 
The work of T. Mohaupt was partially supported by the STFC consolidated
grant ST/G00062X/1.
G.L.C. would like to thank the Max-Planck-Institut f\"ur
Gravitationsphysik (Albert-Einstein-Institute), the University of the Witwatersrand and Kvali Institute for the Physics and Mathematics of the Universe (IPMU) for kind hospitality during 
various stages of this
work. T.M. thanks CAMGSD of Instituto Superior T\' ecnico for kind
hospitality,
and Vicente Cort\'es for useful discussions. S.N. thanks Gary Gibbons for useful
discussions.
\begin{appendix}

\section{Symmetric spaces \label{backgroundLie}}

\subsection{Symmetric spaces:  signatures and classification}

We collect some remarks, based on \cite{Helgason,Gilmore}, about the signatures of the semi-Riemannian
metrics that are obtained on simple, real, connected
Lie groups $G$ (with real simple Lie algebra  $\mathfrak{g}$) 
and symmetric
spaces $G/H$, and as well about the classification of semi-Riemannian
symmetric spaces. The Killing form on $\mathfrak{g}$ is negative
definite on the linear span of the compact generators and 
positive definite on the linear span of the non-compact
generators. The difference between the numbers $N_n$ of 
linearly independent non-compact generators and $N_c$ of 
linearly independent compact generators is an invariant which 
is known as  the character $\chi(\mathfrak{g}) = N_n - N_c$
of the real simple Lie algebra $\mathfrak{g}$. Thus the character
encodes the signature of the Killing form and of the corresponding
semi-Riemannian metric on the group. In particular the metric
can only be (negative) definite if the group is compact.\footnote{
One usually multiplies the metric by an overall minus sign to 
have a positive definite metric.} It is also clear that 
symmetric spaces $G/H$ where $G$ is compact have a definite
metric, and the same is true if $G$ is non-compact and $H$
a maximal compact subgroup. Symmetric spaces with indefinite
metrics arise when $H$ is a maximal non-compact subgroup.

All real simple Lie algebras which have the same complexification
$\mathfrak{g}_{\mathbb{C}}$ can be related by `analytic continuation.'
More precisely, given a real simple Lie algebra and an involutive
automorphism, one has an eigenspace decomposition
\[
\mathfrak{g} = \mathfrak{h} \oplus \mathfrak{p} \;,
\]
and 
\[
\mathfrak{g}' = \mathfrak{h} \oplus i \mathfrak{p}
\]
is a real simple Lie algebra with the same complexification.
This `analytic continuation' is also known as the unitary 
trick. Starting from a given real simple Lie algebra all
Lie algebras with the same complexification can be obtained
in this way. Since these eigenspace decompositions are symmetric
decompositions, this is closely related to the classification
of symmetric spaces. Loosely speaking, the classification of
symmetric spaces is `the square' of the procedure which classifies
the real forms of a given complex simple Lie algebra.

One particular class of involutive automorphisms are 
Cartan involutions. An involution $\theta$ is called a
Cartan involution if
\[
B_\theta(Y,Z) : = - B_{\rm Killing}(Y,\theta Z) \;,\;\;\;
Y,Z \in \mathfrak{g}
\]
is positive definite. Therefore the Cartan decomposition,
\[
\mathfrak{g} = \mathfrak{h} \oplus \mathfrak{p} \;,
\]
that is the symmetric decomposition with respect to
a Cartan involution, coincides with an orthogonal decomposition 
of $\mathfrak{g}$ into complementary subspaces, such that the
restriction of the Killing form is negative and positive definite,
respectively. Hence, the symmetric decomposition coincides  
with the decomposition into compact and non-compact
generators.

Every real simple Lie algebra has a 
Cartan involution. For the compact real form, the identity is a Cartan 
involution.
For the real normal form, the Cartan involution acts as follows.
Choose as generators the standard generators
$H_j, E_\alpha, E_{-\alpha}$ of the complex form of the
algebra, and restrict to real linear combinations.
Here $j=1, \ldots, \mbox{rank}(\mathfrak{g})$ and
$\alpha$ are the positive roots. Then the Cartan 
involution acts by 
\begin{equation}
\label{CartanInvolution}
\theta \;: H_j \mapsto - H_j \;,\;\;\; E_\alpha \mapsto - E_{-\alpha}
\;,\;\;\;E_{-\alpha} \mapsto - E_{\alpha} \;.
\end{equation}
Thus generators $H_j, E_{\alpha} + E_{-\alpha}$ of the $(-1)$-eigenspace
span the positive definite, 
non-compact directions $\mathfrak{p}$ , and the generators 
of the $(+1)$-eigenspace $E_\alpha - E_{-\alpha}$
span the positive definite,  compact directions $\mathfrak{h}$ 
in the Cartan decomposition 
\[
\mathfrak{g} = \mathfrak{h} \oplus \mathfrak{p}
\]
of the normal real form $\mathfrak{g}$. Applying the unitary 
trick amounts to multiplying the non-compact generators by $i$ 
and brings us from the normal real 
form to the compact real form
\[
\mathfrak{g}_c = \mathfrak{h} \oplus i \mathfrak{p} \;.
\]

\subsection{The space $SL(2,\mathbb{R})/SO(2)$ \label{slso}}

Dimensional reduction of pure gravity from four to three dimensions
leads to the non-compact, definite Riemannian symmetric space
$M = G/H = SL(2,\mathbb{R})/SO(2)$. The underlying symmetric
pair $(\mathfrak{g}, \mathfrak{h}) = (\mathfrak{sl}_2(\mathbb{R}),
\mathfrak{so}_2)$ of Lie algebras is associated to the Cartan
decomposition
\[
\mathfrak{g} = \mathfrak{h} \oplus \mathfrak{p} 
\]
of the normal real form $\mathfrak{g} = \mathfrak{sl}_2(\mathbb{R})$ 
of the complex simple Lie algebra $A_1$. Since $H=SO(2)$ is
an orthogonal group, the generalized transposition \eqref{gentrans} at the Lie algebra
level 
 is given by
matrix transposition, $Z^\natural = Z^T$, and the Cartan decomposition
coincides with the decomposition of matrices into a symmetric
and an antisymmetric part. Matrices $Z\in \mathfrak{h}$ in the 
$(+1)$-eigenspace of the Cartan involution 
are antisymmetric, generate compact one-parameter subgroups, 
and form the negative definite eigenspace
of the Killing form, while matrices in the $(-1)$-eigenspace
$\mathfrak{p}$ are symmetric, generate non-compact one-parameter subgroups, 
and form the positive
definite eigenspace of the Killing form.

An explicit choice of generators adapted to the Cartan decomposition 
is given by
\begin{eqnarray}
Y^1 = \begin{pmatrix}
1 & 0 \\
0 & -1
\end{pmatrix} \;\;\;,\;\;\;
Y^2 = \begin{pmatrix}
0 & 1 \\
1 & 0
\end{pmatrix} \;\;\;,\;\;\;
Y^3 = \begin{pmatrix}
0 & 1 \\
-1 & 0
\end{pmatrix} \;,
\end{eqnarray}
with $Y^1, Y^2 \in \mathfrak{p}$ and $Y^3 \in \mathfrak{h}$.
The commutation relations
\begin{eqnarray}
[Y^1, Y^2] = 2 Y^3\in \mathfrak{h}  \;\;\;,\;\;\; 
[Y^1, Y^3] = 2 Y^2 \in \mathfrak{p}  \;\;\;,\;\;\; 
[Y^2, Y^3] = - 2 Y^1 \in \mathfrak{p}\;,
\end{eqnarray}
show explicitly that this is a symmetric decomposition
(\ref{Sym_Decomp}).
To check that this is a Cartan decomposition, i.e. 
that we obtain a decomposition into eigenspaces of the Killing form,
we use the relation
\[
B(Y,Z) = 2n \mbox{tr}(YZ)
\]
between the Killing form of $SL(n,\mathbb{R})$ and the 
trace in the fundamental representation, and compute
\[
\mbox{tr}(Y^1 Y^1) = \mbox{tr}(Y^2Y^2) = 2 \;,\;\;\;
\mbox{tr}(Y^3Y^3) = - 2 \;,\;\;\;
\mbox{tr}(Y^iY^j) = 0 \;,\;\;i\not=j\;.
\]

The coset space $G/H=
SL(2,\mathbb{R})/SO(2)$ can be identified with $\exp(\mathfrak{p})$. 
This leads to a parametrization of the form
\[
\exp\left( a Y^1 + b Y^2  \right)
\;, \mbox{where}\;\; a,b \in \mathbb{R} \;
\]
in terms of symmetric matrices. 
As discussed before, a more convenient parametrization is obtained by using
the Iwasawa decompostion $G=HL$ of the simple non-compact real Lie group 
$G=SL(2,\mathbb{R})$ 
into the maximal compact subgroup $H=SO(2)$ and a triangular
subgroup $L$. Since $SL(2,\mathbb{R})$ is the normal real form of 
the complex simple Lie group $SL(2,\mathbb{C})$, $L$ is in fact
an Iwasawa subgroup. 
To find $L$ explicitly, we have to identify a maximal triangular Lie subalgebra
$\mathfrak{l} \subset \mathfrak{g}$ and exponentiate it,
$L = \exp(\mathfrak{l})$. 
By inspection of the basis $(Y^1, Y^2, Y^3)$, we find that the 
subalgebra generated 
$Y^1$ and $Y^2-Y^3$ is triangular, and obviously maximal. 
Note in particular that $(Y^2-Y^3)^2=0$, which simplifies 
exponentiation.\footnote{
Since $Y^2-Y^3$ is triangular with vanishing diagonal, it generates
a nilpotent Lie subalgebra $\mathfrak{n}$ with associated 
nilpotent Lie subgroup $N$. Together with the abelian Lie subgroup
$A$ generated by $Y^1$, and a compact generator for $H$ we obtain
the refinement $G=HAN$ of the Iwasawa decomposition.
} 
Using this
we obtain the following convenient parametrisation 
of $SL(2,\mathbb{R})/SO(2)$ by the Iwasawa subgroup $L$:
\begin{equation}
V = e^{\tfrac12 \ln \Delta \, Y^1} \, e^{\tfrac{B}{2 \Delta} (Y^2 - Y^3)} =
\begin{pmatrix}
\Delta^{1/2} & \quad 0 \\
B \, \Delta^{-1/2} & \quad \Delta^{-1/2}
\end{pmatrix} \in L\subset SL(2,\mathbb{R}) \;.
\end{equation}
As discussed before, we can then obtain a description in 
terms of elements of $\exp(\mathfrak{p})$, which for the
case at hand are symmetric matrices, by the standard
embedding
\[
L \ni V \mapsto M :\ V^T V \in \exp \mathfrak{p} = G/H \subset G \;.
\]
Here we use that $V^TV \in G$ is $\natural$-symmetric and
therefore lies in $\exp(\mathfrak{p})$. 
For $G=SL(2,\mathbb{R})$ we obtain the $\natural$-symmetric representative:
\begin{equation}
M= \begin{pmatrix}
\Delta + B^2/\Delta & \quad B/\Delta\\
B/\Delta & \quad \Delta^{-1}
\end{pmatrix} \;.
\label{Msl2}
\end{equation}
 
\subsection{The space $SL(3,\mathbb{R})/SO(2,1)$}

Dimensional reduction of pure five-dimensional gravity to three
dimensions leads to the Riemannian
symmetric spaces $SL(3,\mathbb{R})/SO(3)$ or  
$SL(3,\mathbb{R})/SO(2,1)$, depending on whether the reduction 
is purely spacelike, or includes time. According to \cite{Chakrabarty:2014ora}
Appendix A1, one can choose a Cartan-Weyl basis for
the normal real form $\mathfrak{sl}_3(\mathbb{R})$ of the simple
complex Lie algebra $A_2$, which takes the following form:
\begin{equation}
\label{A_2}
H_1 = \frac{1}{\sqrt{3}}
\left( \begin{array}{ccc}
1 & 0 & 0 \\
0 & -2& 0 \\
0 & 0 & 1 \\
\end{array} \right) \;,\;\;\;
H_2 = 
\left( \begin{array}{ccc}
1 & 0 & 0 \\
0 & 0 & 0 \\
0 & 0 & -1 \\
\end{array} \right) \;,\;\;\;
E_1 = 
\left( \begin{array}{ccc}
0 & 0 & 0 \\
0 & 0 & 1 \\
0 & 0 & 0 \\
\end{array} \right) \;,\;\;\;
\end{equation}
\[
E_2 = 
\left( \begin{array}{ccc}
0 & 1 & 0 \\
0 & 0 & 0 \\
0 & 0 & 0 \\
\end{array} \right) \;,\;\;\;
E_3 = [E_2, E_1] \;,\;\;\; F_i = E_i^T \;,\;\;i=1,2,3\;.
\]
This choice of generators corresponds to the following choice 
of simple roots of $A_2$:
\[
\alpha_1 = (-\sqrt{3}, 1) \;,\;\;
\alpha_2 = (\sqrt{3}, 1 ) \;,
\]
with highest root $\alpha_3 = \alpha_1+\alpha_2 = (0,2)$. 
The normal real form $\mathfrak{sl}_3(\mathbb{R})$ of $A_2$,
is spanned by the real linear combinations of $H_1, H_2, E_1, E_2,
E_3, F_1, F_2, F_3$.

As for any group of the form $SL(n,\mathbb{R})$,
the Cartan involution acts by transposition times $(-1)$:
\[
H_i \mapsto - H_i \;,\;\;\;i=1,2 \;,\;\;\;
E_k \mapsto - F_k \;,\;\;\; F_k \mapsto - E_k \;, k=1,2,3\;.
\]
The associated $\natural$-transposition is again matrix transposition.
The Cartan decomposition is 
\[
\mathfrak{sl}_3(\mathbb{R}) = \mathfrak{so}(3) \oplus \mathfrak{p} \;.
\]
The three generators of the $(+1)$-eigenspace can be chosen
to be the antisymmetric matrices
\[
E_k - F_k \;, \;\;\;k=1,2,3\;.
\]
As generators for the $(-1)$-eigenspace $\mathfrak{p}$ 
we can choose the symmetric matrices
\[
H_i \;, E_k + F_k  \;,\;\;\;i=1,2, k=1,2,3\;.
\]
We can then parametrize $SL(3,\mathbb{R})/SO(3)$ in the same
way as $SL(2,\mathbb{R})/SO(2)$, which we discussed in \ref{slso}.

However, when we want to generate stationary solutions of 
five-dimensional gravity, one of the dimension we reduce over
is time, and we have instead to deal with the 
indefinite symmetric space $SL(3,\mathbb{R})/SO(2,1)$.
This can be done by choosing a different involution and 
associated $\natural$-transposition. 
The difference between
$\mathfrak{so}_{3}$ and $\mathfrak{so}_{2,1}$ is that one preserves 
a positive definite symmetric bilinear form, while the other 
preserves a symmetric bilinear form of signature $(2,1)$:
\[
Z \in \mathfrak{so}_3 \Rightarrow Z^T + Z = 0 \;,\;\;\;
Z \in \mathfrak{so}_{2,1} \Rightarrow Z^T \eta + \eta Z = 0
\]
where $\eta$ is symmetric $3\times 3$ matrix of signature $(2,1)$.
Following \cite{Chakrabarty:2014ora}, one can take the standard diagonal Gram matrix
$\eta=\mbox{diag}(1,\epsilon_2,\epsilon_1)$ where $\epsilon_i=\pm 1$. 
In the main part of the text we took $\eta=\mbox{diag}(1,-1,1)$.
The condition defining 
$\mathfrak{so}_{2,1}$ can be viewed as a generalized
antisymmetry condition, $(\eta Z)^T = - \eta Z$, and the 
corresponding involutive  Lie algebra automorphisms is
\[
\theta \;:\;\;\; Z \rightarrow  - \eta Z^T \eta \;,
\]
because Lie algebra elements with $(\eta Z)^T = \pm \eta Z$
have eigenvalues $\mp 1$ under $\theta$. Upon 
exponentiation one obtains the involutive Lie group automorphism 
\[
\Theta\;: g = \exp(Z) \mapsto \Theta(g) = \exp( - \eta Z^T \eta) \;.
\]
The corresponding generalized transposition at group level is 
\[
\natural\;: g \mapsto g^\natural= \Theta( g^{-1} ) = \exp( \eta Z^T \eta )
= \exp( \eta Z \eta)^T \;,
\]
while at Lie algebra level it is
\[
Z^\natural = - \theta(Z) = \eta Z^T \eta \;.
\]
Taking $\eta = \mbox{diag}(1, \epsilon_2, \epsilon_1)$, with
any choice of  
$\epsilon_i$ such that $\eta$ has indefinite signature,  
the involution induces a decomposition 
\[
\mathfrak{sl}_3(\mathbb{R}) = \mathfrak{so}_{2,1} \oplus \mathfrak{p}'
\]
with $\exp(\mathfrak{p}') \simeq SL(3,\mathbb{R})/SO(2,1)$. 
The explicit operation of the involution  
on the Cartan-Weyl generators of $\mathfrak{sl}_3(\mathbb{R})$ is 
\[
H_i \mapsto - H_i \;,\;\;\;
E_1 \mapsto -\epsilon_1 \epsilon_2 F_1 \;,\;\;\;
E_2 \mapsto - \epsilon_2 F_2 \;,\;\;\;
E_3 \mapsto -\epsilon_1 F_3 \;. 
\]
It is convenient to parametrize the symmetric spaces using
an Iwasawa subgroup of $SL(3,\mathbb{R})$. The standard
Iwasawa subgroup $L \subset SL(3,\mathbb{R})$ is generated 
by $H_1, H_2, E_1, E_2, E_3$ (which obviously is upper triangular).
Then elements of $L$ can be parametrized as
\[
V = g_1 g_2 g_3 g_4 g_5 = 
\exp(\frac{1}{2} \phi_1 H_1 ) 
\exp( \frac{1}{2} \phi_2 H_2 ) 
\exp(\chi_1 E_1)
\exp(\chi_2 E_2)
\exp(\chi_3 E_3) \;,
\]
where $g_i$ denote the elements of 
the five one-parameter subgroups associated with the generators.
With the explicit choice of generators of \cite{Chakrabarty:2014ora}
we obtain:
\begin{eqnarray}
V = \begin{pmatrix}
e^{\Sigma_1}  & \quad e^{\Sigma_1} \, \chi_2 & \quad e^{\Sigma_1} \, \chi_3 \\
0 & \quad e^{\Sigma_2} & \quad e^{\Sigma_2}\, \chi_1 \\
0 & \quad 0 & \quad e^{\Sigma_3}
\end{pmatrix} \;,
\end{eqnarray}
with
\begin{eqnarray}
\Sigma_1 &=& \frac12 \left( \frac{1}{\sqrt{3}} \phi_1 + \phi_2 \right) \;, \nonumber\\
\Sigma_2 &=& - \frac{1}{\sqrt{3}} \phi_1 \;, \nonumber\\
\Sigma_3 &=& \frac12 \left( \frac{1}{\sqrt{3}} \phi_1 - \phi_2 \right) \;, 
\label{sigphi}
\end{eqnarray}
which satisfies $\Sigma_1 + \Sigma_2 + \Sigma_3 =0$.

Note that $V$ can be used to parametrize both $SL(3,\mathbb{R})/SO(3)$ 
and (an open subset of) $SL(3,\mathbb{R})/SO(2,1)$:
\begin{enumerate}
\item
$L$ acts simply transitively on $SL(3,\mathbb{R})/SO(3)$ and thus
can be globally identified with it. By
\[
L \ni V \mapsto M = V^T V \in G
\]
we obtain a group element invariant under matrix transposition, and
odd under the Cartan involution, which therefore lives in the
subspace $\exp \mathfrak{p} \subset SL(3,\mathbb{R})$. 
\item
$L$ acts with open orbit on $SL(3,\mathbb{R})/SO(2,1)$, and thus can
be identified with an  open part of the coset. By 
\[
L \ni V \mapsto M = V^\natural V \in G \;,
\]
where $V^\natural$ is the generalized transposition under which
$\mathfrak{so}_{2,1}$ is antisymmetric, we obtain an element of
$SL(3,\mathbb{R})$ which is an exponential of an element of the
subspace $\mathfrak{p}'$ complementary to $\mathfrak{so}_{2,1}$.  
\end{enumerate}

We are interested in the coset $SL(3,\mathbb{R})/SO(2,1)$ and
therefore compute $V^{\natural} = g_5^{\natural} \, g_4^{\natural} \, g_3^{\natural} \, g_2^{\natural} \, g_1^{\natural} $
to obtain
\begin{eqnarray}
V^{\natural} = \begin{pmatrix}
e^{\Sigma_1}  & \quad 0 & \quad 0 \\
- e^{\Sigma_1} \, \chi_2  & \quad  e^{\Sigma_2} & \quad 0 \\
e^{\Sigma_1} \, \chi_3 &\quad  - e^{\Sigma_2} \, \chi_1  &\quad  e^{\Sigma_3} 
\end{pmatrix} \;.
\end{eqnarray}
The corresponding $\natural$-symmetric representative is:
\begin{eqnarray}
M = V^{\natural} \, V = \begin{pmatrix}
e^{2\Sigma_1}   & \quad  e^{2\Sigma_1} \, \chi_2  & \quad e^{2\Sigma_1} \, \chi_3 \\
- e^{2\Sigma_1} \, \chi_2  & \quad - e^{2\Sigma_1} \, \chi_2^2 +  e^{2\Sigma_2}
& \quad  - e^{2\Sigma_1} \, \chi_2 \, \chi_3 +  e^{2\Sigma_2} \, \chi_1\\
e^{2\Sigma_1} \, \chi_3 & \quad e^{2\Sigma_1} \, \chi_2 \, \chi_3 -  e^{2\Sigma_2} \, \chi_1
 & \quad - e^{2\Sigma_2} \, \chi_1^2 +  e^{2\Sigma_1} \, \chi_3^2 + e^{2\Sigma_3}
\end{pmatrix} \;.
\label{cosetrepM}
\end{eqnarray}

\subsection{The space $SU(2,1)/(   SL(2,\mathbb{R}) \times U(1)    )$   }

Dimensional reduction of four-dimensional Einstein-Maxwell theory
to three dimensions leads to the Riemannian symmetric
spaces $SU(2,1)/(SU(2)\times U(1))$ for space-like and
$SU(2,1)/(SL(2,\mathbb{R})\times U(1))$ for time-like reduction.
The Lie algebra $\mathfrak{g}=\mathfrak{su}(2,1)$ is a non-compact
real form of $A_2$, which is not isomorphic to the normal real
from $\mathfrak{sl}(3,\mathbb{R})$. It can be obtained starting 
from the compact form $\mathfrak{su}(3)$ or from the normal 
real form by making the appropriate symmetric decomposition and
applying the unitary trick.

We will use the parametrisation given by \cite{Chakrabarty:2014ora}
Appendix A2. The pseudo-unitary group $SU(2,1)$ 
is characterized by preserving
a hermitian sesquilinear form of signature $(++-)$. Denoting 
the Gram matrix of the hermitian form by $\kappa$, this means that
\be
g^\dagger \kappa g = \kappa
\label{gkgk}
\ee
at the group level and 
\[
Z^\dagger \kappa + \kappa Z = 0 \Leftrightarrow (\kappa Z)^\dagger = -
\kappa Z
\]
at the Lie algebra level. The explicit choice made in \cite{Chakrabarty:2014ora}
is
\[
\kappa = \left( \begin{array}{ccc}
0&0&-1\\
0&1&0\\
-1&0&0\\
\end{array} \right) \;.
\]
With this choice, a basis of $\mathfrak{su}(2,1)$ is obtained
by taking the following linear combinations of standard generators
$H_1, H_2, E_1, E_2, E_3, F_1, F_2, F_3$ 
of $A_2$, which were given explicity in (\ref{A_2}):
\[
i\sqrt{3} H_1, H_2, E_1 + E_2, F_1 + F_2, i(E_2 - E_1), i(F_2-F_1), iE_3,
iF_3 \;.
\]
Here $i\sqrt{3}H_1, H_2$ are Cartan generators of $\mathfrak{su}(2,1)$,
while $E_1+E_2, i(E_2-E_1), iE_3$ correspond to the positive roots
and generate a nilpotent subalgebra $\mathfrak{n}_+$, 
and $F_1+F_2, i(F_2-F_1), i F_3$ correspond to the negative roots,
and generate a nilpotent subalgebra $\mathfrak{n}_-$.

Now one can look at symmetric decompositions of
$\mathfrak{su}_{2,1}$. The maximal compact subalgebra is 
$\mathfrak{su}_2 \oplus \mathfrak{u}_1$, resulting from 
the Cartan decomposition
\[
\mathfrak{su}_{2,1} = (\mathfrak{su}_2 \oplus \mathfrak{u}_1) 
\oplus \mathfrak{p} \;.
\]
Since the Cartan-invariant subalgebra is a unitary Lie algebra, 
the generalized transposition acts as hermitian conjugation on matrices:
\[
Z \in  \mathfrak{p} \Leftrightarrow 
(\kappa Z)^\dagger = - \kappa Z \;,\;\;\;\mbox{and}\;\;\;
Z^\dagger = Z \;.
\]
The decomposition 
\[
\mathfrak{su}_{2,1} = (\mathfrak{sl}_2(\mathbb{R}) \oplus \mathfrak{u}_1) 
\oplus \mathfrak{p}'
\]
is obtained by replacing Hermitian conjugation by 
the generalized transposition
$Z^\natural = \eta Z^\dagger \eta^{-1}$, where $\eta$ is the diagonal
matrix with entries $(1,-1,1$)  \cite{Chakrabarty:2014ora}:
\[
Z \in  \mathfrak{p}' \Leftrightarrow
(\kappa Z)^\dagger = - \kappa Z \;,\;\;\;\mbox{and}\;\;\;
(\eta Z)^\dagger = \eta Z \;.
\]
We will not need explicit generators for $\mathfrak{p}'$. Instead
we again work with the Lie algebra $\mathfrak{l}$ of the
triangular subgroup $L$ appearing in the Iwasawa decomposition of 
$SU(2,1)$. One possible choice is
the standard Borel subalgebra
$\mathfrak{b}_+ = \langle H_2 \rangle \oplus \mathfrak{n}_+$,
obtained by combining the non-compact Cartan generator $H_2$ with 
the generators  corresponding to the positive roots \cite{Houart:2009ed}. 
In the chosen basis, the generators are
\[
H_2 = \left( \begin{array}{ccc}
1&0&0\\
0&0&0\\
0&0&-1\\
\end{array} \right) \;,\;\;\;
E_1 + E_2 = \left(\begin{array}{ccc}
0&1&0\\
0&0&1\\
0&0&0\\
\end{array} \right) 
\;,\;\;\;
\]
\[
i(E_2-E_1) = \left( \begin{array}{ccc}
0&i&0\\
0&0&-i\\
0&0&0\\
\end{array} \right) \;,\;\;\;
i E_3 = \left( \begin{array}{ccc}
0&0&i\\
0&0&0\\
0&0&0\\
\end{array} \right) \;.
\]
By exponentiation we obtain the following one-parameter subgroups:
\begin{eqnarray}
g_1 &=&\exp(\frac{1}{2} \varphi H_2) = \left( \begin{array}{ccc}
e^{\varphi/2} & 0 & 0 \\
0 & 1 & 0 \\
0 & 0 & e^{-\varphi/2} \nonumber\\
\end{array} \right) \;, \\
g_2 &=& \exp(\sqrt{2} \chi_2 (E_1 + E_2) ) = \left( \begin{array}{ccc}
1 & \sqrt{2} \chi_e & \chi_e^2 \\
0 & 1 & \sqrt{2} \chi_e \\
0 & 0 & 1 \\
\end{array} \right) \;, \nonumber\\
g_3 &=& \exp(i \sqrt{2} \chi_m (E_2 - E_1)) = \left( \begin{array}{ccc}
1 & i \sqrt{2} \chi_m & \chi_m^2 \\
0 & 1 & -i \sqrt{2} \chi_m \\
0 & 0 & 1 \\
\end{array} \right) \;, \nonumber\\
g_4 &=& \exp(\sqrt{2} \lambda i E_3) = \left( \begin{array}{ccc}
1 & 0 & i \sqrt{2} \lambda \\
0 & 1 & 0 \\
0 & 0 & 1 \\
\end{array} \right) \;.
\label{g14su2}
\end{eqnarray}
From this we obtain a triangular parametrisation of coset elements by 
computing $V = g_1 \, g_2 \, g_3 \, g_4$ and obtain
\begin{eqnarray}
V = \begin{pmatrix}
e^{\varphi/2}  &\quad  \sqrt{2} \,e^{\varphi/2} Z &\quad  \,e^{\varphi/2} \, \Sigma \\
0 &\quad  1 & \quad \sqrt{2} \, \bar{Z} \\
0 & \quad 0 &\quad  e^{-\varphi/2} 
\end{pmatrix} \;,
\end{eqnarray}
with
\begin{eqnarray}
Z &=& \chi_e + i \chi_m \;, \nonumber\\
\Sigma &=& \chi_e^2 + \chi_m^2  + i \left( \sqrt{2} \lambda - 2 \chi_e \chi_m \right) \;.
\label{sigmZ}
\end{eqnarray}
Next we compute $g_i^\natural = \eta g_i^\dagger \eta^{-1}$:
\[
g_1^\natural= \left( \begin{array}{ccc}
e^{\varphi/2} & 0 & 0 \\
0 & 1 & 0 \\
0 & 0 & e^{-\varphi/2} \\
\end{array} \right) \;,\;\;\;
g_2^\natural = \left( \begin{array}{ccc}
1&0&0\\
-\sqrt{2}\chi_e &1 & 0 \\
\chi_e^2 & - \sqrt{2} \chi_e & 1 \\
\end{array} \right) \;,\;\;\;
\]
\[
g_3^\natural = \left( \begin{array}{ccc}
1&0&0\\
i\sqrt{2}\chi_m &1 & 0 \\
\chi_m^2 & - i \sqrt{2} \chi_m & 1 \\
\end{array} \right) \;,\;\;\;
g_4^\natural = \left( \begin{array}{ccc}
1 & 0 & 0 \\
0 & 1 & 0 \\
-i \sqrt{2} \lambda & 0 & 1 \\
\end{array} \right) \;.
\]
From this we obtain 
$V^{\natural} = g_4^{\natural} \, g_3^{\natural} \, g_2^{\natural} \, g_1^{\natural} $:
\begin{eqnarray}
V^{\natural} = \begin{pmatrix}
e^{\varphi/2}  & \quad 0 & \quad 0 \\
-\sqrt{2} \,e^{\varphi/2} \bar{Z}  & \quad 1 & \quad 0 \\
e^{\varphi/2} \, \bar{\Sigma} &\quad  - \sqrt{2} \, Z  & \quad e^{-\varphi/2} 
\end{pmatrix} \;.
\end{eqnarray}
The $\natural$-symmetric coset representative is therefore
\begin{eqnarray}
M = V^{\natural} \, V = \begin{pmatrix}
e^{\varphi}  & \quad \sqrt{2} \,e^{\varphi} {Z} & \quad e^{\varphi} \, {\Sigma} \\
-\sqrt{2} \,e^{\varphi} \bar{Z}  & \quad 1 - 2 e^{\varphi} \, |Z|^2 & \quad  \sqrt{2} \, \bar{Z} \left(1 - e^{\varphi} \, \Sigma \right)  \\
e^{\varphi} \, \bar{\Sigma} &\quad  - \sqrt{2} \, Z \left(1 - e^{\varphi} \, \bar{\Sigma} \right)  & \quad e^{\varphi} \, |\Sigma|^2 - 2 
|Z|^2 + e^{-\varphi} 
\end{pmatrix} \;.
\label{cosetrepMsu21}
\end{eqnarray}
Note that our $V$ is different from the representative
\[
V' = \exp( \frac{\varphi}{2} H_2 ) \exp ( \sqrt{2} \chi_e (E_1+E_2)
+ \sqrt{2} \chi_m i (E_2-E_1) + \sqrt{2} \lambda i E_3) 
\]
used by \cite{Houart:2009ed} and
\cite{Chakrabarty:2014ora},
which has $\Sigma$ replaced by $\Sigma' = \chi_e^2 + \chi_m^2 + i 
\sqrt{2} \lambda$ (without the additional term $-2i \chi_e \chi_m$).

\end{appendix}


\providecommand{\href}[2]{#2}\begingroup\raggedright\endgroup

\end{document}